\documentclass[twocolumn,notitlepage,superscriptaddress,tightenlines,nofootinbib]{revtex4-1}
\usepackage{amsmath,verbatim,latexsym,amssymb,graphicx,indentfirst,mathrsfs,mathtools,amsthm,bbm,bm,url,cancel,hyperref,subcaption,enumerate,enumitem}
\usepackage[font=small,labelfont=bf,format=plain,justification=raggedright,singlelinecheck=false]{caption}

\usepackage{verbatim,indentfirst}
\usepackage[title,titletoc]{appendix}

\usepackage{graphicx}
\usepackage{epstopdf}
\newcommand{\caphead}[1]{{\bf #1}}

\newtheoremstyle{indented}{5pt}{3pt}{\addtolength{\leftskip}{3.5em}}{}{\bfseries}{.}{.5em}{}
\theoremstyle{indented}

\makeatletter
\newcommand\footnoteref[1]{\protected@xdef\@thefnmark{\ref{#1}}\@footnotemark}
\makeatother

\usepackage{soul}

\usepackage{color}
\usepackage[dvipsnames]{xcolor}
\usepackage[normalem]{ulem}

\newcommand{\VeryLoc}{<} 
\newcommand{\Loc}{>} 
\newcommand{\HDim}{\mathcal{N}} 
\newcommand{\Sites}{N}

\newcommand{\Power}{ \mathscr{P} } 
\newcommand{\In}{ {\rm in} } 
\newcommand{\diab}{{\rm diab}}

\newcommand{\MBL}{{\rm MBL}}
\newcommand{\ETH}{{\rm GOE}}
\newcommand{\Anderson}{ {\rm And} }

\newcommand{\Otto}{{\rm Otto}}
\newcommand{\QHO}{{\rm QHO}}

\newcommand{\therm}{{\rm th}}
\newcommand{\coupling}{g}

\newcommand{\Carnot}{\text{Carnot}} 
\newcommand{\Wb}{W_{\rm b}}  
\newcommand{\HTemp}{{\rm H}} 
\newcommand{\CTemp}{ {\rm C} }  
\newcommand{\THot}{T_\HTemp}  
\newcommand{\TCold}{T_\CTemp}  
\newcommand{\betaH}{\beta_\HTemp}  
\newcommand{\betaC}{\beta_\CTemp}  
\newcommand{\gap}{ \Delta }
\newcommand{\HScale}{\mathcal{E}}

\newcommand{\HalfLZ}{\text{frac-LZ}}

\newcommand{\disp}{ {\rm displ} }  
\newcommand{\qubit}{{\rm qubit}}  
\newcommand{\meso}{{\rm meso}}  
\newcommand{\Sim}{{\rm sim}}  
\newcommand{\ZH}{Z}    

\newcommand{\dAvg}{\expval{\delta}}

\newcommand{\macro}{{\rm macro}}
\newcommand{\deltaMBL}{\delta_-}
\newcommand{\Sys}{S}
\newcommand{\J}{ \mathcal{J} }
\newcommand{\JFar}{\J_{L \gg \xi}}
\newcommand{\JClose}{\J_{L \leq \xi}}
\newcommand{\Cp}{C_{\rm P}} 
\newcommand{\Cv}{C_{\rm v}}  

\newcommand{\DOS}{\mu}  
\newcommand{\bath}{{\rm bath}}  
\newcommand{\cycle}{{\rm cycle}}  
\newcommand{\HighOrd}{{\rm high-ord.}}  
\newcommand{\cold}{{\rm cold}}

\newcommand{\inter}{ {\rm int} }   
\newcommand{\hc}{ {\rm h.c.} }
\newcommand{\tot}{{\rm tot}}
\newcommand{\Tr}{{\rm Tr}}   
\def\id{\mathbbm{1}}   
\newcommand{\kB}{k_\mathrm{B}}  

\newcommand{\worst}{\mathrm{worst}}

\newcommand{\LParen}{ \bm{(} }
\newcommand{\RParen}{ \bm{)} }
\newcommand*{\Set}[1]{\left\{  #1  \right\}}
\newcommand{\1}{ { (1) } }  
\newcommand{\2}{ { (2) } }  
\newcommand{\3}{ { (3) } }

\newcommand{\ParenJ}{ { (j) } } 
\newcommand{\ParenE}{ { (E) } } 
\newcommand{\LL}{ { (L) } } 

\renewcommand\th{ {\rm th} }

\newcommand*{\bra}[1]{\langle #1\rvert}
\newcommand*{\ket}[1]{\lvert #1 \rangle}
\newcommand*{\braket}[2]{\langle #1 \lvert #2 \rangle}
\newcommand*{\ketbra}[2]{\lvert #1 \rangle\!\langle #2 \rvert}
\newcommand*{\expval}[1]{\left\langle  #1  \right\rangle}

\renewcommand{\thesection}{\Roman{section}}
\renewcommand{\thesubsection}{\thesection \Alph{subsection}}
\renewcommand{\thesubsubsection}{\thesubsection \arabic{subsubsection}}

\makeatletter
\def\p@subsection{}
\makeatother
\makeatletter
\def\p@subsubsection{}
\makeatother

\begin{document}
\title{MBL-mobile: Quantum engine based on many-body localization}

\author{Nicole~Yunger~Halpern}
\email{Current email and address: nicoleyh@g.harvard.edu,
Harvard-Smithsonian ITAMP, Cambridge, MA 02138, USA}
\affiliation{Institute for Quantum Information and Matter, California Institute of Technology, Pasadena, CA 91125, USA}

\author{Christopher~David~White}
\email{cdwhite@caltech.edu}
\affiliation{Institute for Quantum Information and Matter, California Institute of Technology, Pasadena, CA 91125, USA}

\author{Sarang~Gopalakrishnan}
\email{sarang.gopalakrishnan@gmail.com}
\affiliation{Institute for Quantum Information and Matter, California Institute of Technology, Pasadena, CA 91125, USA}
\affiliation{Department of Physics, California Institute of Technology, Pasadena, CA 91125, USA}
\affiliation{Walter Burke Institute, California Institute of Technology, Pasadena, CA 91125, USA}
\affiliation{College of Staten Island, City University of New York, Staten Island, NY 10314, USA}

\author{Gil~Refael}
\email{refael@caltech.edu}
\affiliation{Institute for Quantum Information and Matter, California Institute of Technology, Pasadena, CA 91125, USA}
\affiliation{Department of Physics, California Institute of Technology, Pasadena, CA 91125, USA}
\affiliation{Walter Burke Institute, California Institute of Technology, Pasadena, CA 91125, USA}
\date{\today}

%
%
\keywords{
Many-body localization,
quantum thermal machines,
statistical mechanics,
thermodynamics,
Eigenstate Thermalization Hypothesis,
quantum many-body systems}

%
%
\begin{abstract}

Many-body-localized (MBL) systems do not thermalize under their intrinsic dynamics. The athermality of MBL, we propose, can be harnessed for thermodynamic tasks. We illustrate this ability by formulating an Otto engine cycle for a quantum many-body system. The system is ramped between a strongly localized MBL regime and a thermal (or weakly localized) regime. The difference between the energy-level correlations of MBL systems and of thermal systems enables mesoscale engines to run in parallel in the thermodynamic limit, enhances the engine's reliability, and suppresses worst-case trials. We estimate analytically and calculate numerically the engine's efficiency and per-cycle power. The efficiency mirrors the efficiency of the conventional thermodynamic Otto engine. The per-cycle power scales linearly with the system size and inverse-exponentially with a localization length. This work introduces a thermodynamic lens onto MBL, which, having been studied much recently, can now be considered for use in thermodynamic tasks.

\end{abstract}
{\let\newpage\relax\maketitle}


%
%
Many-body localization (MBL) has emerged as a unique phase in which
an isolated interacting quantum system resists internal thermalization
for long times.
MBL systems are integrable and have local integrals of motion~\cite{Huse_14_phenomenology},
which retain information about initial conditions
for long times or even indefinitely~\cite{KjallIsing}. 
This and other aspects of MBL were recently observed experimentally \cite{Schreiber_15_Observation,Kondov_15_Disorder,Ovadia_15_Evidence,Choi_16_Exploring,Luschen_17_Signatures,Kucsko_16_Critical,Smith_16_Many,Bordia_17_Probing}.
In contrast, in thermalizing isolated quantum systems, information and energy can diffuse easily. Such systems obey the eigenstate thermalization hypothesis (ETH)~\cite{Deutsch_91_Quantum,Srednicki_94_Chaos,Rigol_07_Relaxation,rigol-dunjko-olshanii}.

A tantalizing question is whether the unique properties of MBL could be utilized. So far, MBL has been proposed to be used as robust quantum memories~\cite{Nandkishore_15_MBL}. We believe, however, that the potential of MBL is much greater.  MBL systems behave athermally, and athermality (lack of thermal equilibrium) facilitates thermodynamic tasks~\cite{Janzing_00_Thermodynamic,Dahlsten_11_Inadequacy,Aberg_13_Truly,Brandao_13_Resource,Horodecki_13_Fundamental,Egloff_15_Measure,Goold_15_review,Gour_15_Resource,YungerHalpern_16_Beyond,Deffner_16_Quantum,Wilming_17_Third}. 
When a cold bath is put in contact with a hot environment, for instance, 
work can be extracted from the heat flow. 
Could MBL's athermality have thermodynamic applications?

\begin{figure}[tb]
\centering
\includegraphics[width=.45\textwidth, clip=true]{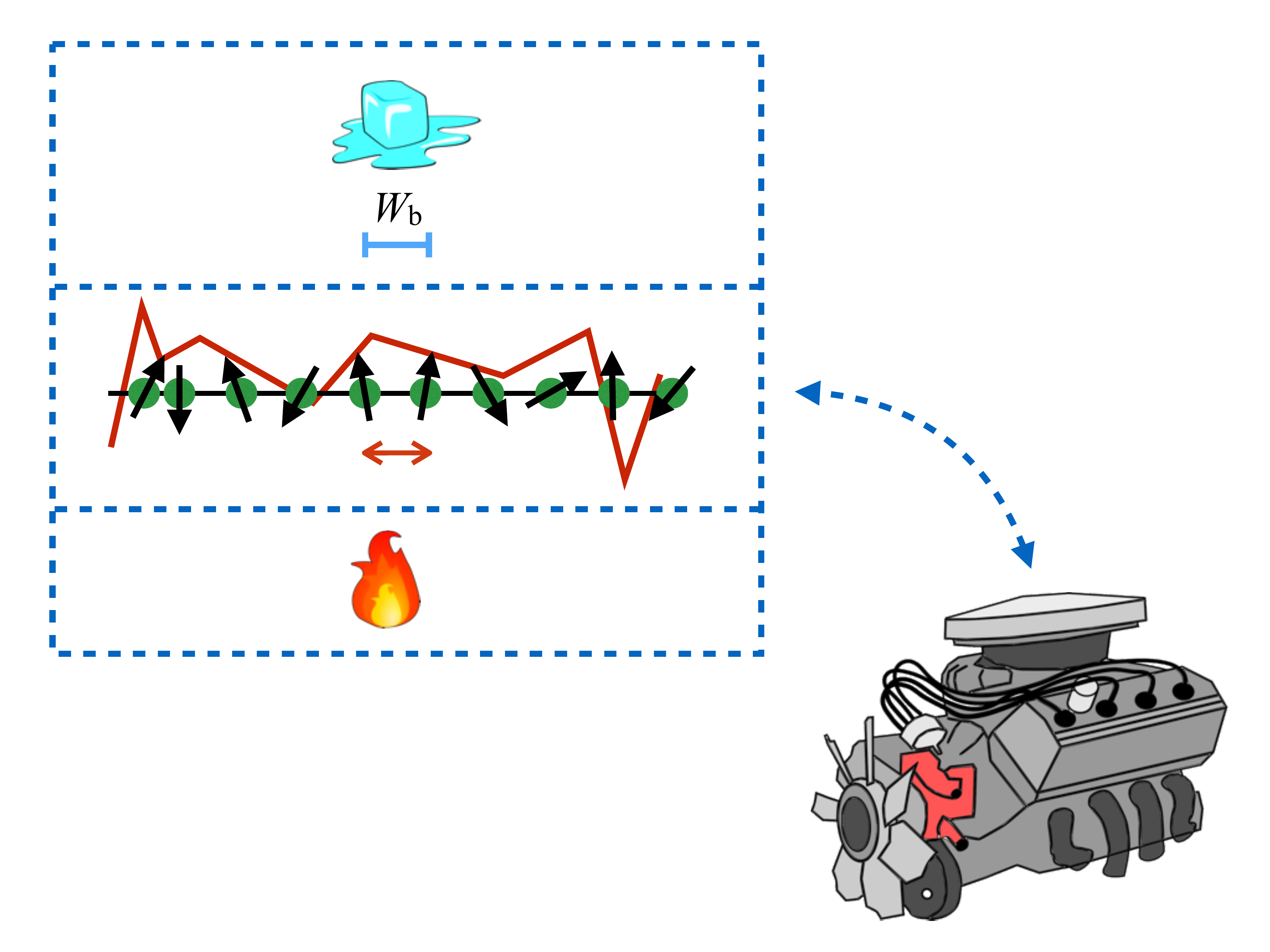}
\caption{\caphead{Schematic of MBL engine:}
We formulate an Otto engine cycle for a many-body quantum system
that exhibits an MBL phase.
We illustrate the engine with a spin chain (green dots and black arrows).
A random disorder potential (jagged red line) localizes the particles.
Particles interact and hop between sites
(horizontal red arrows).
Consider strengthening the interactions and the hopping frequency.
The system transitions from strong localization
to a thermal phase or to weak localization.
The engine thermalizes with a hot bath (flames)
and with a cold bath (ice cube).
The cold bath has a small bandwidth, $\Wb$,
to take advantage of small energy gaps' greater prevalence
in the highly localized regime.}
\label{fig:Artist_conceptn}
\end{figure}

We present one by formulating, analyzing, 
and numerically simulating an Otto engine cycle
for a quantum many-body system that has an MBL phase.
The engine contacts a hot bath and a narrow-bandwidth cold bath, 
as sketched in Fig.~\ref{fig:Artist_conceptn}.
This application unites the growing fields of 
quantum thermal machines~\cite{Geusic_maser_67,del_Campo_14_Super,Brunner_15_Ent_fridge,Binder_15_Quantacell,Woods_15_Maximum,Gelbwaser_15_Strongly,Song_16_polariton_engine,Tercas_16_Casimir,PerarnauLlobet_16_Work,Kosloff_17_QHO,Lekscha_16_Quantum,Jaramillo_16_Quantum,Gelbwaser_18_Single}
and MBL~\cite{BAA,Oganesyan_07_Level_stats,Pal_Huse_10_MBL,Huse_14_phenomenology,Nandkishore_15_MBL,serbynmoore}. 
Our proposal could conceivably be explored in ultracold-atom~\cite{Schreiber_15_Observation,Kondov_15_Disorder,Choi_16_Exploring,Luschen_17_Signatures,Bordia_17_Probing},
nitrogen-vacancy-center~\cite{Kucsko_16_Critical},
trapped-ion~\cite{Smith_16_Many}, and possibly
doped-semiconductor~\cite{Kramer_93_Localization} experiments.

%
%

Our engine relies on two properties that distinguish MBL from thermal systems: its spectral correlations~\cite{Sivan_87_Energy,serbynmoore} and its localization. The spectral-correlation properties enable us to build a mesoscale level-statistics engine. The localization enables us to link mesoscale engines together, 
creating a large engine with an extensive work output.
  
Take an interacting finite spin chain as an example.
Consider the statistics of the gaps 
between consecutive energy eigenvalues 
far from the energy band's edges.
A gap distribution $P (\delta)$ encodes
the probability that any given gap
has size $\delta$.
The MBL gap distribution enables small (and large) gaps to appear
much more often than in ETH spectra~\cite{D'Alessio_16_From}. 
This difference enables MBL to enhance 
our quantum many-body Otto cycle.

Let us introduce the MBL and ETH distributions in greater detail.
Let $\dAvg_E$ denote the average gap at the energy $E$.
MBL gaps approximately obey Poisson statistics~\cite{Oganesyan_07_Level_stats,D'Alessio_16_From}:
\begin{align}
   \label{eq:P_MBL_Main}
   P_\MBL^\ParenE(\delta)  
   \approx  \frac{1}{\dAvg_E}  e^{-\delta / \dAvg_E }  \, .
\end{align}
Any given gap has a decent chance of being small: 
As $\delta  \to  0$,
$P_\MBL^\ParenE (\delta)  \to  \frac{1}{\dAvg_E}  >  0$.
Neighboring energies have finite probabilities of lying close together: 
MBL systems' energies do not repel each other, 
unlike thermal systems' energies.
Thermalizing systems governed by real Hamiltonians 
obey the level statistics of random matrices drawn from 
the Gaussian orthogonal ensemble (GOE)~\cite{Oganesyan_07_Level_stats}: 
\begin{align}
   \label{eq:P_ETH_Main}
   P_\ETH^\ParenE (\delta)  
   \approx   \frac{\pi}{2}   \frac{\delta}{\dAvg_E^2}  \:  
   e^{-\frac{\pi}{4}  \delta^2  /  \dAvg_E^2}  \, .
\end{align}
Unlike in MBL spectra, small gaps rarely appear:
As $\delta\to0$,  $P_\ETH^\ParenE (\delta)  \to  0$.

%

MBL's athermal gap statistics should be construed as 
a thermodynamic resource, we find, as athermal quantum states have been~\cite{Janzing_00_Thermodynamic,Dahlsten_11_Inadequacy,Aberg_13_Truly,
Brandao_13_Resource,Horodecki_13_Fundamental,Egloff_15_Measure,
Goold_15_review,Gour_15_Resource,YungerHalpern_16_Beyond,
Deffner_16_Quantum,Wilming_17_Third}.  
In particular, MBL's athermal gap statistics improve our engine's reliability:
The amount $W_\tot$ of work extracted by our engine 
fluctuates relatively little from successful trial to successful trial.
Athermal statistics also lower the probability of worst-case trials,
in which the engine outputs net negative work, $W_\tot < 0$.
Furthermore, MBL's localization 
enables the engine to scale robustly: 
Mesoscale ``subengines'' can run in parallel 
without disturbing each other much,
due to the localization inherent in MBL. 
Even in the thermodynamic limit, 
an MBL system behaves like an ensemble of finite,  
mesoscale quantum systems, due to its  
\emph{local level correlations}~\cite{Sivan_87_Energy,imryma,Syzranov_17_OTOCs}. 
Any local operator can probe only
a discrete set of sharp energy levels, 
which emerge from its direct environment.



This paper is organized as follows.
Section~\ref{section:Thermo_backgrnd_main} contains background about
the Otto cycle and about quantum work and heat.
In Sec.~\ref{section:Meso_main}, we introduce 
the mesoscopic MBL engine.
In Sec.~\ref{section:Qubit_main}, we introduce the basic idea 
with a qubit (two-level quantum system). 
In Sec.~\ref{section:Meso_setup}, we scale the engine up to 
a mesoscopic chain tuned between MBL and ETH regimes.
In Sec.~\ref{section:Quant_main}, we calculate 
the engine's work output and efficiency.
In Sec.~\ref{section:Thermo_limit_main}, we argue that 
the mesoscopic segments can be combined into 
a macroscopic MBL system while operating in parallel.
In Sec.~\ref{section:Diab_main}, we discuss limitations on the speed at which the engine can be run and, consequently, the engine's power.
This leads us to a more careful consideration of diabatic corrections to the  work output, communication amongst subengines, and the cold bath's nature.
We test our analytic calculations in Sec.~\ref{section:Numerics_main},
with numerical simulations of disordered spin chains.
In Sec.~\ref{section:Order_main}, we provide order-of-magnitude 
estimates for a localized semiconductor engine's power and power density.

%
%
%
\section{Thermodynamic background}
\label{section:Thermo_backgrnd_main}

The classical Otto engine consists of a gas that 
expands, cools, contracts, and heats~\cite{MIT_Otto}.
During the two isentropic (constant-entropy) strokes,  
the gas's volume is tuned 
between values $V_1$ and $V_2 < V_1$. 
The \emph{compression ratio} is defined as $r := \frac{ V_1 }{ V_2 }$ . 
The heating and cooling are isochoric (constant-volume). 
 The engine outputs a net amount $W_\tot$
of work per cycle, absorbing heat $Q_\In > 0$
during the heating isochore.

A general engine's thermodynamic efficiency is
\begin{align}
   \label{eq:Eff_main}
   \eta  :=  \frac{ W_\tot }{ Q_\In }  \, .
\end{align}
The Otto engine operates at the efficiency
\begin{align}
   \label{eq:Otto_eff_main}
   \eta_\Otto  =  1  -  \frac{1}{  r^{ \gamma  -  1 }  }
   <  \eta_\Carnot  \, .
\end{align}
A ratio of the gas's 
constant-pressure and constant-volume specific heats
is denoted by $\gamma:=\frac{\Cp}{\Cv}$.
The Carnot efficiency $\eta_\Carnot$ upper-bounds
the efficiency of every thermodynamic engine that involves just two heat baths.

An Otto cycle for quantum harmonic oscillators (QHOs) was discussed in Refs.~\cite{Scully_02_Quantum,Abah_12_Single,Deng_13_Boosting,del_Campo_14_Super,Zheng_14_Work,Karimi_16_Otto,V_Anders_15_review,Kosloff_17_QHO}.
The QHO's gap plays the role
of the classical Otto engine's volume.
Let $\omega$ and $\Omega > \omega$ denote 
the values between which
the angular frequency is tuned.
The ideal QHO Otto cycle operates at the efficiency
\begin{align}
   \label{eq:Eff_QHO}
   \eta_\QHO  =  1  -  \frac{ \omega }{ \Omega }  \, .
\end{align}
This oscillator model resembles the qubit toy model
that informs our MBL Otto cycle (Sec.~\ref{section:Qubit_main}).
The energy eigenbasis changes in our model, however, 
and the engine scales robustly to macroscopically many qubits.

Consider tuning an open system, slowly, 
between times $t = 0$ and $t = \tau$. 
The heat and work absorbed are defined as
\begin{align}
   & \label{eq:Work_def_main}
   W:=\int_0^\tau dt\;\Tr   \left(  \rho  \:  \frac{dH}{dt}  \right)
   \quad \text{and} \\ 
   &  \label{eq:Heat_def_main}
   Q:=\int_0^\tau dt\;\Tr\left(  \frac{d\rho}{dt}  \:  H  \right)  
\end{align}
in quantum thermodynamics~\cite{V_Anders_15_review}.
This $Q$ definition is narrower than the definition
prevalent in the MBL literature~\cite{Lin_16_Ginzburg,Corboz_16_Variational,Gopalakrishnan_16_Regimes,D'Alessio_16_From}: 
Here, all energy exchanged during unitary evolution 
counts as work.

%
%
%

%
%
\begin{figure}[tb]
\centering
\includegraphics[width=.45\textwidth, clip=true]{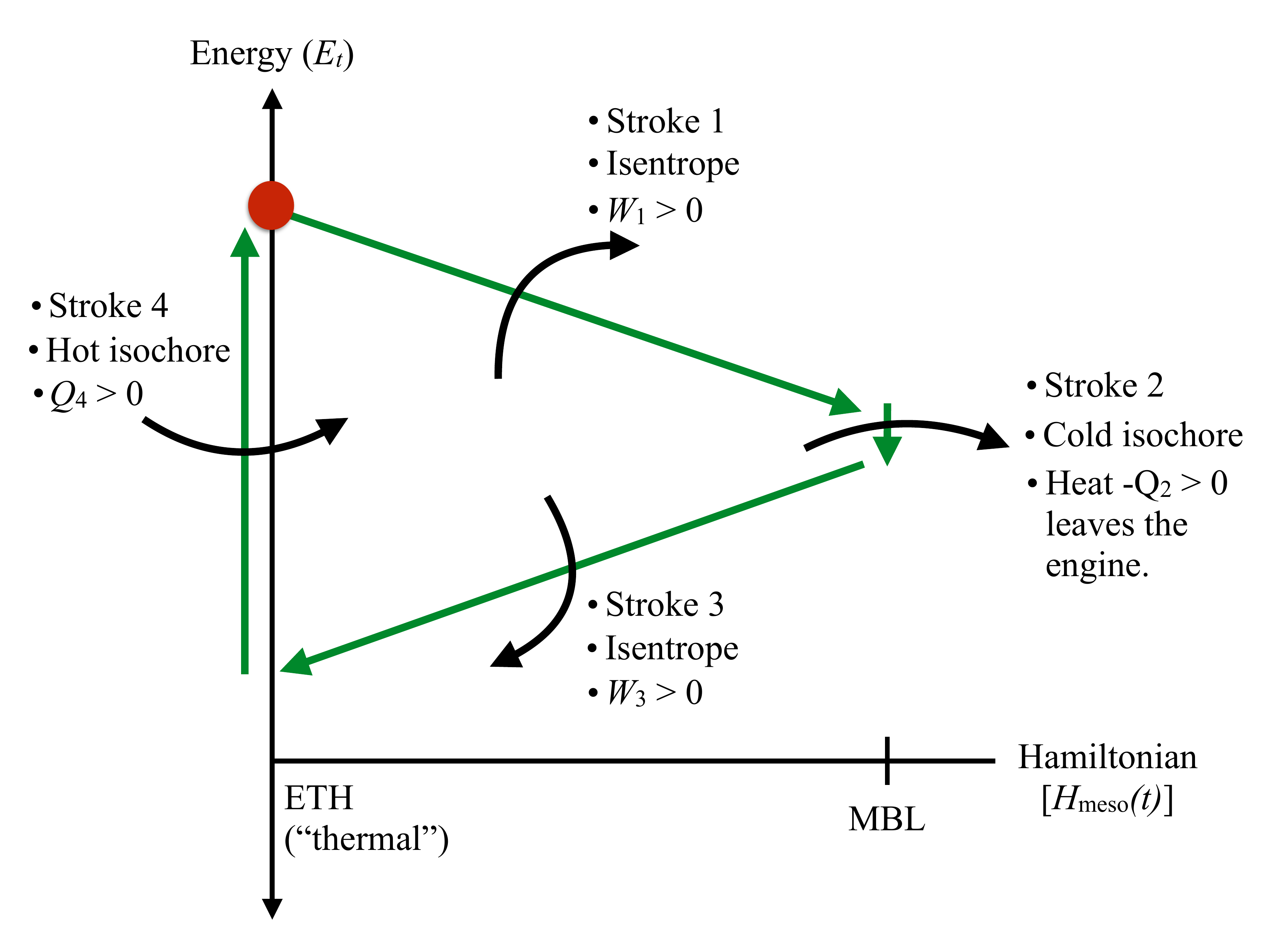}
\caption{\caphead{Otto engine cycle for
a mesoscale MBL system:}
Two energies in the many-body spectrum 
capture the cycle's basic physics.
The engine can be regarded as beginning each trial
in an energy eigenstate drawn from a Gibbs distribution.
The red dot represents the engine's starting state 
in some trial of interest.
During stroke 1, $H_\meso(t)$ is tuned from ``thermal'' to MBL.
During stroke 2, the engine thermalizes with a cold bath.
$H_\meso(t)$ returns from MBL to thermal during stroke 3.
Stroke 4 resets the engine, which thermalizes with a hot bath.
The tunings (strokes 1 and 3) map onto
the thermodynamic Otto cycle's isentropes.
The thermalizations (strokes 2 and 4) map onto isochores.
The engine outputs work $W_1$ and $W_3$
during the tunings
and absorbs heat $Q_2$ and $Q_4$ during thermalizations.
MBL gap statistics' lack of level repulsion
enhances the cycle:
The engine ``slides down'' the lines that represent tunings,
losing energy outputted as work.}
\label{fig:Compare_thermo_Otto_fig}
\end{figure}
\section{A mesoscale MBL engine}
\label{section:Meso_main}

We aim to formulate an MBL engine cycle 
for the thermodynamic limit.
Our road to that goal runs through 
a finite-size, or mesoscale, MBL engine.
In Sec.~\ref{section:Qubit_main}, 
we introduce the intuition behind the mesoscale engine 
via a qubit toy model.
Then, we describe (Sec.~\ref{section:Meso_setup}) 
and quantitatively analyze (Sec.~\ref{section:Quant_main}) 
the mesoscale MBL engine.
Table~\ref{table:Notation} offers a spotter's guide to notation.

\subsection{Qubit toy model}
\label{section:Qubit_main} 

At the MBL Otto engine's core lies a qubit Otto engine
whose energy eigenbasis transforms during the cycle~\cite{Kosloff_02_Discrete,Kieu_04_Second,Kosloff_10_Optimal,Cakmak_17_Irreversible}.
Consider a two-level system evolving under 
the time-varying Hamiltonian 
\begin{align}
   H_\qubit(t)  :=   (1-\alpha_t)  h\sigma^x
   +   \alpha_t h'\sigma^z  \, .
\end{align}
$\sigma^x$and $\sigma^z$ denote the Pauli $x$- and $z$-operators.
$\alpha_t$ denotes a parameter tuned between 0 and 1.

\begin{figure}[h]
  \centering
  \includegraphics[width=0.45\textwidth]{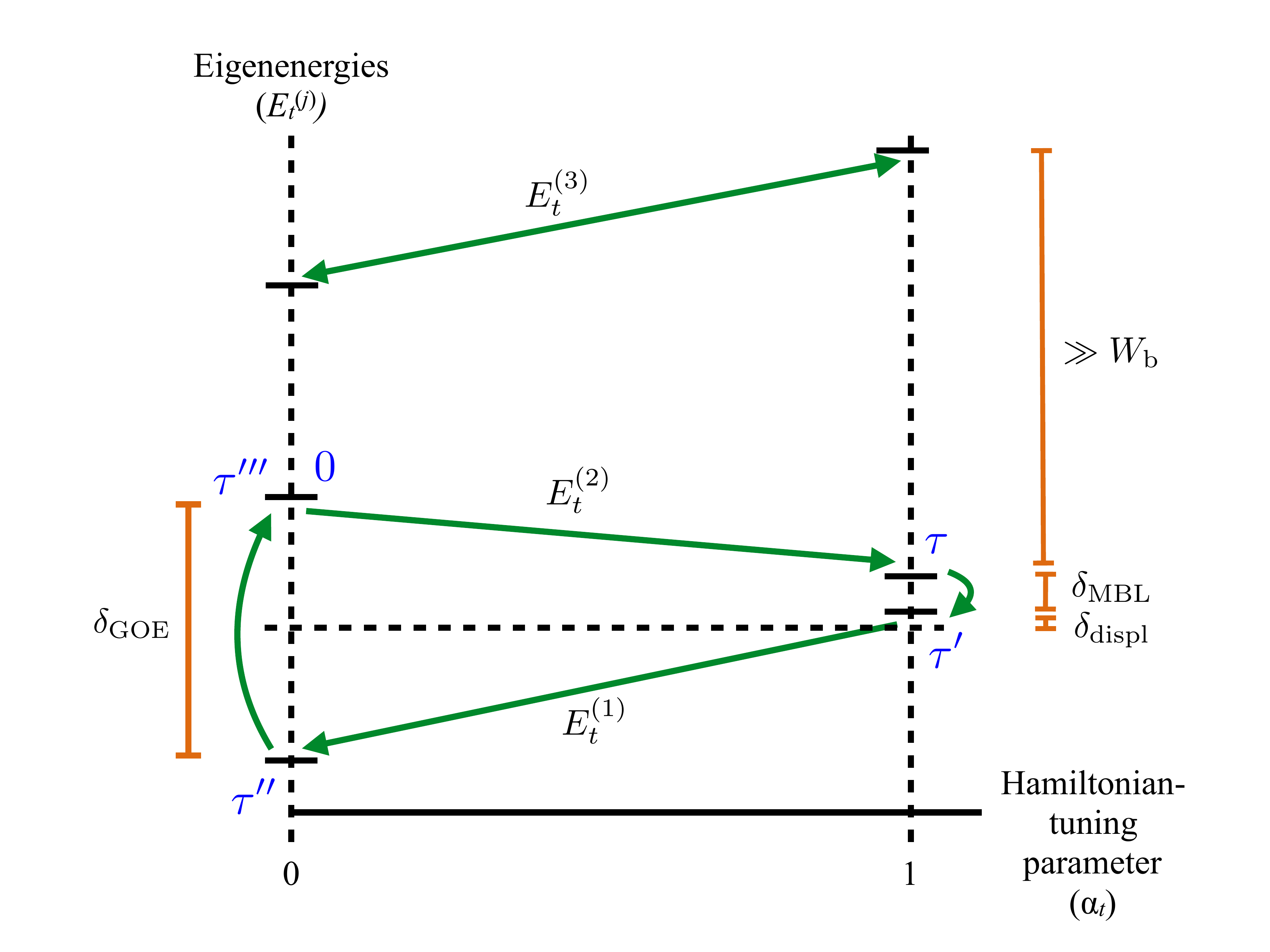}
  \caption{\caphead{Qubit toy model for the MBL Otto cycle:}
A qubit models two ``working levels'' in
the MBL Otto engine's many-body spectrum.
The energy eigenstates $\ket{ E_t^\1 }$ and $\ket{ E_t^\2 }$ 
span the ``working subspace.''
The gap $E_t^\2 - E_t^\1$ begins at size $\delta_\ETH$
during a successful trial.
The gap shrinks to $\delta_\MBL$, then returns to $\delta_\ETH$.
In addition to changing the gap,
each Hamiltonian tuning 
changes the eigenstates' functional forms.
The displacement $\delta_\disp$ is included for generality. 
The blue text marks the times 
$t = 0, \tau, \ldots, \tau'''$ at which the strokes begin and end
during a work-outputting trial.
The spectator level $\ket{ E_t^\3 }$ fails to impact the engine's efficiency.
The cold bath has too narrow a bandwidth $\Wb$
to couple $\ket{ E_t^\3 }$ to any other level.
If the engine begins any trial on the top green line,
the engine remains on that line throughout the trial.
Zero net work is outputted.}
\label{fig:2Level_v3}
\end{figure}

Figure~\ref{fig:2Level_v3} illustrates the cycle.
The engine begins in thermal equilibrium 
at a high temperature $\THot$.
During stroke 1, the engine is thermally isolated,
and $\alpha_t$ is tuned from 0 to 1. 
During stroke 2, the engine thermalizes 
to a temperature $\TCold  \ll  \THot$. 
During stroke 3, the engine is thermally isolated, 
and $\alpha_t$ returns from 1 to 0.
During stroke 4, the engine resets 
by thermalizing with the hot bath.

Let us make two simplifying assumptions
(see~\cite[App.~C]{NYH_17_MBL} for a generalization):
First, let $\THot=\infty$ and $\TCold=0$.
Second, assume that the engine is tuned slowly enough
to satisfy the quantum adiabatic theorem.
We also choose\footnote{
The gaps' labels are suggestive: A qubit, having only one gap,
obeys neither $\ETH$ nor $\MBL$ gap statistics.
But, when large, the qubit gap apes a typical $\ETH$ gap;
and, when small, the qubit gap apes a useful $\MBL$ gap.
This mimicry illustrates how the mesoscopic engine benefits from
the greater prevalence of small gaps in MBL spectra
than in $\ETH$ spectra.}
\[
h=  \frac{ \delta_\ETH }{2}  \, ,  \quad
h'=  \frac{ \delta_\MBL }{2}\, , \]  
and $\delta_\ETH\gg\delta_\MBL$.

Let us analyze the cycle's energetics.
The system begins with $\expval{ H_\qubit (t) }  =  0$. 
Stroke 1 preserves the infinite-temperature state $\id / 2$.
The energy drops to $- \delta_\MBL / 2$ 
during stroke 2 and to $- \delta_\ETH / 2$ during stroke 3.
During stroke 4, the engine resets to zero average energy,
absorbing heat
$\expval{Q_4}=  \frac{ \delta_\ETH }{ 2 }$, on average.

The energy exchanged during the tunings (strokes 1 and 3)
constitutes work [Eq.~\eqref{eq:Work_def_main}], 
while the energy exchanged during the thermalizations
(strokes 2 and 4) is heat
[Eq.~\eqref{eq:Heat_def_main}].
The engine outputs the \emph{per-cycle power,} 
or average work performed per cycle,
$\expval{W_\tot}=\frac{1}{2}(\delta_\ETH-\delta_\MBL)$.

The efficiency is 
$\eta_\qubit   =   \frac{\expval{W_\tot}}{\expval{Q_4}}
=   1-\frac{\delta_\MBL}{\delta_\ETH}$. 
This result is equivalent to
the efficiency $\eta_\Otto$ of
a thermodynamic Otto engine [Eq.~\eqref{eq:Otto_eff_main}].
The gap ratio $\frac{\delta_\MBL}{\delta_\ETH}$
plays the role of $\frac{1}{ r^{\gamma-1} }$.
$\eta_\qubit$ equals also $\eta_\QHO$
[Eq.~\eqref{eq:Eff_QHO}]
if the frequency ratio $\omega / \Omega$ is chosen to equal
$\delta_\MBL / \delta_\ETH$.
As shown in Sections~\ref{section:Meso_main}-\ref{section:Thermo_limit_main}, however,
the qubit engine can scale to 
a large composite engine
of densely packed qubit subengines operating in parallel.
The dense packing is possible if
the qubits are encoded in
the MBL system's localized degrees of freedom 
(l-bits, roughly speaking~\cite{Huse_14_phenomenology}).

\subsection{Set-up for the mesoscale MBL engine}
\label{section:Meso_setup}

%
%
%

%
%
\begin{table*}[t]
\begin{center}
\begin{tabular}{|c|c|}
   \hline
        Symbol
   &  Significance
   \\   \hline \hline
        $\Sites$
   &    Number of sites per mesoscale engine (in Sec.~\ref{section:Meso_main})
         or per mesoscale subengine  
   \\    
         
   &   (in the macroscopic engine, in Sec.~\ref{section:Thermo_limit_main}).
        Chosen, in the latter case, to equal $\xi_\Loc$. 
   \\    \hline
        $\HDim$
   &   Dimensionality of one mesoscale (sub)engine's Hilbert space.
   \\    \hline
        $\HScale$
   &   Unit of energy, average energy density per site.
   \\    \hline
   &
         Hamiltonian parameter tuned from 0 
         (in the mesoscale engine's ETH regime,
  \\  
         $\alpha_t$
  &
         or the macroscopic engine's shallowly localized regime) 
  \\
  &
         to 1 (in the engine's deeply MBL regime).
   \\    \hline
         $\dAvg$
   &    
         Average gap in the energy spectrum of a length-$\Sites$ MBL system.
   \\    \hline
         $\Wb$
   &
         Bandwidth of the cold bath. Small: $\Wb \ll \dAvg$.
   \\    \hline
         $\betaH = 1 / \THot$
   &
         Inverse temperature of the hot bath.
   \\    \hline
         $\betaC = 1 / \TCold$
   &
         Inverse temperature of the cold bath.
   \\    \hline
         $\deltaMBL$
   &    Level-repulsion scale of a length-$\Sites$ MBL system.
         Minimal size reasonably attributable to
   \\
   &
         any energy gap. 
         Smallest gap size at which a Poissonian~\eqref{eq:P_MBL_Main}
         approximates 
   \\    
   &
         the MBL gap distribution well.
   \\    \hline
         $v$
   &    Speed at which the Hamiltonian is tuned:
          $v  :=  \HScale  \frac{  d \alpha_t }{ t }$.
   \\
   &
          Has dimensions of $1/\text{time}^2$,
          in accordance with part of~\cite{DeGrandi_10_APT}.
   \\    \hline
         $\xi_\Loc$
   &    Localization length of macroscopic MBL engine when 
         shallowly localized.
   \\   
   &    Length of mesoscale subengine.
   \\    \hline
         $\xi_\VeryLoc$
   &    Localization length of macroscopic MBL engine 
         when deeply localized.
         Satisfies $\xi_\VeryLoc < \xi_\Loc$.
   \\    \hline
         $X_\macro$
  &    Characteristic $X$ of the macroscopic MBL engine (e.g., $X = \Sites, \dAvg$).
  \\    \hline
         $g$
   &    Strength of coupling between engine and cold bath.
   \\    \hline
         $\tau_\cycle$
   &    Time required to implement one cycle.
   \\    \hline
         $\dAvg^\LL$ 
   &    Average energy gap of a length-$L$ MBL system.
   \\    \hline
\end{tabular}
\caption{\caphead{Parameters of
the mesoscopic and macroscopic MBL engines:}
Introduced in Sections~\ref{section:Meso_main} and~\ref{section:Thermo_limit_main}.
Boltzmann's and Planck's constants are set to one: $\kB = \hbar = 1$.}
\label{table:Notation}
\end{center}
\end{table*}

The next step is an interacting finite-size system 
tuned between MBL and ETH phases. 
Envision a mesoscale engine as a one-dimensional (1D) system 
of $\Sites  \approx  10$ sites.
This engine will ultimately model one region in 
a thermodynamically large MBL engine. 
We will analyze 
the mesoscopic engine's per-trial power $\expval{ W_\tot }$, 
the efficiency $\eta_\MBL$,
and work costs $\expval{ W_\diab }$ of undesirable diabatic transitions.

The mesoscopic engine evolves under the Hamiltonian
\begin{align}
   \label{eq:H_meso_main}
   H_\meso(t)  :=  \frac{ \HScale }{ Q ( \alpha_t ) }  \left[
   (1-\alpha_t )H_\ETH   +   \alpha_t  \, H_\MBL  \right]  \, .
\end{align}
The unit of energy, or average energy density per site,
is denoted by $\HScale$.
The tuning parameter $\alpha_t  \in  [0, 1]$.
When $\alpha_t  =  0$, the system evolves under
a random Hamiltonian $H_\ETH$
whose gaps $\delta$ are distributed according to 
$P^\ParenE_\ETH ( \delta )$
[Eq.~\eqref{eq:P_ETH_Main}].
When $\alpha_t = 1$, $H_\meso(t) = H_\MBL$, 
a Hamiltonian whose gaps are distributed according to 
$P^\ParenE_\MBL ( \delta )$
[Eq.~\eqref{eq:P_MBL_Main}].
For a concrete example, take a random-field Heisenberg model 
whose disorder strength is tuned.
$H_\ETH$ and $H_\MBL$ have the same bond term, 
but the disorder strength varies in time.
We simulate (a rescaled version of) this model 
in Sec.~\ref{section:Numerics_main}.


The mesoscale engine's cycle
is analogous to the qubit cycle,
including initialization at $\alpha_t = 0$, 
tuning of $\alpha_t$ to one, 
thermalization with a temperature-$\TCold$ bath,
tuning of $\alpha_t$ to zero,
and thermalization~\cite{Huse_15_Localized,DeLuca_15_Dynamic,Levi_16_Robustness,Fischer_16_Dynamics} 
with a temperature-$\THot$ bath.
To highlight the role of level statistics in the cycle,
we hold the average energy gap, $\dAvg$, constant.\footnote{
\label{footnote:dAvg}
$\dAvg$ is defined as follows.
The density of states at the energy $E$ has the form
$\DOS(E)  \approx  
\frac{ \HDim }{  \sqrt{ 2 \pi \Sites }  \,  \HScale  }  \,
e^{ - E^2 / 2 \Sites  \HScale^2 }$
(see Table~\ref{table:Notation} for the symbols' meanings).
Inverting $\DOS (E)$ yields
the \emph{local average gap}:
$\dAvg_E  :=  \frac{1}{  \DOS (E) }$.
Inverting the average of $\DOS (E)$ yields
the \emph{average gap},
\begin{align}
   \label{eq:dAvg_def}
   \dAvg  :=  \frac{ 1 }{ \expval{  \DOS (E) }_{\rm energies} } 
   =  \frac{ \HDim }{   \int_{ - \infty }^\infty  dE  \;  \DOS^2 (E)  }
   =   \frac{ 2 \sqrt{ \pi \Sites }  }{ \HDim }  \,  \HScale    \, .
\end{align}
}
We do so using the renormalization factor $Q ( \alpha_t )$.\footnote{
Imagine removing $Q(\alpha_t)$ from Eq.~\eqref{eq:H_meso_main}.
One could increase $\alpha_t$---could 
tune the Hamiltonian from ETH to MBL~\cite{serbynmoore}---by
strengthening a disorder potential.
This strengthening would expand the energy band;
tuning oppositely would compress the band.
By expanding and compressing, in accordion fashion,
and thermalizing, one could extract work.
This engine would benefit little from properties of MBL,
whose thermodynamic benefits we wish to highlight.
Hence we ``zero out'' the accordion motion, 
by fixing $\dAvg$ through $Q( \alpha_t )$.
For a brief discussion of the accordion-like engine, 
see App.~\ref{section:Bandwidth_engine_app}. }
Section~\ref{section:Numerics_main} details how we define $Q (\alpha_t)$
in numerical simulations.

The key distinction between 
GOE level statistics~\eqref{eq:P_ETH_Main} 
and Poisson (MBL) statistics~\eqref{eq:P_MBL_Main}
is that small gaps (and large gaps)
appear more often in Poisson spectra.
A toy model illuminates these level statistics' physical origin:
An MBL system can be modeled as 
a set of noninteracting quasilocal qubits~\cite{Huse_14_phenomenology}.
Let $g_j$ denote the $j^\th$ qubit's gap.
Two qubits, $j$ and $j'$, may have nearly equal gaps: $g_j  \approx  g_{j'}$. 
The difference $| g_j  -  g_{j'} |$ equals a gap
in the many-body energy spectrum.
Tuning the Hamiltonian from MBL to ETH
couples the qubits together,
producing matrix elements between the nearly degenerate states.
These matrix elements force energies apart.

To take advantage of the phases' distinct level statistics,
we use a cold bath that has a small bandwidth $\Wb$.
According to Sec.~\ref{section:Qubit_main},
net positive work is extracted from the qubit engine because 
$\delta_\MBL  <  \delta_\ETH$.
The mesoscale analog of $\delta_\ETH$ is $\sim \dAvg$,
the typical gap ascended during hot thermalization. 
The engine must not emit energy on this scale
during cold thermalization. 
Limiting $\Wb$ ensures that cold thermalization
relaxes the engine only across gaps 
$\delta\leq \Wb  \ll  \dAvg$.
Such anomalously small gaps appear more often 
in MBL energy spectra than in ETH spectra
~\cite{Khaetskii_02_Electron,Gopalakrishnan_14_Mean,Parameswaran_17_Spin}.


This level-statistics argument holds only 
within superselection sectors.
Suppose, for example, that $H_\meso(t)$ conserves the particle number.
The level-statistics arguments apply only if 
the particle number remains constant 
throughout the cycle~\cite[App.~F]{NYH_17_MBL}.
Our numerical simulations (Sec.~\ref{section:Numerics_main}) 
take place at half-filling,
in a subspace of dimensionality $\HDim$ 
of the order of magnitude of
the whole space's dimensionality:
$\HDim  \sim  \frac{ 2^\Sites }{ \sqrt{ \Sites } }$.


%
%
%

We are now ready to begin analyzing the mesoscopic-engine Otto cycle. 
The engine begins in the thermal state 
$\rho(0)=e^{-\betaH H_\ETH}/\ZH$, wherein 
$\ZH:=\Tr\left(e^{-\betaH H_\ETH}\right)$.
The engine can be regarded as starting each trial
in some energy eigenstate $j$
drawn according to the Gibbs distribution
(Fig.~\ref{fig:Compare_thermo_Otto_fig}).
During stroke 1, $H_\meso(t)$ is tuned from $H_\ETH$ to $H_\MBL$.
We approximate the tuning as quantum-adiabatic
(diabatic corrections are modeled in Sec.~\ref{section:Diab_main}).
Stroke 2, cold thermalization, depends on 
the gap $\delta'_j$ between the $j^\th$ and $(j-1)^\th$ MBL levels.
$\delta'_j$ typically exceeds $\Wb$.
If it does, cold thermalization preserves the engine's energy,
and the cycle outputs $W_\tot=0$. 
With probability $\sim  \frac{\Wb}{\dAvg}$,
the gap is small enough to thermalize: $\delta'_j  <  \Wb$. 
In this case, cold thermalization drops the engine to level $j-1$.
Stroke 3 brings the engine to level $j-1$ of $H_\ETH$. 
The gap $\delta_j$ between the $(j-1)^\th$ and $j^\th$ $H_\ETH$ levels 
is $\dAvg  \gg  \Wb$, with the high probability
$\sim  1 -  ( \Wb / \dAvg )^2$.
Hence the engine likely outputs $W_\tot>0$.
Hot thermalization (stroke 4) returns the engine to $\rho(0)$.

%
%
%
\subsection{Quantitative analysis of the mesoscale engine in the adiabatic limit}
\label{section:Quant_main}

How well does the mesoscale Otto engine perform?
We calculate average work $\expval{ W_\tot }$
outputted per cycle
and the efficiency $\eta_\MBL$.
Details appear in App.~\ref{section:PowerApp}.

We focus on the parameter regime in which 
the cold bath is very cold, 
the cold-bath bandwidth $\Wb$ is very small, and
the hot bath is very hot:
$\TCold   \ll   \Wb  \ll  \dAvg$, and
$\sqrt{ \Sites }  \:  \betaH \HScale  \ll  1$.
The mesoscale engine resembles a qubit engine
whose state and gaps are averaged over.
The gaps, $\delta_j$ and $\delta'_j$,
obey the distributions $P_\ETH^\ParenE(\delta_j)$ and 
$P_\MBL^\ParenE(\delta'_j)$
[Eqs.~\eqref{eq:P_ETH_Main} and~\eqref{eq:P_MBL_Main}].
Correlations between the $H_\ETH$ and $H_\MBL$ spectra 
can be neglected.

We make three simplifying assumptions, generalizing later:
(i) The engine is assumed to be tuned quantum-adiabatically.
Diabatic corrections are estimated in Sec.~\ref{section:Diab_main}.
(ii) The hot bath is at $\THot = \infty$.
We neglect finite-temperature corrections, which scale as 
$\Sites ( \betaH \HScale )^2  \frac{ ( \Wb )^2 }{ \dAvg }$
and are calculated numerically in Suppl.~Mat.~\ref{section:PowerApp}.
(iii) The gap distributions vary negligibly with energy:
$P_\ETH^\ParenE(\delta_j)  \approx  P_\ETH ( \delta_j )$, and 
$P_\MBL^\ParenE(\delta'_j)  \approx  P_\MBL ( \delta'_j )$,
wherein $\dAvg_E  \approx  \dAvg$.


%
%
%
\textbf{Average work $\expval{ W_\tot }$ per cycle:} 
The key is whether the cold bath relaxes the engine 
downwards across the MBL-side gap $\delta'  \equiv  \delta'_j$,
distributed as $P_\MBL ( \delta' )$,
during a given trial. 
If $\delta' < \Wb$, the engine has a probability
$1  /  (1  +  e^{-  \betaC  \delta}  )$
of thermalizing.
Hence the overall probability of relaxation by the cold bath is 
\begin{align}
   \label{eq:PCold}
   p_\cold  \approx  \int\limits_{0}^{\Wb}  d\delta' \;
   \frac{1}{\dAvg }   \frac{  e^{-\delta'  /  \dAvg}  }{ 1  +  e^{- \betaC \delta' }  } \approx\frac{1}{\dAvg}\left(\Wb- T_C \ln 2\right),
\end{align}
wherein we neglected $\Wb/\dAvg$ by setting 
$e^{-\delta'  /  \dAvg}\approx 1$. 

Alternatively, the cold bath could excite the engine
to a level a distance $\delta'$ above the initial level. 
Such an upward hop occurs with a probability
\begin{align}
   \label{eq:PbarCold}
   \bar{p}_\cold  \approx  \int\limits_{0}^{\Wb}  d\delta' \;
   \frac{ e^{-\delta'  /  \dAvg}}{\dAvg } \frac{e^{-\betaC \delta'}}{ 1  +  e^{- \betaC \delta' }  } \approx \frac{T_C \ln 2}{\dAvg}  \, .
\end{align}

If the engine relaxed downward during stroke 2, 
then upon thermalizing with the hot bath during stroke 4, 
the engine gains heat $\expval{Q}_4  \approx  \dAvg$, 
on average. 
If the engine thermalized upward during stroke 2, 
then the engine loses $\dAvg$ during stroke 4,
on average. 
Therefore, the cycle outputs average work
\begin{equation}
   \expval{ W_\tot } 
   \approx \left( p_\cold -\bar{p}_\cold\right) \dAvg  +  \expval{ Q_2 }  
   \label{eq:WTotApprox2_Main}
   \approx   \Wb  -  \frac{ 2 \ln 2 }{ \betaC }  \, .
\end{equation} 
$\expval{ Q_2 }$ denotes 
the average heat absorbed by the engine during cold thermalization:
\begin{equation}
   \expval{ Q_2 }   \approx   - \int\limits_{0}^{\Wb}   d\delta' \;
    \frac{ \delta' }{\dAvg}   
   \frac{  e^{-\delta' / \dAvg}  }{ 1  +  e^{-  \betaC  \delta' }}
   \approx  - \frac{ ( \Wb )^2}{2 \dAvg}  \, ,
\end{equation}
which is $\ll  \expval{ Q_4 }$.
This per-cycle power scales with the system size $\Sites$ as\footnote{
The \emph{effective bandwidth} is defined as follows.
The many-body system has a Gaussian density of states:
$\DOS(E)  \approx  
\frac{ \HDim }{  \sqrt{ 2 \pi \Sites }  \,  \HScale  }  \,
e^{ - E^2 / 2 \Sites  \HScale^2 }$.
The states within a standard deviation $\HScale \sqrt{ \Sites }$
of the mean obey 
Eqs.~\eqref{eq:P_MBL_Main} and~\eqref{eq:P_ETH_Main}.
These states form the effective band,
whose width scales as $\HScale  \sqrt{ \Sites }$.}
$\Wb  \ll  \dAvg  
\sim  \frac{ \text{effective bandwidth} }{ \text{\# energy eigenstates} }
\sim  \frac{ \HScale \sqrt{ \Sites } }{\HDim} $. 
\textbf{Efficiency $\eta_\MBL$:}
The efficiency is
\begin{align}
   \label{eq:Eff_MBL}  
    \eta_\MBL 
    = \frac{ \expval{ W_\tot } }{ \expval{ Q_4 } }
    =  \frac{ \expval{ Q_4 }  +  \expval{ Q_2 } }{ \expval{ Q_4 } }
    \approx  1  -  \frac{ \Wb }{ 2 \dAvg } \, .
\end{align}
The imperfection is small, $\frac{\Wb}{ 2 \dAvg}  \ll  1$,
because the cold bath has a small bandwidth. 
This result mirrors the qubit-engine efficiency $\eta_\qubit$.\footnote{
$\eta_\MBL$ is comparable also to $\eta_\QHO$ [Eq.~\eqref{eq:Eff_QHO}].
Imagine operating an ensemble 
of independent QHO engines.
Let the $j^\th$ QHO frequency be tuned between 
$\Omega_j$ and $\omega_j$, distributed according to 
$P_\ETH ( \Omega_j )$ and
$P_\MBL ( \omega_j )$.
The average MBL-like gap $\omega_j$,
conditioned on $\omega_j  \in  [ 0 , \Wb ]$, is
$\expval{ \omega_j }     \sim \frac{ 1 }{ \Wb / \dAvg }
   \int_0^{ \Wb }  d \omega_j  \,  \omega_j  \,
   P_\MBL ( \omega_j )
   \approx  \frac{1}{ \Wb }  \int_0^{ \Wb }  d \omega_j  \;   \omega_j  
   =  \frac{ \Wb }{ 2 }  \, .$
Averaging the efficiency over the QHO ensemble yields
$\expval{ \eta_\QHO }
   :=  1  -  \frac{ \expval{ \omega } }{ \expval{ \Omega} }
   \approx  1 - \frac{ \Wb }{ 2 \dAvg }  
   \approx  \eta_\MBL  \, .$
The mesoscale MBL engine operates at the ideal average efficiency
of an ensemble of QHO engines.
But MBL enables qubit-like engines to pack together densely
in a large composite engine.}
But our engine is a many-body system
of $\Sites$ interacting sites.
MBL will allow us to employ
segments of the system as 
independent qubit-like subengines, despite interactions.
In the absence of MBL, each subengine's effective $\dAvg = 0$.
With $\dAvg$ vanishes the ability to extract $\expval{ W_\tot } > 0$.
Whereas the efficiency is nearly perfect, 
an effective engine requires also a finite power. 
The MBL engine's power will depend on dynamics, 
as discussed below.


\section{MBL engine in the thermodynamic limit}
\label{section:Thermo_limit_main}

The MBL engine's advantage lies in having 
a simple thermodynamic limit 
that does not compromise efficiency or power output. 
A nonlocalized Otto engine would suffer from 
a suppression of the average level spacing:
$\dAvg  \sim  \frac{  \HScale  \sqrt{ \Sites }  }{  2^\Sites  }$, 
which suppresses the average output per cycle,
$\expval{W_\tot}  \sim  \Wb  \ll  \dAvg$,
exponentially in the system size. 
Additionally, the tuning speed $v$ must shrink exponentially:
$H_\meso(t)$ is ideally tuned quantum-adiabatically.
The time per tuning stroke must far exceed 
$\dAvg^{ - 1 }$.
The mesoscale engine scales poorly, but properties of MBL offer a solution.

A thermodynamically large MBL Otto engine consists of 
mesoscale subengines that operate mostly independently. 
This independence hinges on \emph{local level correlations} of the MBL phase~\cite{Sivan_87_Energy,imryma,Syzranov_17_OTOCs}: 
Subsystems separated by a distance $L$
evolve roughly independently 
until times exponential in $L$, due to the localization~\cite{Nandkishore_15_MBL}.
%

Particularly important is the scaling of 
the typical strength of 
a local operator in an MBL phase. 
Let $O$ denote 
a generic local operator
that has support on just a size-$L$ region.
$O$ can connect only energy eigenstates 
$\ket{ \psi_1 }$ and $\ket{ \psi_2 }$ 
that differ just in their local integrals of motion 
in that region. 
Such states are said to be ``close together,'' 
or ``a distance $L$ apart .'' 
Let $\xi$ denote the system's localization length.
If the eigenfunctions lie far apart ($L \gtrsim \xi$), 
the matrix-element size scales as 
\begin{align}
   \label{eq:O_Far}
   | O_{21} |  \sim  2^{ -L }    e^{ - L / \xi } \, .
\end{align}
(All lengths appear in units of the lattice spacing, set to one.)
This scaling determines the typical level spacing, 
since such matrix elements give rise to level repulsion:
\begin{align}
   \label{eq:delta}
   \delta  \sim  \HScale  2^{ -L }    e^{ - L / \xi } \, 
\end{align}
(possibly to within  a power-law correction). 
The localization-induced exponential suppresses 
long-distance communication
(see~\cite{Anderson_58_Absence,Sivan_87_Energy,BAA,Nandkishore_15_MBL}
and App.~\ref{section:ThermoLimitApp}).


Let us apply this principle to a chain of 
$\Sites$-site mesoscale engines
separated by $\Sites$-site buffers.
The engine is cycled between
a shallowly localized ($H_\ETH$-like) Hamiltonian, 
which has a localization length  $\xi_\Loc$,
and a deeply localized ($H_\MBL$-like) Hamiltonian, 
which has $\xi_\VeryLoc  \ll  \xi_\Loc$.

The key element in the construction is that 
the cold bath acts through local operators 
confined to $< \Sites \sim\xi_\Loc$ sites. 
This defines the subengines of 
the thermodynamic MBL Otto engine.
Localization guarantees that ``what happens in a subengine stays in a subengine'':
Subengines do not interfere much with each other's operation.

This subdivision boosts the engine's power.
A length-$\Sites$ mesoscale engine operates at
the average per-cycle power
$\expval{ W_\tot }_\meso 
\sim  \Wb  
\ll  \frac{ \HScale \sqrt{\Sites} }{ 2^\Sites }$
(Sec.~\ref{section:Quant_main}).  
A subdivided length-$\Sites_\macro$ MBL engine 
outputs average work
$\sim  \frac{\Sites_\macro}{2 \Sites}  
\expval{W_\tot}_\meso$. 
In contrast, if the length-$\Sites_\macro$ engine were not subdivided,
it would output average work
$\sim   \frac{ \HScale \sqrt{\Sites_\macro} }{ 
2^{\Sites_\macro} }$,
 which vanishes in the thermodynamic limit.

\section{Time-scale restrictions on 
the MBL Otto engine's operation}
\label{section:Diab_main}

\begin{figure}[tb]
\centering
\includegraphics[width=.45\textwidth, clip=true]{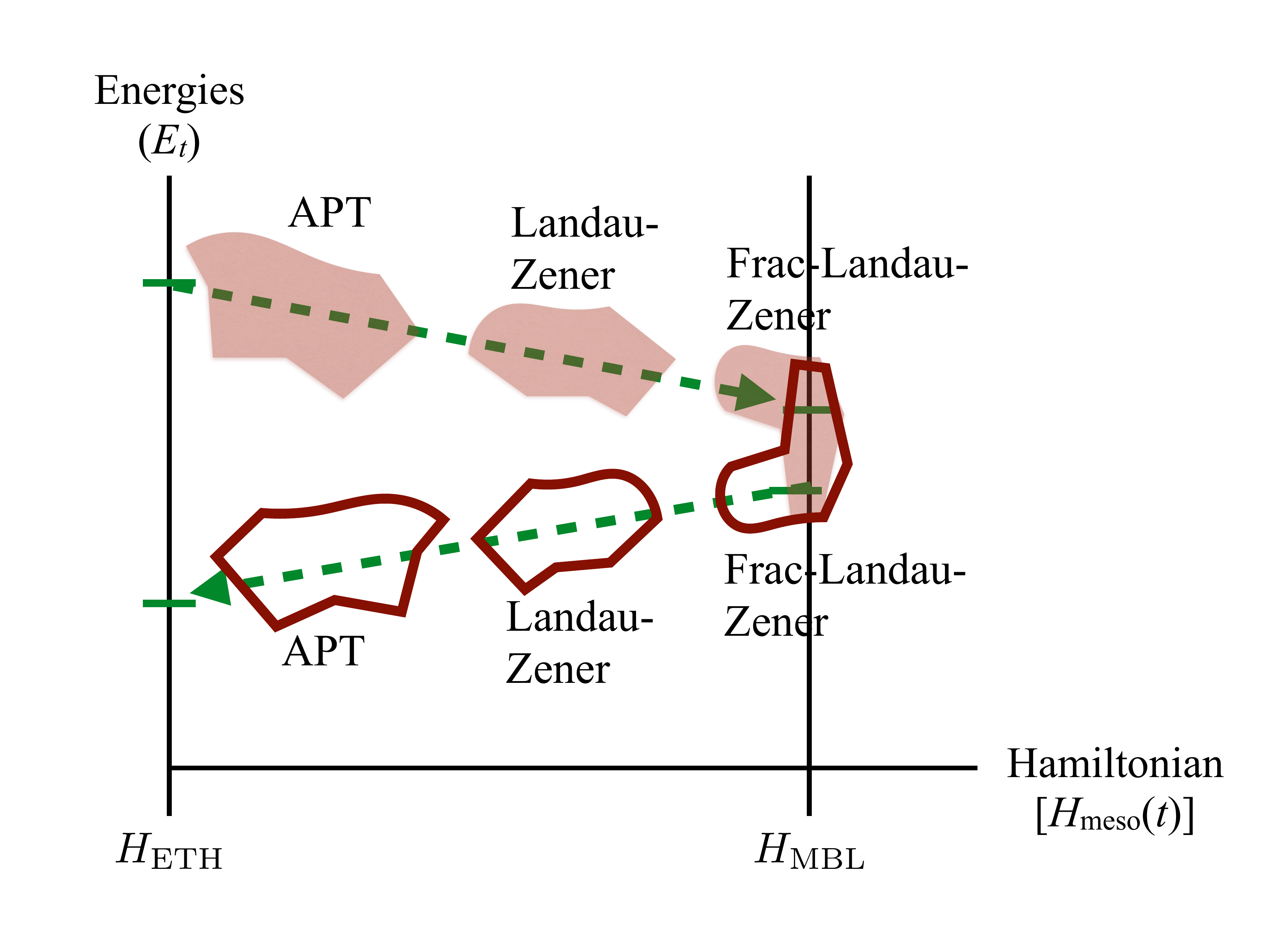}
\caption{\caphead{Three (times two) classes of diabatic transitions:}
Hops to arbitrary energy levels, modeled with
general adiabatic perturbation theory (APT),
plague the ETH regime.
Landau-Zener transitions and fractional-Landau-Zener transitions
plague the many-body-localized regime.}
\label{fig:Diab_Types}
\end{figure}

\label{section:Times_main}

We estimate the restrictions on 
the speed with which the Hamiltonian must be tuned 
to avoid undesirable diabatic transitions
and intersubengine communication.
Most importantly, we estimate the time required
for cold thermalization (stroke 2). 

\subsection{Diabatic corrections}
\label{ss:diabatic-correction}
We have modeled the Hamiltonian tuning
as quantum-adiabatic, but realistic tuning speeds 
$v  :=  \HScale \left\lvert  \frac{ d \alpha_t }{ dt }  \right\rvert$ 
are finite.
To understand diabatic tuning's effects, 
we distinguish the time-$t$ density matrix $\rho(t)$ 
from the corresponding diagonal ensemble,
  \begin{align} 
  \label{eq_Diag_Ens}
  \begin{split}
    & \rho_{\mathrm{diag}} (t)
    = \sum_j \ket{E_j(t)}   \varepsilon_j(t)  \bra{E_j(t)}  
    \, , \quad \text{wherein} \\
    & \varepsilon_j(t) 
    = \bra {E_j(t)} \rho \ket{E_j(t)}
    \end{split} \end{align}
and $\ket{E_j(t)}$ denotes an instantaneous energy eigenbasis of
$H_{\meso}(t) = \sum_j \ket{E_j(t)} E_j(t) \bra{E_j(t)}$.
The average energy depends on $\rho(t)$ only through
$\rho_{\mathrm{diag}}(t)$.
[More generally, the state's off-diagonal elements 
dephase under the dynamics.
$\rho_{\rm diag}(t)$ is ``slow'' and captures
most of the relevant physics~\cite{D'Alessio_16_From}.]

In the adiabatic limit,
$\varepsilon_j(t) = \varepsilon_j(0)$.
We seek to understand how this statement breaks down 
when the tuning proceeds at a finite speed $v$.
It is useful to think of ``infinite-temperature thermalization'' 
in the sense of this diagonal ensemble: 
Fast tuning may push the diagonal-ensemble weights 
$\varepsilon_j(t)$ towards uniformity---even 
though the process is unitary 
and the entropy $S = -\rho (t) \ln\rho (t)$ remains constant---thanks 
to the off-diagonal elements.

The effects of diabatic tuning appear in three distinct regimes, 
which we label ``fractional-Landau-Zener,'' ``Landau-Zener,'' and ``APT'' 
(Fig.~\ref{fig:Diab_Types}).
We estimate the average per-cycle work costs 
$\expval{ W_\diab}$ of diabatic jumps,
guided by the numerics in Sec.~\ref{section:Numerics_main}. 
We focus on $\THot = \infty$ and $\TCold = 0$, for simplicity. 
Since $\THot = \infty$, diabatic hops cannot bring 
$\rho_{\rm diag}(t)$ closer to $\id / 2^\Sites$---cannot
change the average energy---during stroke 1.
Hence we focus on stroke 3.

\subsubsection{Fractional-Landau-Zener transitions}
\label{sss:frac-LZ}
At the beginning of stroke 3, 
nonequilbrium effects could excite the system 
back across the small gap to energy level $j$. 
The transition would cost work
and would prevent the trial from outputting $W_\tot > 0$.
We dub this excitation a \emph{fractional-Landau-Zener (frac-LZ) transition.}
It could be suppressed by 
a sufficiently slow drive \cite{DeGrandi_10_APT}.
The effects, and the resultant bound on $v$,
are simple to derive.
  
Let the gap start stroke 3 at size $\delta$
and grow to a size $\Delta > \delta$.
Because the two energy levels begin close together, 
one cannot straightforwardly apply the Landau-Zener formula.
One must use the fractional-Landau-Zener result of De Grandi and Polkovnikov~\cite{DeGrandi_10_APT},
\begin{align}
   \label{eq:HalfLZ_mainP}
   p_{ \HalfLZ }(\delta)  
   \approx   \frac{ v^2 ( \deltaMBL )^2 }{ 16}    
    \left(   \frac{1}{ \delta^6 }  +  \frac{1}{\Delta^6}  \right)
    \approx    \frac{  v^2 ( \deltaMBL )^2  }{ 16 \delta^6 }  \, .
\end{align}
$\deltaMBL$ denotes the MBL level-repulsion scale,
the characteristic matrix element 
introduced by a perturbation between 
eigenstates of an unperturbed Hamiltonian.
We suppose that energy-level pairs with $p_{\HalfLZ} \lesssim 1$ 
are returned to the infinite-temperature state 
from which the cold bath disturbed them.
These pairs do not contribute to $\expval{ W_\tot }$. 
Pairs that contribute have $p_\HalfLZ < 1$, i.e.,
  \begin{equation}
    \delta > (v \delta_-)^{1/3}  \, .
  \end{equation}
If the rest of the stroke is adiabatic, 
the average work performed during the cycle is
\begin{align}
   \expval{ W_\tot }  
   \sim  \expval{Q_4}  -  \expval{Q_2}  -   (v \delta_-)^{1/3}  \, ,
\end{align}
which results immediately in the correction
\begin{equation}
    \expval{ W_{\mathrm{diab},\HalfLZ} } 
    \sim (v\delta_-)^{1/3}  \, .
\end{equation}
This correction is negligible at speeds low enough that
\begin{equation}
    \label{eq:fraclz-small}
    v \ll  \frac{ ( \Wb )^3 }{ \delta_- }  \, .
\end{equation}

\subsubsection{Landau-Zener transitions}
\label{LZtrans}
While the system is localized, the disturbances induced by the tuning 
$\frac{ dH(t) }{dt}$ can propagate only a short distance $l_v$.
The tuning effectively reduces the mesoscale engine 
to a length-$l_v$ subengine. 
To estimate $l_v$, we compare 
the minimum gap of a length-$l_v$ subsystem to the speed $v$:
\begin{equation}
    \HScale 2^{-l_v}e^{-l_v/\xi_<} \sim \sqrt v\;.
    \label{vscale}
\end{equation}
The left-hand side comes from Eq.~\eqref{eq:delta}.
This minimum gap---the closest that two levels are likely to approach---is
given by the smallest level-repulsion scale, $\deltaMBL$.
$\deltaMBL$ characterizes the deeply localized system,
whose $\xi = \xi_\VeryLoc$. Consequently,
\begin{equation}
    l_v 
    \sim \frac{\ln (\HScale^2 / v)}{2\Big(\ln 2 + \frac 1 \xi_<\Big)}  \, .
\end{equation}

Suppose that $l_v \le \Sites$, 
and consider a length-$l_v$ effective subengine. 
In the adiabatic limit, $\expval{ W_\tot }$ 
does not depend on the engine's size. 
($\expval{ W_\tot }$ depends only on 
the bath bandwidth $\Wb \ll \dAvg$.) 
To estimate how a finite $v$ changes $\expval{ W_\tot }$,
we consider the gaps $\delta < \Wb$
of the size-$l_v$ subengine.
We divide the gaps into two classes:
\begin{enumerate}

   \item \label{shortgap} 
   Gaps connected by flipping l-bits 
   on a region of diameter $l < l_v$. 
   The tuning is adiabatic with respect to these gaps, 
   so they result in work output.
   
   %
   \item \label{longgap} 
   Gaps connected by flipping l-bits on a region of diameter $l = l_v$. 
   The tuning is resonant with these gaps
   and so thermalizes them, in the sense of the diagonal ensemble
   [Eq.~\eqref{eq_Diag_Ens}]:
   The tuning makes the instantaneous-energy-eigenvector weights 
   $\varepsilon_j$ uniform, on average.
\end{enumerate}
Type-\ref{shortgap} gaps form
a $v$-independent $O(1)$ fraction $\theta$
of the length-$l_v$ subengine's short-length-scale gaps.\footnote{
We can estimate $\theta$ crudely.
For a given diameter-$l_v$ subset, 
each gap connected by a diameter-$(l_v - 1)$ operator 
can be made into a diameter-$l_v$ gap:
One flips the last $(l_v)^\th$ l-bit. 
Adding a qubit to the system doubles 
the dimensionality of the system's Hilbert space. 
The number of levels doubles, 
so the number of gaps approximately doubles, 
so $\theta \approx 1/2$. 
This estimate neglects several combinatorial matters.
A more detailed analysis would account for 
the two different diameter-$(l_v-1)$ regions 
of a given length-$l_v $ subengine,
gaps connected by l-bit flips in the intersections of those subengines,
the number of possible diameter-$l_v$ subengines 
of an $\Sites$-site system, etc.}
Type-\ref{longgap} gaps therefore make up a fraction $1 - \theta$.
Hence Landau-Zener physics leads to 
a $v$-independent $O(1)$ diabatic correction $(1 - \theta) \Wb$
to $\expval{ W_\tot }$, 
provided that $v$ is high enough that $l_v < \Sites$.

\subsubsection{Adiabatic-perturbation-theory (APT) transitions}

When the system is in the ETH phase 
(or has correlation length $\xi \sim \Sites$), 
typical minimum gaps (points of closest approach) 
are still given by the level-repulsion scale, 
which is now $\dAvg$. 
Hence one expects the tuning to be adiabatic if
\begin{equation}
   v \ll \dAvg^2 \, .
\end{equation}
This criterion could be as stringent 
(depending on the system size and localization lengths)
as the requirement~\eqref{eq:fraclz-small} that 
fractional Landau-Zener transitions occur rarely.
The numerics in Sec.~\ref{ss:numerics:diabatic} indicate that 
fractional-Landau-Zener transitions limit the power 
more than APT transitions do.

Both fractional Landau-Zener transitions and APT transitions
bound the cycle time $\tau_\cycle$ less stringently than 
thermalization with the cold bath;
hence a more detailed analysis of APT transitions would be gratuitous.
Such an analysis would rely on
the general adiabatic perturbation theory of De Grandi and Polkovnikov~\cite{DeGrandi_10_APT}; hence the moniker ``APT transitions.''

%
%
%
\subsection{Precluding communication between subengines}
\label{ss:preventing-communication}

To maintain the MBL engine's advantage, 
we must approximately isolate subengines. 
The subengines' (near) independence implies
a lower bound on the tuning speed $v$:
The price paid for scalability is the impossibility of adiabaticity. 
Suppose that $H_\macro(t)$ were tuned infinitely slowly.
Information would have time to propagate 
from one subengine to every other.
The slow spread of information through MBL~\cite{Khemani_15_NPhys_Nonlocal}
lower-bounds $v$. This consideration, however, 
does not turn out to be the most restrictive constraint on the cycle time. 
Therefore, we address it only qualitatively.  

As explained in Sec. \ref{LZtrans}, $v$ determines 
the effective size of an MBL subengine. 
Ideally, $v$ is large enough to prevent adiabatic transitions between 
configurations extended beyond the mesoscale $\Sites$. 
For each stage of the engine's operation, 
$v$ should exceed the speed given in Eq.~\eqref{vscale}
for the localization length $\xi$ 
of a length-$(\Sites + 1)$ chain:
\begin{equation}
   \label{eq_v_GG_Help}
   v   \gg [ \delta_{-}(\Sites+1, \xi) ]^2
   \sim   \HScale^2 2^{-2(\Sites+1)}e^{-2(\Sites+1)/\xi}.
\end{equation}
(We have made explicit the dependence 
of the level-repulsion scale $\delta_-$ 
on the mesoscale-engine size $\Sites$ 
and on the localization length $\xi$.)
During stroke 1, $\xi$ drops, so the RHS of~\eqref{eq_v_GG_Help}
decays quickly.
Hence the speed should interpolate between 
$[  \delta_{-}(\Sites  +  1,   \xi_\Loc)  ]^2$ and 
$\frac{ ( \Wb )^3 }{ \delta_-(\Sites,  \xi_\VeryLoc) }$
[from Ineq.~\eqref{eq:fraclz-small}].


\subsection{Lower bound on the cycle time $\tau_\cycle$ from cold thermalization:}
Thermalization with the cold bath (stroke 2) 
bounds $\tau_\cycle$ more stringently
than the Hamiltonian tunings do.
The reasons are (i) the slowness with which MBL thermalizes and
(ii) the restriction $\Wb  \ll  \dAvg$ on the cold-bath bandwidth.
We elaborate after introducing our cold-thermalization model
(see~\cite[App.~I]{NYH_17_MBL} for details).

We envision the cold bath as a bosonic system
that couples to the engine locally, as via the Hamiltonian
\begin{align}
   \label{eq:H_Inter_Main}
   H_\inter  & =  \coupling  
   \int_{ - \Wb /  \xi_\Loc }^{ \Wb / \xi_\Loc } d \omega   
   \sum_{j \in {\rm subengine}}
   \left(  c_j^\dag c_{ j + 1 }   +  \hc  \right)
   \left(  b_\omega  +  b_\omega^\dag  \right)
   \nonumber \\ & \qquad \times
   \delta \LParen  \langle 0 | c_j  H_\macro ( \tau ) c_{j + 1}^\dag | 0 \rangle
                            -  \omega  \RParen  \, .
\end{align}
The sum runs over the sites in the subengines, 
excluding the sites in the buffers between subengines.
The coupling strength is denoted by $\coupling$.
We have switched from spin notation 
to fermion notation via a Jordan-Wigner transformation.
$c_j$ and $c_j^\dag$ denote
the annihilation and creation of a fermion at site $j$.
$H_\macro (t)$ denotes the Hamiltonian that would govern the engine
at time $t$ in the bath's absence.
Cold thermalization lasts from $t = \tau$ to $t = \tau'$ (Fig.~\ref{fig:2Level_v3}).
$b_\omega$ and $b_\omega^\dag$ represent
the annihilation and creation of a frequency-$\omega$ boson in the bath.
The Dirac delta function is denoted by $\delta ( . )$.

The bath couples locally, e.g., 
to pairs of nearest-neighbor spins.
This locality prevents subengines from 
interacting with each other much through the bath.
The bath can, e.g., flip spin $j$ upward while flipping spin $j + 1$ downward.
These flips likely change a subengine's energy by an amount $E$.
The bath can effectively absorb only energy quanta of size
$\leq  \Wb$ from any subengine.
The cap is set by the bath's speed of sound~\cite{KimPRL13},
which follows from microscopic parameters in 
the bath's Hamiltonian~\cite{Lieb_72_Finite}.
The rest of the energy emitted during the spin flips, $| E - \Wb |$, 
is distributed across the subengine
as the intrinsic subengine Hamiltonian flips more spins.

Let $\tau_\therm$ denote the time required for stroke 2.
We estimate $\tau_\therm$ from Fermi's Golden Rule,
\begin{align}
   \label{eq:FGR_Main}
   \Gamma_{fi}  =  \frac{2 \pi }{ \hbar }  
   | \langle f | V | i \rangle |^2  \,  \DOS_\bath  \, .
\end{align}
Cold thermalization transitions the engine
from an energy level $\ket{ i }$ to a level $\ket{ f }$.
The bath has a density of states
$\DOS_\bath  \sim  1 / \Wb$.
$V$ denotes the operator, defined on the engine's Hilbert space, 
induced by the coupling to the bath.

We estimate the matrix-element size
$| \langle f | V | i \rangle |$ as follows.
Cold thermalization transfers energy $E_{if}  \sim  \Wb$
from the subengine to the bath.
$\Wb$ is very small.
Hence the energy change 
rearranges particles across a large distance 
$L  \gg  \xi  =  \xi_\VeryLoc$,
due to local level correlations~\eqref{eq:delta}.
$V$ nontrivially transforms just a few subengine sites.
Such a local operator rearranges particles 
across a large distance $L$
at a rate that scales as~\eqref{eq:delta},
$\HScale e^{ - L / \xi }  \;  2^{ - L }
\sim  \deltaMBL$.
Whereas $\HScale$ sets the scale of the level repulsion $\deltaMBL$,
$\coupling$ sets the scale of $| \langle f | V | i \rangle |$.
The correlation length $\xi = \xi_\VeryLoc$ during cold thermalization.
We approximate $L$ with the subengine length $\xi_\Loc$. Hence
$| \langle f | V | i \rangle |  \sim  \frac{ \coupling \deltaMBL }{ \HScale}$.

We substitute into Eq.~\eqref{eq:FGR_Main}.
The transition rate $\Gamma_{fi}  =  \frac{1}{ \tau_\therm }$.
Inverting yields
\begin{align}
   \label{eq:Tau_therm_main}
   \tau_\cycle  \sim  \tau_\therm  \sim
   \Wb   \left(  \frac{  \HScale }{ \coupling \deltaMBL }  \right)^2  \, .
\end{align}

To bound $\tau_\cycle$, we must bound the coupling $\coupling$.
The interaction is assumed to be Markovian:
Information leaked from the engine dissipates throughout the bath quickly.
Bath correlation functions must decay much more quickly
than the coupling transfers energy.
If $\tau_\bath$ denotes the correlation-decay time,
$\tau_\bath  <   \frac{1}{ \coupling }$.
The small-bandwidth bath's $\tau_\bath  \sim  1 / \Wb$, 
so  $\coupling  <  \Wb$.
This inequality, with Ineq.~\eqref{eq:Tau_therm_main}, implies
\begin{align}
   \label{eq:Markov_main}
   \tau_\cycle
   =  \tau_\therm  
   >  \frac{ \HScale^2 }{  \Wb  ( \deltaMBL )^2  }
   \sim  \frac{ 10 }{ \HScale }  \:
   e^{ 2 \xi_\Loc  / \xi_\VeryLoc }  \:  2^{ 3 \xi_\Loc }  \, .
\end{align}
The final expression follows if $\Wb  \sim \frac{ \dAvg }{ 10 }$.

Like Markovianity, higher-order processes bound $\tau_\therm$.
Such processes transfer energy $E > \Wb$ between
the engine and the cold bath.
These transfers must be suppressed.
$g^a$, wherein $a > 1$,
determine the rates at which these processes occur.
The resulting bound on $\tau_\therm$ 
is less stringent than Ineq.~\eqref{eq:Markov_main}
(App.~\ref{section:Tau_therm_virtual_app}).

\section{Numerical simulations}
\label{section:Numerics_main}
We use numerical exact diagonalization 
to check our analytical results.
In Sec.~\ref{ss:hamiltonian}, we describe 
the Hamiltonian used in our numerics.
In Sec.~\ref{ss:numerics:adiabatic}, we study
engine performance in the adiabatic limit
(addressed analytically in Sec.~\ref{section:Quant_main}).
In Sec.~\ref{ss:numerics:diabatic}, we study diabatic corrections
(addressed analytically in Sec.~\ref{ss:diabatic-correction}).
We numerically study the preclusion of 
communication between mesoscale subengines 
(addressed analytically in Sec.~\ref{ss:preventing-communication})
only insofar as these results follow from diabatic corrections:
Limitations on computational power restricted
the system size to 12 sites.
Details about the simulation appear in App.~\ref{section:MBLNumApp}.
Our code is available at \url{https://github.com/christopherdavidwhite/MBL-mobile}.

\subsection{Hamiltonian}
\label{ss:hamiltonian}
The engine can be implemented with a disordered Heisenberg model.
A similar model's MBL phase has been realized 
with ultracold atoms~\cite{Schreiber_15_Observation}.
We numerically simulated a 1D mesoscale chain governed by a Hamiltonian
\begin{align} 
      \label{eq:SpinHam}
      H_\Sim(t)  = \frac{\HScale}{ Q \LParen h ( \alpha_t ) \RParen }   \Bigg[  
      \sum_{j = 1}^{\Sites - 1}   
      \bm{\sigma}_j  \cdot  \bm{\sigma}_{j+1}
      +  h ( \alpha_t )  \sum_{j = 1}^\Sites
      h_j     \sigma_j^z  \Bigg] \, ;
\end{align}
this is a special case of the general mesoscopic Hamiltonian~\eqref{eq:H_meso_main} described in Sec.~\ref{section:Meso_setup}. 
Equation~\eqref{eq:SpinHam} describes spins 
equivalent to interacting spinless fermions.
Energies are expressed in units of $\HScale$,
the average per-site energy density.
For $\gamma=x,y,z$, the $\gamma^\th$ Pauli operator 
that operates nontrivially on the $j^\th$ site is denoted by 
$\sigma_j^\gamma$.
The Heisenberg interaction
$\bm{\sigma}_j \cdot \bm{\sigma}_{j+1}$ 
encodes nearest-neighbor hopping and repulsion.

The tuning parameter $\alpha_t\in[0,1]$
determines the phase occupied by $H_\Sim(t)$.
The site-$j$ disorder potential depends on 
a random variable $h_j$ distributed uniformly across $[-1,1].$
The disorder strength $h(\alpha_t)$ varies as
$h(\alpha_t)=\alpha_t\,h_\ETH+(1-\alpha_t)h_\MBL$.
When $\alpha_t = 0$, the disorder is weak, $h = h_\ETH$,
and the engine occupies the ETH phase. 
When $\alpha_t = 1$, the disorder is strong, $h = h_\MBL  \gg  h_\ETH$,
and the engine occupies the MBL phase.


The normalization factor $Q\LParen h(\alpha_t)\RParen$
preserves the width of the density of states (DOS)
and so preserves $\dAvg$.
$Q \LParen h(\alpha_t)\RParen$ prevents 
the work extractable via change of bandwidth 
from polluting
the work extracted with help from level statistics (see App.~\ref{section:Bandwidth_engine_app} for a discussion of work extraction from bandwidth change).
$Q \LParen h(\alpha_t)\RParen$ is defined 
and calculated in App.~\ref{section:MBLNumApp:scale}.

The ETH-side field had a magnitude $h(0) = 2.0$, 
and the MBL-side field had a magnitude $h(1) = 20.0$. 
These $h( \alpha_t )$ values fall squarely on opposite sides 
of the MBL transition at $h \approx 7$.

\subsection{Adiabatic engine}
\label{ss:numerics:adiabatic}

\begin{figure}
  \begin{subfigure}{0.45\textwidth}
    \includegraphics[width=\textwidth]{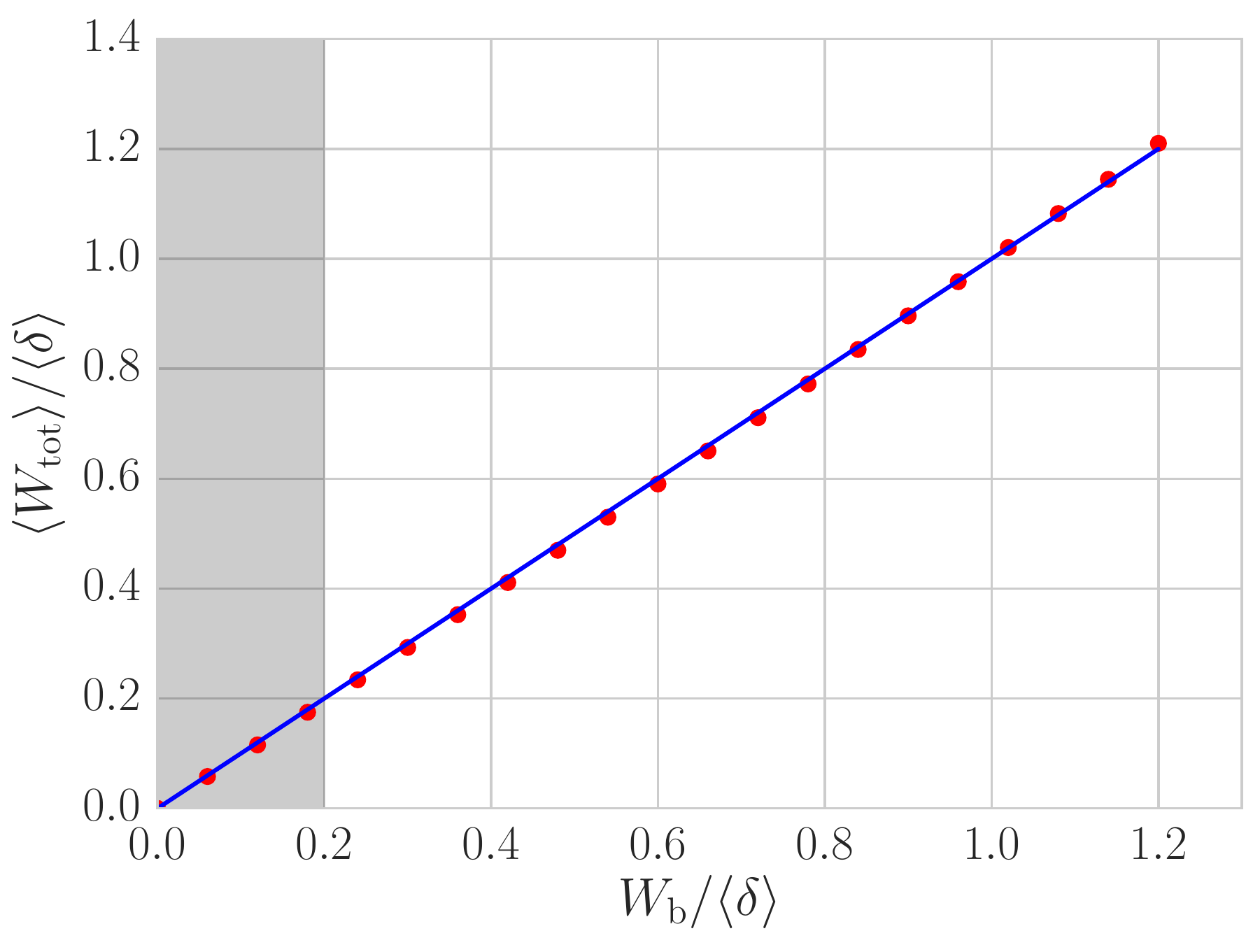}
    \caption{}
    \label{fig:Numerics_main_WTOT}
  \end{subfigure}
  %
  \begin{subfigure}{0.45\textwidth}
    \includegraphics[width=\textwidth]{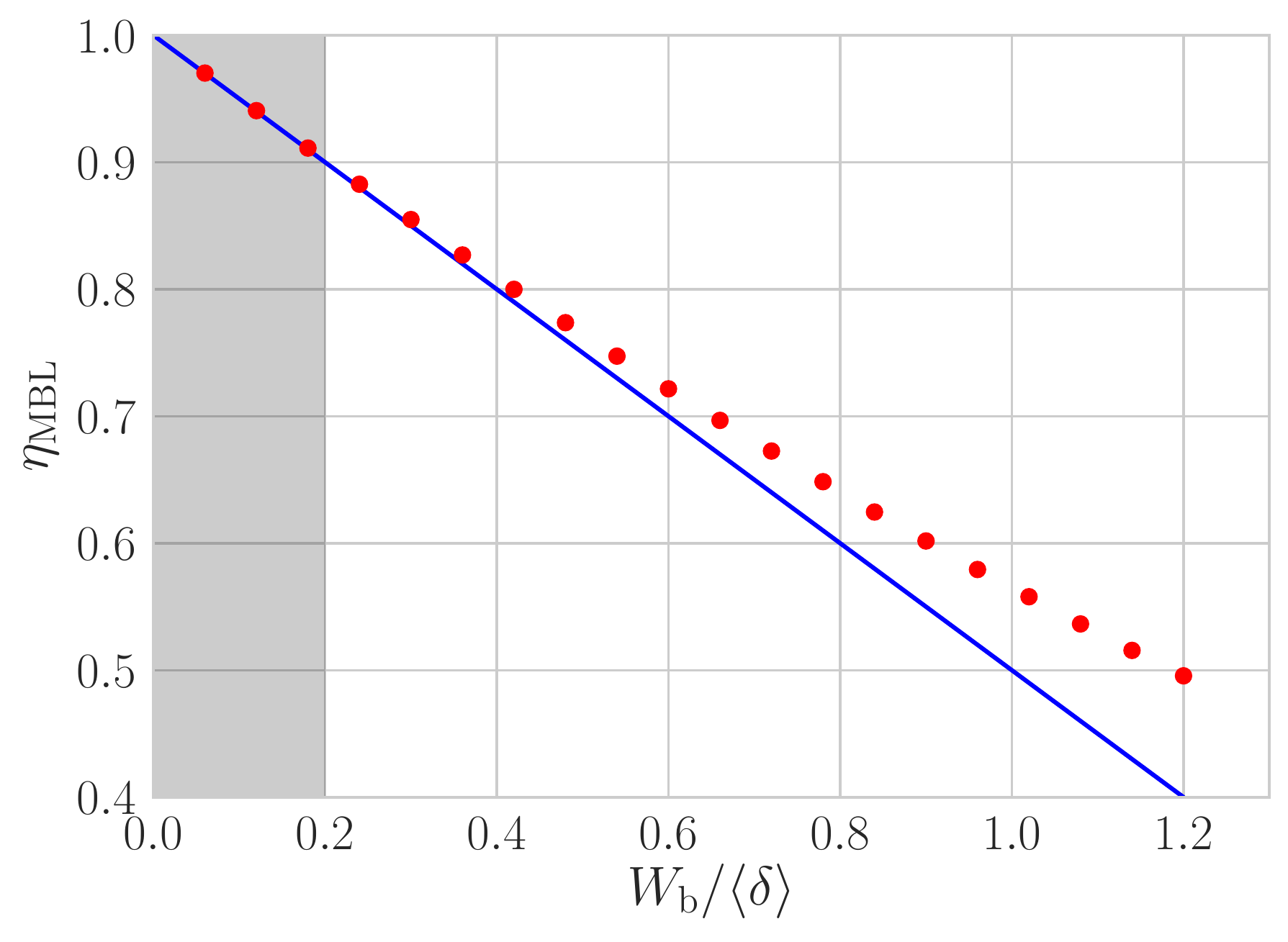}
    \caption{}
    \label{fig:Numerics_main_eta}
  \end{subfigure}
  \caption{\caphead{
    Average per-cycle power $\expval{ W_\tot }$ (top) 
    and efficiency  $\eta_\MBL$ (bottom) as functions of 
    the cold-bath bandwidth $\Wb$:}
    Each red dot represents an average over 1,000 disorder realizations of the random-field Heisenberg Hamiltonian~\eqref{eq:SpinHam}.
    The blue lines represent the analytical predictions~\eqref{eq:WTotApprox2_Main} and~\eqref{eq:Eff_MBL} of Sec.~\ref{section:Quant_main}.
    When $\Wb \ll \dAvg$ (in the gray shaded region), 
    the engine operates in the regime of interest.
    Here, $\expval{ W_\tot }$ and $\eta_\MBL$ vary linearly with $\Wb$, 
    as predicted. The error bars are smaller than the numerical-data points.}
  \label{fig:Numerics_main}
\end{figure} 

We compare the analytical predictions of of Sec.~\ref{section:Quant_main} and App.~\ref{section:PowerApp} to numerical simulations of a 12-site engine governed by the Hamiltonian \eqref{eq:SpinHam}. 
During strokes 1 and 3,
the state was evolved
as though the Hamiltonian were tuned adiabatically.
We index the energies $E_j ( t )$
from least to greatest at each instant:
$E_j ( t )  <  E_k ( t )  \;  \forall j < k$.
Let $\rho_j$ denote the state's weight on eigenstate $j$ 
of the initial Hamiltonian, whose $\alpha_t = 0$. 
The engine ends stroke 1 with weight $\rho_j$ 
on eigenstate $j$ of 
the post-tuning Hamiltonian, whose $\alpha_t = 1$.

The main results appear in Fig.~\ref{fig:Numerics_main}.
Figure~\ref{fig:Numerics_main_WTOT} shows 
the average work extracted per cycle, $\expval{ W_\tot }$.
Figure~\ref{fig:Numerics_main_eta} shows 
the efficiency, $\eta_\MBL$.

In these simulations, the baths had
the extreme temperatures $\THot = \infty$ and $\TCold = 0$.
This limiting case elucidates the $\Wb$-dependence 
of $\expval{ W_\tot }$ and of $\eta_\MBL$:
Disregarding finite-temperature corrections,
on a first pass, builds intuition.
Finite-temperature numerics appear alongside 
finite-temperature analytical calculations
in App.~\ref{section:PowerApp}.

Figure~\ref{fig:Numerics_main} shows how 
the per-cycle power and the efficiency
depend on the cold-bath bandwidth $\Wb$.
As expected, $\expval{ W_\tot }  \approx  \Wb$. 
The dependence's linearity, and the unit proportionality factor,
agree with Eq.~\eqref{eq:WTotApprox2_Main}.
Also as expected, the efficiency declines as the cold-bath bandwidth rises:
$\eta_\MBL  \approx  1  -  \frac{ \Wb }{ 2 \dAvg } \, .$
The linear dependence and the proportionality factor
agree with Eq.~\eqref{eq:Eff_MBL}.

The gray columns in Fig.~\ref{fig:Numerics_main} highlight the regime
in which the analytics were performed, where $\frac{ \Wb }{ \dAvg }  \ll  1$.
If the cold-bath bandwidth is small, $\Wb < \dAvg$, 
the analytics-numerics agreement is close.
But the numerics agree with the analytics even outside this regime.
If $\Wb \gtrsim \dAvg$, the analytics slightly underestimate $\eta_\MBL$:
The simulated engine operates more efficiently than predicted.
To predict the numerics' overachievement,
one would calculate higher-order corrections 
in App.~\ref{section:PowerApp}:
One would Taylor-approximate to higher powers,
modeling subleading physical processes.
Such processes include the engine's dropping across
a chain of three small gaps, $\delta'_1 \, ,  \delta'_2 \, ,  \delta'_3 < \Wb$,
during cold thermalization.

The error bars are smaller than the numerical-data points.
Each error bar represents the error in the estimate of a mean
(of $\expval{ W_\tot }$ or of
$\eta_\MBL  :=  1  -  \frac{ \expval{ W_\tot } }{ \expval{ Q_\In } }$)
over 1,000 disorder realizations.
Each error bar extends a distance
$(\text{sample standard deviation})/\sqrt{\text{\# realizations}}$ 
above and below that mean.

\subsection{Diabatic engine}
\label{ss:numerics:diabatic}

We then simulated strokes 1 and 3 as though 
$H_\Sim(t)$ were tuned at finite speed $v$.
Computational limitations restricted the engine to 8 sites.
(That our upper bounds on $v$ scale as powers of 
$\dAvg \sim 2^{-\Sites}$ implies that 
these simulations quickly become slow to run.)
We simulate a stepwise tuning, taking
  \begin{equation}
    \alpha_t 
    =  \frac{  \delta t  \,  \lfloor t/\delta t \rfloor  }{T}  \, .
  \end{equation}
$\delta t$ denotes a time-step size, and 
$T  \propto  \frac{ h_\MBL - h_{\mathrm{GOE}} }{ v }$ 
denotes the time for which one tuning stroke lasts. 
This protocol is more violent than the protocols treated analytically: 
$v$ is assumed to remain finite in the diabatic analytics.
In the numerics, we tune by sudden jumps
(for reasons of numerical convenience). 
We work at $\THot = \infty$ and $\TCold = 0$---again,
to capture the essential physics without 
the complication of finite-temperature corrections.

\begin{figure}
    \includegraphics[width=0.45\textwidth]{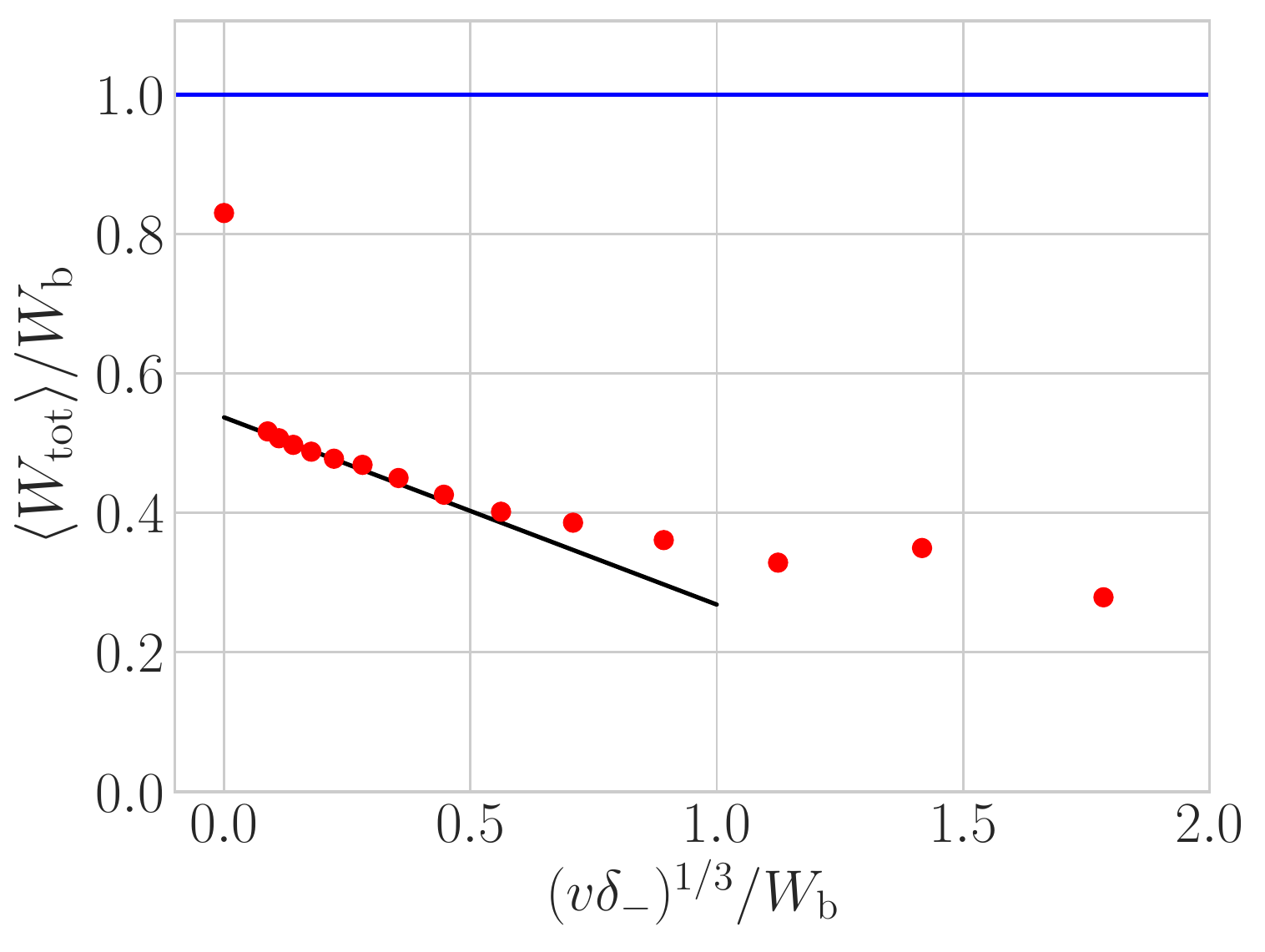}
    \caption{\caphead{Average per-cycle work as a function of tuning speed:} 
    We numerically simulated 995 disorder realizations of 
    the random-field Heisenberg Hamiltonian~\eqref{eq:SpinHam} 
    for a system of $\Sites = 8$ sites (red dots).
    The results are compared to the analytical estimate~\eqref{eq:WTotApprox2_Main} 
    for the adiabatic work output (blue line) 
    and an empirical straight-line fit 
    $W_\tot  = W_0 - W_1 (v \delta_-)^{1/3}/ \Wb$
    (black line). 
    Errors in the estimate of the mean, computed as 
    $(\text{sample standard deviation})/\sqrt(\#\ \text{realizations})$, 
    lead to error bars smaller than the numerical-data points.
    }
    \label{fig:Numerics_main_diabatic-work}
\end{figure}

Figure~\ref{fig:Numerics_main_diabatic-work} shows
the average work output, $\expval{ W_\tot }$,
as a function of $v$. 
Despite the simulated protocol's violence, 
both a fractional-Landau-Zener correction $W_{\HalfLZ} \sim (v \delta_-)^3$, 
explained in Sec.~\ref{sss:frac-LZ}, 
and a $v$-independent $O(1)$ Landau-Zener correction, 
explained in Sec.~\ref{LZtrans}, are visible.
We believe that the adiabatic numerics 
($v = 0$ red dot) differ from the analytics (blue line) 
due to finite-size effects: 
For small systems away from the spectrum's center, 
the average gap estimated from the density of states 
can vary appreciably over one gap.
These numerics confirm the analytics
and signal the MBL Otto engine's robustness 
with respect to changes in the tuning protocol.

\section{Order-of-magnitude estimates}
\label{section:Order_main}

How well does the localized engine perform?
We estimate the engine's power and power density,
in addition to comparing the engine with three competitors.

\textbf{Localized engine:}
Localization has been achieved in solid-state systems.\footnote{
This localization is single-particle, or Anderson~\cite{Anderson_58_Absence}, 
rather than many-body.
Suppl.~Mat.~\ref{section:Anderson_engine_main} extends
the MBL Otto engine to an Anderson-localized Otto engine.}
Consider silicon doped with phosphorus~\cite{Kramer_93_Localization}.
A distance of $\sim  10  \text{ nm}$
may separate phosphorus impurities.
Let our engine cycle's shallowly localized regime
have a localization length of $\xi_\Loc \sim 10$ sites, 
or $100  \text{ nm}$.
The work-outputting degrees of freedom will be electronic.
The localized states will correspond to energies 
$\HScale \sim 1 \text{ eV}$. 
Each subengine's half-filling Hilbert space has dimensionality 
$\HDim  =  {10 \choose 5}  \sim  10^2$.
Hence each subengine has an effective average gap
$\dAvg  \sim  \frac{ \HScale  \sqrt{\Sites} }{ \HDim }  
\sim  \frac{ 1  \text{ eV}  }{ 10^2 }  \sim  10  \text{ meV}$.
The cold-bath bandwidth must satisfy $\dAvg \gg \Wb \, .$
We set $\Wb$ to be an order of magnitude down from $\dAvg$:
$\Wb   \sim  1   \text{ meV}  \sim 10  \text{ K}$. 
The cold-bath bandwidth approximates 
the work outputted by one subengine per cycle:\footnote{
The use of semiconductors would require corrections to our results.
(Dipolar interactions would couple the impurities' spins.
Energy eigenfunctions would decay as power laws with distance.)
But we aim for just a rough estimate.}
$\expval{ W_\tot }  \sim  \Wb
\sim  1   \text{ meV}$
[Eq.~\eqref{eq:WTotApprox2_Main}].

What volume does a localized subengine fill?
Suppose that the engine is three-dimensional (3D).\footnote{
Until now, we have supposed that the engine is 1D.
Anderson localization, which has been realized in semiconductors,
exists in all dimensionalities.
Yet whether MBL exists in dimensionalities $D > 1$ 
remains an open question.
Some evidence suggests that MBL exists in $D \geq 2$~\cite{Choi_16_Exploring,Kucsko_16_Critical,Bordia_17_Probing}.
But attributing a 3D volume to the engine
facilitates comparisons with competitors.
We imagine 10-nm-long 1D strings of sites.
Strings are arrayed in a plane, separated by 10 nm.
Planes are stacked atop each other, separated by another 10 nm.}
A little room should separate the subengines.
Classical-control equipment requires more room.
Also, the subengine needs space to connect to the baths.
We therefore associate each subengine with a volume of
$V  \approx  (100 \text{ nm})^3$.

The last element needed is the cycle time, $\tau_\cycle$. 
We choose for $\deltaMBL$ to be 
a little smaller than $\Wb$---of the same order: 
$\deltaMBL  \sim  \Wb  \sim  1 \text{ meV}$.
In the extreme case allowed by Ineq.~\eqref{eq:Markov_main},
$\tau_\cycle  \sim  \frac{ \hbar \HScale^2 }{ \Wb ( \deltaMBL )^2 }
\sim  \frac{ \hbar  \HScale^2 }{ ( \Wb )^3 }
\sim  \frac{ ( 10^{ -15} \text{ eV s}  )  ( 1 \text{ eV} )^2 }{
         ( 1 \text{ meV} )^3 }
\sim  1 \text{ $\mu$s}$.

The localized engine therefore operates with a power 
$\Power  \sim  \frac{ \Wb }{ \tau_\cycle }
\sim  \frac{ 1 \text{ meV}  }{  1 \text{ $\mu$s} }
\approx 10^{-16}  \text{ W}$. 
Interestingly, this $\Power$ is one order of magnitude greater than 
a flagellar motor's~\cite{Brown_13_Bacterial} power,
according to our estimates.

We can assess the engine by calculating not only its power, 
but also its power density.
The localized engine packs a punch at 
$\frac{ \Power }{ V }  
\sim  \frac{ 10^{-16} \text{ W}  }{  ( 10^{-7} \text{ m} )^3 }
=  100 \text{ kW}/\text{m}^3$.

\textbf{Car engine:}
The quintessential Otto engine powers cars.
A typical car engine outputs
$\Power \sim 100 \text{ horsepower}  \sim  100 \text{ kW} \, .$
A car's power density is $\frac{ \Power }{ V }  
\sim  \frac{ 100 \text{ kW} }{ 100 \text{ L} }
=  1  \text{ MW} / \text{ m}^3$
(wherein L represents liters).
The car engine's $\frac{ \Power }{ V }$ exceeds 
the MBL engine's
by only an order of magnitude,
according to these rough estimates.

\textbf{Array of quantum dots:}
MBL has been modeled with
quasilocal bits~\cite{Huse_14_phenomenology,Chandran_15_Constructing}.
A string of ideally independent bits or qubits, such as quantum dots,
forms a natural competitor.
Each quantum dot would form a qubit Otto engine 
whose gap is shrunk, widened, and shrunk~\cite{Geva_92_Quantum,Geva_92_On,Feldmann_96_Heat,He_02_Quantum,Alvarado_17_Role}. 

A realization could consist of double quantum dots~\cite{Petta_05_Coherent,Petta_06_Charge}. 
The scales in~\cite{Petta_05_Coherent,Petta_06_Charge} suggest that 
a quantum-dot engine could output an amount 
$W_\tot  \sim 10  \text{ meV}$  of work per cycle per dot. 
We approximate the cycle time $\tau_\cycle$ with
the spin relaxation time: $\tau_\cycle  \sim  1  \: \mu \text{s}$.
(The energy eigenbasis need not rotate,
unlike for the MBL engine.
Hence diabatic hops do not lower-bound
the ideal-quantum-dot $\tau_\cycle$.)
The power would be 
$\Power  \sim  \frac{ W_\tot }{ \tau_\cycle }  
\sim  \frac{ 10  \text{ meV}  }{  1  \:  \mu \text{s} }
\sim 10^{-15} \text{ W}$.
The quantum-dot engine's power exceeds the MBL engine's
by an order of magnitude.

However, the quantum dots must be separated widely.
Otherwise, they will interact, as an ETH system.
(See~\cite{Kosloff_10_Optimal} for disadvantages of interactions
in another quantum thermal machine.
Spin-spin couplings cause ``quantum friction,''
limiting the temperatures to which a refrigerator can cool.)
We compensate by attributing a volume 
$V  \sim (1 \:  \mu \text{m})^3$ to each dot.
The power density becomes 
$\frac{ \Power  }{  V } \sim 1  \text{ kW} / \text{m}^3$,
two orders of magnitude less than the localized engine's.
Localization naturally implies
near independence of the subengines.

In Suppl.~Mat.~\ref{section:CompetitorApp}, we compare 
the MBL Otto engine to four competitors:
a bandwidth engine, a variant of the MBL engine 
that is tuned between two disorder strengths, 
an engine of quantum dots (analyzed partially above), 
and an Anderson-localized engine.
We argue that the MBL Otto engine is 
more robust against perturbations than 
the bandwidth, Anderson, and quantum-dot engines.
We also argue that our MBL engine is more reliable than 
the equal-disorder-strength engine:
Our MBL engine's $W_\tot$ varies less from trial to trial
and suppresses worst-case trials, in which $W_\tot < 0$.
This paper's arguments go through 
almost unchanged for an Anderson-localized medium. 
Such a medium would lack robustness against interactions, though: 
Even if the interactions do not delocalize the medium---which 
would destroy the engine---they 
would turn the Anderson engine into an MBL engine. 
One can view our MBL engine as 
an easy generalization of the Anderson engine.

%
%
\section{Outlook}
\label{section:Outlook}

The realization of thermodynamic cycles
with quantum many-body systems
was proposed very recently~\cite{PerarnauLlobet_16_Work,Lekscha_16_Quantum,Jaramillo_16_Quantum,Campisi_16_Power,Modak_17_Work,Verstraelen_17_Unitary,Ferraro_17_High,Ma_17_Quantum}.
MBL offers a natural platform, due to its ``athermality''
and to athermality's resourcefulness in thermodynamics.
We designed an Otto engine 
that benefits from the discrepancy between 
many-body-localized and ``thermal'' level statistics.
The engine illustrates how MBL
can be used for thermodynamic advantage.

Realizing the engine may provide a near-term challenge 
for existing experimental set-ups.
Possible platforms include ultracold atoms~\cite{Schreiber_15_Observation,Kondov_15_Disorder,Choi_16_Exploring,Luschen_17_Signatures,Bordia_17_Probing}; nitrogen-vacancy centers~\cite{Kucsko_16_Critical};
trapped ions~\cite{Smith_16_Many}; and
doped semiconductors~\cite{Kramer_93_Localization},
for which we provided order-of-magnitude estimates.
Realizations will require platform-dependent corrections
due to, e.g., variable-range hopping induced by particle-phonon interactions.
As another example, semiconductors' impurities
suffer from dipolar interactions.
The interactions extend particles' wave functions
from decaying exponentially across space to decaying as power laws.

Reversing the engine should pump heat from the cold bath to the hot,
lowering the cold bath's temperature.
Low temperatures facilitate quantum computation
and low-temperature experiments.
An MBL engine cycle might therefore facilitate state preparation
and coherence preservation
in quantum many-body experiments:
A quantum many-body engine would cool
quantum many-body systems.

We have defined as work
the energy outputted during Hamiltonian tunings.
Some battery must store this energy.
We have refrained from specifying the battery's physical form,
using an \emph{implicit battery model}.
An equivalent \emph{explicit battery model} could depend on 
the experimental platform.
Quantum-thermodynamics batteries have been modeled abstractly with 
ladder-like Hamiltonians~\cite{Skrzypczyk_13_Extracting}.
An oscillator battery for our engine could manifest as 
the mode of an electromagnetic field in cavity quantum electrodynamics.

MBL is expected to have thermodynamic applications
beyond this Otto engine.
A localized ratchet, for example, could leverage information
to transform heat into work.
Additionally, the paucity of transport in MBL may have 
technological applications beyond thermodynamics.
Dielectrics, for example, prevent particles from flowing
in undesirable directions.
But dielectrics break down in strong fields.
To survive, a dielectric must insulate well---as does MBL.

In addition to suggesting applications of MBL,
this work identifies an opportunity within quantum thermodynamics.
Athermal quantum states (e.g., $\rho  \neq e^{-H/T}/Z$) are usually 
regarded as resources in quantum thermodynamics~\cite{Janzing_00_Thermodynamic,Dahlsten_11_Inadequacy,Brandao_13_Resource,Horodecki_13_Fundamental,Goold_15_review,Gour_15_Resource,YungerHalpern_16_Beyond,LostaglioJR14,Lostaglio_15_Thermodynamic,YungerHalpern_16_Microcanonical,Guryanova_16_Thermodynamics,Deffner_16_Quantum,Wilming_17_Third}.
Not only athermal states, we have argued,
but also athermal energy-level statistics,
offer thermodynamic advantages.
Generalizing the quantum-thermodynamics definition 
of ``resource''
may expand the set of goals
that thermodynamic agents can achieve.

Optimization offers another theoretical opportunity.
We have shown that the engine works,
but better protocols could be designed.
For example, we prescribe 
nearly quantum-adiabatic tunings.
Shortcuts to adiabaticity (STA) avoid both
diabatic transitions and exponentially slow tunings~\cite{Chen_10_Fast,Kosloff_10_Optimal,Torrontegui_13_Shortcuts,Deng_13_Boosting,del_Campo_14_Super,Abah_16_Performance}.
STA have been used to reduce other quantum engines' cycle times~\cite{Deng_13_Boosting,del_Campo_14_Super,Abah_16_Performance}.
STA might be applied to the many-body Otto cycle,
after being incorporated into MBL generally.

%
%
\section*{Acknowledgements}
This research was supported by NSF grant PHY-0803371. 
The Institute for Quantum Information and Matter (IQIM) 
is an NSF Physics Frontiers Center supported by 
the Gordon and Betty Moore Foundation.
NYH is grateful for partial support from the Walter Burke Institute for Theoretical Physics at Caltech, for a Barbara Groce Graduate Fellowship,
and for an NSF grant for the Institute for Theoretical Atomic, Molecular, and Optical Physics at Harvard University and the Smithsonian Astrophysical Observatory.
This material is based on work supported by the National Science Foundation Graduate Research Fellowship under Grant No. DGE-1144469.
SG acknowledges support from the Walter Burke Foundation
and from the NSF under Grant No. DMR-1653271.
GR acknowledges support from the Packard Foundation.
NYH thanks Nana Liu and \'Alvaro Mart\'in Alhambra 
for discussions.

%
%
%

\bibliographystyle{h-physrev}
\bibliography{MBL_engine,MBL}

%
%
\begin{appendices}

\onecolumngrid

\renewcommand{\thesection}{\Alph{section}}
\renewcommand{\thesubsection}{\Alph{section} \arabic{subsection}}
\renewcommand{\thesubsubsection}{\Alph{section} \arabic{subsection} \roman{subsubsection}}

\makeatletter\@addtoreset{equation}{section}
\def\theequation{\thesection\arabic{equation}}

\section{Analysis of the mesoscopic MBL Otto engine}
\label{section:PowerApp}

In this appendix, we assess the mesoscopic engine 
introduced in Sec.~\ref{section:Meso_main}.
Section~\ref{section:Notation_app} reviews and introduces notation.
Section~\ref{section:Small_params} introduces
small expansion parameters.
Section~\ref{section:PSWAP} reviews the partial swap~\cite{Ziman_01_Quantum,Scarani_02_Thermalizing},
used to model cold thermalization (stroke 2).
The average heat $\expval{ Q_2 }$ absorbed during stroke 2
is calculated in Sec.~\ref{section:Q2};
the average heat $\expval{ Q_4 }$ absorbed during stroke 4,
in Sec.~\ref{section:Q4};
the average per-trial power $\expval{ W_\tot }$,
in Sec.~\ref{section:WTot};
and the efficiency $\eta_\MBL$, in Sec.~\ref{section:AdiabaticEta}.
These calculations rely on adiabatic tuning of the Hamiltonian.

\subsection{Notation and definitions for the mesoscopic engine}
\label{section:Notation_app}

We focus on one mesoscopic engine of $\Sites$ sites.
The engine corresponds to a Hilbert space
of dimensionality $\HDim \sim \frac{ 2^\Sites }{ \sqrt{\Sites} }$.
The Hamiltonian, $H(t)  \equiv  H_\meso (t)$, 
is tuned between $H_\ETH$, which obeys the ETH, 
and $H_\MBL$, which governs an MBL system.
Though the energies form a discrete set,
they can approximated as continuous.
ETH and MBL Hamiltonians have Gaussian DOSs:
\begin{align}
   \label{eq:DOS_App}
   \DOS(E)  =  \frac{ \HDim }{ \sqrt{ 2 \pi  \Sites }   \;  \HScale }  \: 
   e^{ - E^2  /  (2 \Sites  \HScale^2 ) }  \, ,
\end{align}
normalized to 
$\int_{-\infty}^\infty  dE  \;  \DOS ( E )   =  \HDim$.
The unit of energy, or energy density per site, is $\HScale$.
We often extend energy integrals' limits to $\pm \infty$,
as the Gaussian peaks sharply about $E = 0$.

The local average gap is $\dAvg_E  =  \frac{1}{ \DOS(E) }$,
and the average gap is $\dAvg  
:=  \frac{ \HDim }{ \int_{-\infty}^\infty  dE  \;  \DOS^2(E) }
=  \frac{2 \sqrt{ \pi \Sites }  \:  \HScale }{ \HDim }$
(footnote~\ref{footnote:dAvg}).
The average $H_\ETH$ gap, $\dAvg$, 
equals the average $H_\MBL$ gap, by construction.
$\dAvg$ sets the scale for work and heat quantities.
Hence we cast $Q$'s and $W$'s as
$(\text{number})(\text{function of small parameters}) \dAvg$.

The system begins the cycle in the state 
$\rho ( 0 )  =  e^{ - \betaH H_\ETH } / Z$, wherein
$Z :=  \Tr \left( e^{ - \betaH H_\ETH } \right)$
denotes the partition function.
$\Wb$ denotes the cold bath's bandwidth.
We set $\hbar  =  \kB  =  1 \, .$

$H(t)$ is tuned at a speed 
$v  :=  \HScale \left\lvert  \frac{ d \alpha_t }{ dt }  \right\rvert$,
wherein $\alpha_t$ denotes the dimensionless tuning parameter.
$v$ has dimensions of 
$\text{energy}^2$, 
as in~\cite{Landau_Zener_Shevchenko_10}.
Though our $v$ is not defined identically to
the $v$ in~\cite{Landau_Zener_Shevchenko_10},
ours is expected to behave similarly.

\subsection{Small parameters of the mesoscopic engine}
\label{section:Small_params}

We estimate low-order contributions 
to $\expval{ W_\tot }$ and to $\eta_\MBL$
in terms of small parameters:
\begin{enumerate}[leftmargin=*]

   \item The cold bath has a small bandwidth:
   $\frac{ \Wb }{ \dAvg }  \ll  1$.

   \item  The cold bath is cold: $\betaC \Wb \gg 1$. 
   Therefore, 
   $1  \gg  e^{ - \betaC  \Wb } \approx  0$, 
   and $\betaC \dAvg  \gg  1$.

   \item
   The hot bath is hot: $\sqrt{ \Sites }  \:  \betaH \HScale \ll 1$.
   This assumption lets us neglect $\betaH$ 
   from leading-order contributions to heat and work quantities.
   ($\betaH$ dependence manifests in factors of
   $e^{ - \Sites ( \betaH \HScale )^2 / 4 } \, .$)
   Since $\betaH \HScale  \ll  \frac{1}{ \sqrt{ \Sites } }$ and 
   $\frac{ \dAvg }{ \HScale }  \ll 1  \, ,$
   $\betaH \dAvg  
      \ll  \frac{1}{ \sqrt{ \Sites } } \, .$
\end{enumerate}

We focus on the parameter regime in which
\begin{align}
   \label{eq:Regime}
   \TCold  \ll  \Wb  \ll \dAvg
   \qquad \text{and} \qquad
   \sqrt{ \Sites }  \:  \betaH \HScale  \ll  1  \, ,
\end{align}
the regime explored in the numerical simulations of Sec.~\ref{section:Numerics_main}.

\subsection{Partial-swap model of thermalization}
\label{section:PSWAP}

Classical thermalization can be modeled with 
a \emph{probabilistic swap}, or \emph{partial swap,}
or \emph{$p$-SWAP}~\cite{Ziman_01_Quantum,Scarani_02_Thermalizing}.
Let a column vector $\vec{v}$ represent the state.
The thermalization is broken into time steps.
At each step, a doubly stochastic matrix $M_p$ operates on $\vec{v}$.
The matrix's fixed point is a Gibbs state $\vec{g}$.

$M_p$ models a probabilistic swapping out of $\vec{v}$ for $\vec{g}$:
At each time step, the system's state has a probability $1 - p$ of being preserved
and a probability $p \in [0, \: 1]$ of being replaced by $\vec{g}$.
This algorithm gives $M_p$ the form
$M_p  =  (1 - p) \id  +  p \vec{g} (1,1)$.

We illustrate with thermalization across two levels.
Let $0$ and $\Delta$ label the levels, such that
$\vec{g} = \left( \frac{ e^{ - \beta \gap } }{ 1 + e^{ - \beta \gap } } \, ,
\frac{1}{ 1 + e^{ - \beta \gap } } \right)$:
\begin{align}
   M_p  =  \begin{bmatrix}
   1 - p  \;  \frac{ 1 }{ 1 + e^{ - \beta \gap } }  &
   p  \;  \frac{ e^{ - \beta \gap } }{ 1 + e^{ - \beta \gap } }  \\
   p  \;  \frac{ 1 }{ 1 + e^{ - \beta \gap } }   &
   1  -  p  \;  \frac{ e^{ - \beta \gap } }{ 1 + e^{ - \beta \gap } } 
   \end{bmatrix} \, .
\end{align}
The off-diagonal elements, or transition probabilities,
obey detailed balance~\cite{YungerHalpern_15_Introducing,Crooks_98}:
$\frac{ P( 0 \to \gap ) }{ P( \gap \to 0 ) }
=  e^{ - \beta \gap }$.

Repeated application of $M_p$ 
maps every state to $\vec{g}$~\cite{YungerHalpern_15_Introducing}:
$\lim_{n \to \infty} \left( M_p \right)^n  \vec{v}  =  \vec{g}$.
The parameter $p$ reflects the system-bath-coupling strength.
We choose $p = 1$: 
The system thermalizes completely at each time step.
(If $p \neq 1$, a more sophisticated model may be needed
for thermalization across $>2$ levels.)

%
%
%
\subsection{Average heat $\expval{ Q_2 }$ absorbed during stroke 2}
\label{section:Q2}

Let $j$ denote the $H_\ETH$ level
in which the engine begins the trial of interest.
We denote by $Q_2^\ParenJ$ the average heat
absorbed during stroke 2, from the cold bath.
($Q_2^\ParenJ$ will be negative and,
provided that $j$ is around the energy band's center, 
independent of $j$.)

The heat absorbed can be calculated easily from the following observation. 
Stroke 1 (adiabatic tuning) preserves the occupied level's index. 
The level closest to $j$ lies a distance $\delta$ away when stroke 3 begins. 
$\delta$ can have either sign,
can lie above or below $j$.
Heat is exchanged only if $|\delta| <W_b$. 
Let us initially neglect the possibility that 
two nearby consecutive gaps are very small, that 
$| E_{j \pm 2} - E_j |  \leq  \Wb$. 
We can write the average (over trials begun in level $j$) heat absorbed as 
\begin{align}
   Q_2^\ParenJ  
   =\int\limits_{-\Wb}^{\Wb}d\delta \; \delta 
   \frac{e^{-\betaC \delta}}{1+ e^{-\betaC \delta}}   \:
   P_\MBL^\ParenE ( \delta)
   +  O  \left(  \Wb^3/\dAvg^2  \right) \, .
\end{align}
This equation assumes a Sommerfeld-expansion form, 
as the Boltzmann factor is 
$\frac{ e^{ - \betaC \delta } }{ 1  +  e^{ - \betaC \delta } }
=  \Theta ( - \delta )
+ {\rm sgn} (\delta)  \:  
\frac{ e^{ - \betaC | \delta | } }{
         1  +  e^{ - \betaC | \delta | } }$.
Hence
\begin{align}
   Q_2^\ParenJ  
   = - \frac{\Wb^2}{2}\mu(E) 
   +\frac{\pi^2}{6}\mu(E) ( \TCold )^2
   +     O  \left(  [ \Wb ]^3 /  \dAvg^2  \right)
   + O  \left(  \mu(E)^2   [ \TCold ]^3  \right)  \, .
\label{eq:Qj_help4}
\end{align}
The first correction accounts for our not considering 
two levels within $\Wb$ of level $j$. 

Next, we need to average this result over all initial states $j$, 
assuming the initial density operator,
$\rho ( 0 )  =  e^{ - \betaH H_\ETH } / \ZH$:
\begin{align}
   \label{eq:Q2_help1}
   \expval{ Q_2 }  
   & :=  \expval{  \expval{  \expval{ Q_2(E) }_{ \substack{ \text{cold} \\ \text{therm.} } }  }_{\text{gaps}}  }_{ \rho(0) } \\
   & =  \left(  - \frac{ ( \Wb )^2 }{ 2 }  
   +  \frac{ \pi^2 }{ 6 }  \: \frac{ 1 }{ ( \betaC )^2 }  \right)
   \int_{ -\infty }^\infty  dE  \;  \DOS^2(E)  \;
   \frac{ e^{ - \betaH E } }{ \ZH }
  +  \dAvg \Bigg\{  
  O  \left(  \left[  \frac{ \Wb }{ \dAvg }  \right]^3  \right)
  +  O  \left(  \frac{ \Wb }{ \dAvg }  \;   e^{ - \betaC \Wb }  \right) 
  +  O \left(  \left[  \frac{ \DOS(E) }{ \betaC }  \right]^3  \right)  
  \Bigg\} \, .
\end{align}
We substitute in for the DOS from Eq.~\eqref{eq:DOS_App}:
\begin{align}
   \label{eq:Q2_help2}
   \expval{ Q_2 }  & =  
   \frac{ \HDim^2 }{ 2 \pi \Sites \HScale^2 }  \;
   \frac{1}{ \ZH }  \:
    \left(  - \frac{ ( \Wb )^2 }{ 2 }  
    +  \frac{ \pi^2 }{ 6 }  \:  \frac{ 1 }{ ( \betaC )^2 }  \right)
    \int_{ -\infty }^\infty  dE  \;  e^{ - E^2 / \Sites \HScale^2 }  \;
    e^{ - \betaH E }
    + O ( . )  \, ,
\end{align}
wherein the correction terms are abbreviated.
The integral evaluates to $\sqrt{ \pi \Sites }  \:  \HScale  \,  
e^{ \Sites ( \betaH \HScale )^2 / 4 }$.
The partition function is
\begin{align}
   \label{eq:ZH}
   \ZH  =  \int_{ -\infty }^\infty  dE  \;  
   \DOS(E)  e^{ - \betaH E } 
   =  \HDim e^{ \Sites ( \betaH \HScale )^2 / 2 }  \, .
\end{align}
Substituting into Eq.~\eqref{eq:Q2_help2} yields
\begin{align}
   \label{eq:Q2_help3}
   \expval{ Q_2 }  
   & =  \left(  - \frac{ ( \Wb )^2 }{ 2 \dAvg }  
                   +  \frac{ \pi^2 }{ 6 }  \:  \frac{ 1 }{ ( \betaC )^2 \dAvg }  \right)  \:
   e^{ - \Sites ( \betaH \HScale )^2 / 4 }
   +  \dAvg \Bigg\{ 
   O  \left(  \left[  \frac{ \Wb }{ \dAvg }  \right]^3  \right) 
   +  O  \left(  [ \DOS (E)  \,  \Wb ]  \,  \frac{ \DOS(E) }{ \betaC }  \:
                     e^{ - \betaC \Wb }  \right)  
   \nonumber \\ & \qquad   
   +  O \left(  \left[  \frac{ \DOS(E) }{ \betaC }  \right]^3  \right)
   +  O  \left(  \left[ \sqrt{ \Sites } \:  \betaH \HScale \right]^4  \right)  
   \Bigg\}   \, .
\end{align}
We have replaced the prefactor with $\frac{1}{ \dAvg }$, 
using Eq.~\eqref{eq:dAvg_def}.

\begin{figure}
  \begin{subfigure}{0.3\textwidth}
    \centering
    \includegraphics[width=\textwidth]{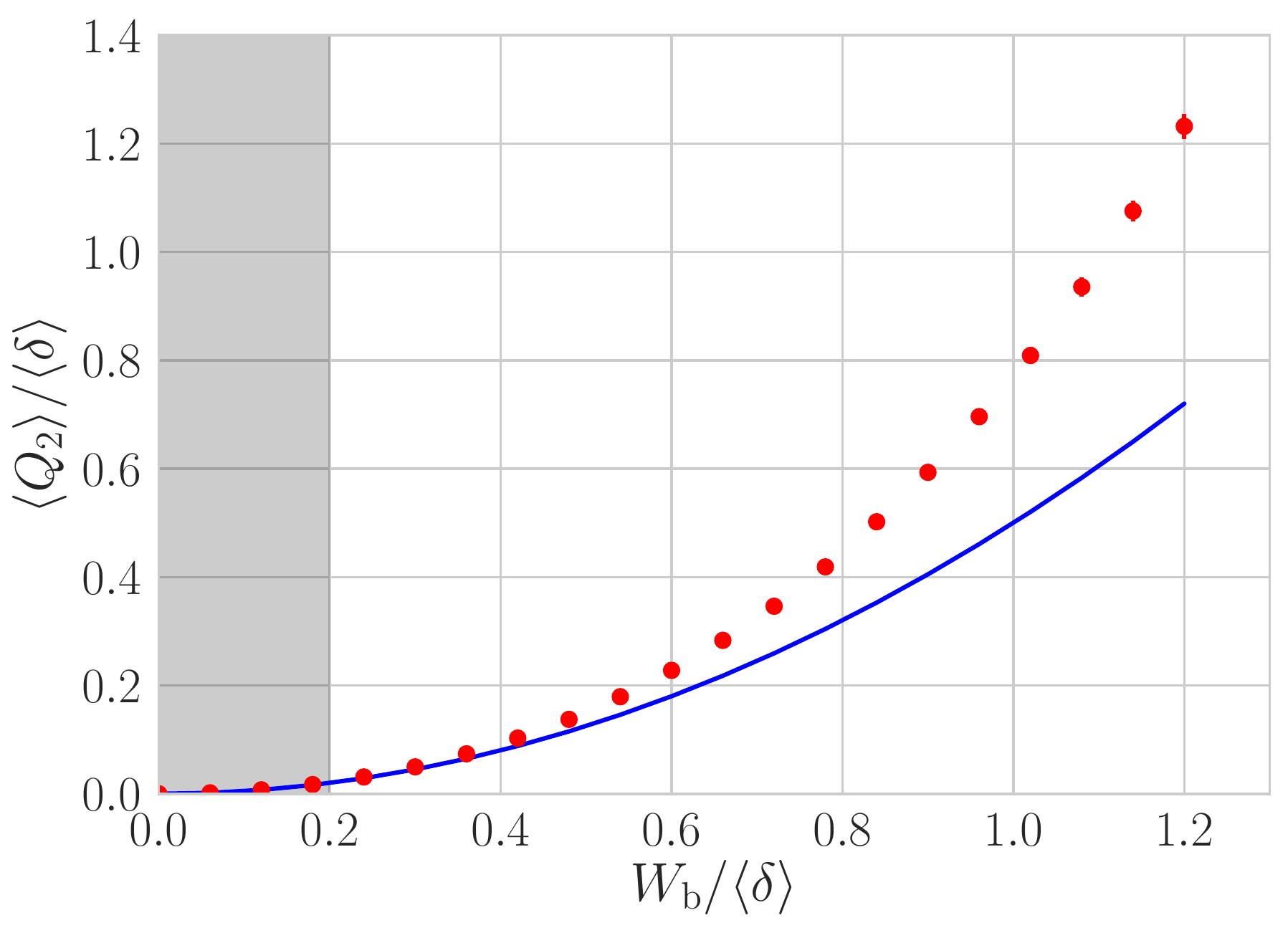}
    \caption{$| \expval{Q_2} |$ vs. $\Wb$    at $\TCold = 0$ and $\THot = \infty$}
    \label{fig:Q2_Wb}
  \end{subfigure}
  \begin{subfigure}{0.3\textwidth}
    \centering
    \includegraphics[width=\textwidth]{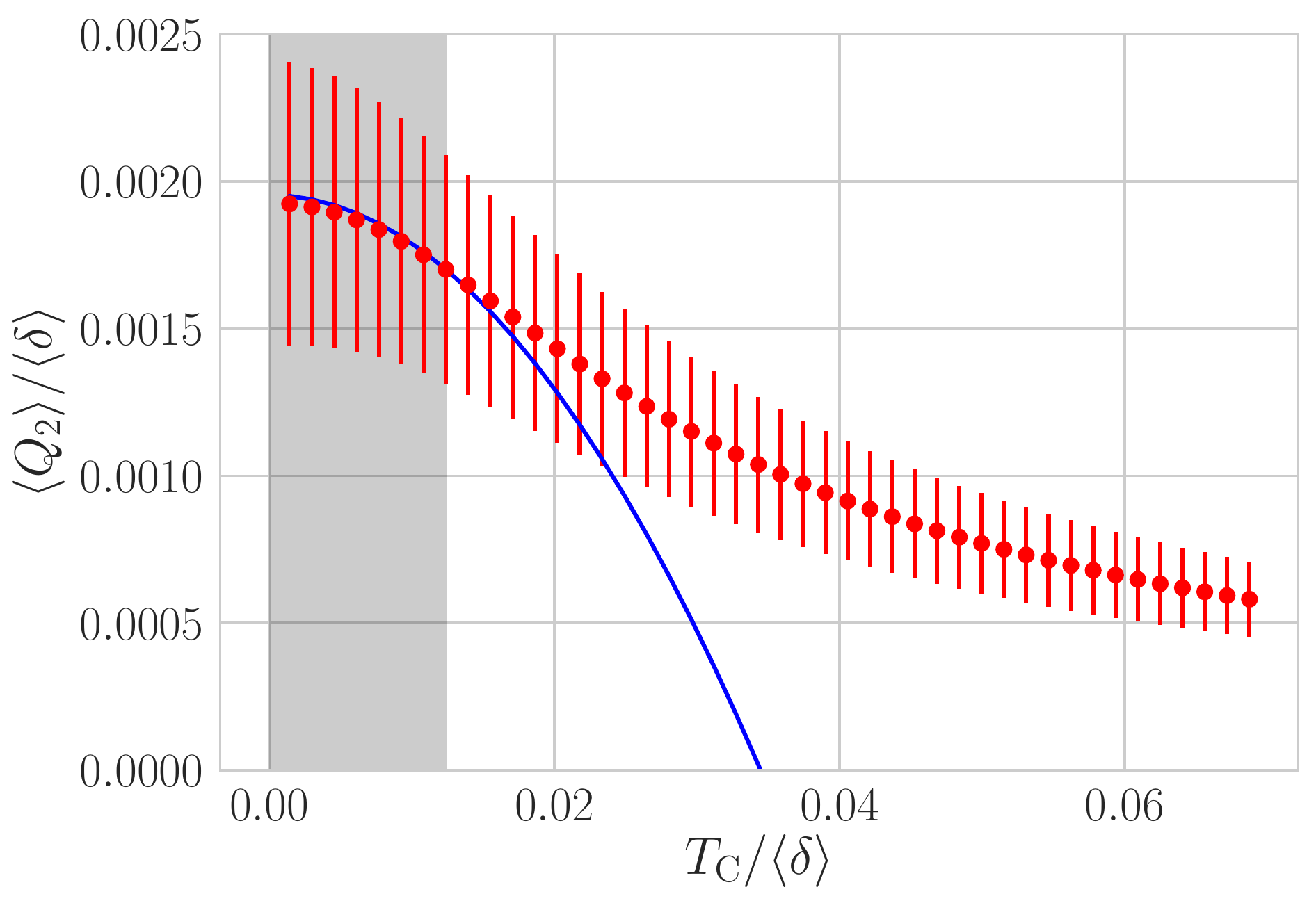}
    \caption{$| \expval{Q_2} |$ vs. $\TCold$ 
    at $\THot = \infty$ and    $\Wb = 2^{-4}\dAvg$}
    \label{fig:Q2_TC}
  \end{subfigure}
  \begin{subfigure}{0.3\textwidth}
    \centering
    \includegraphics[width=\textwidth]{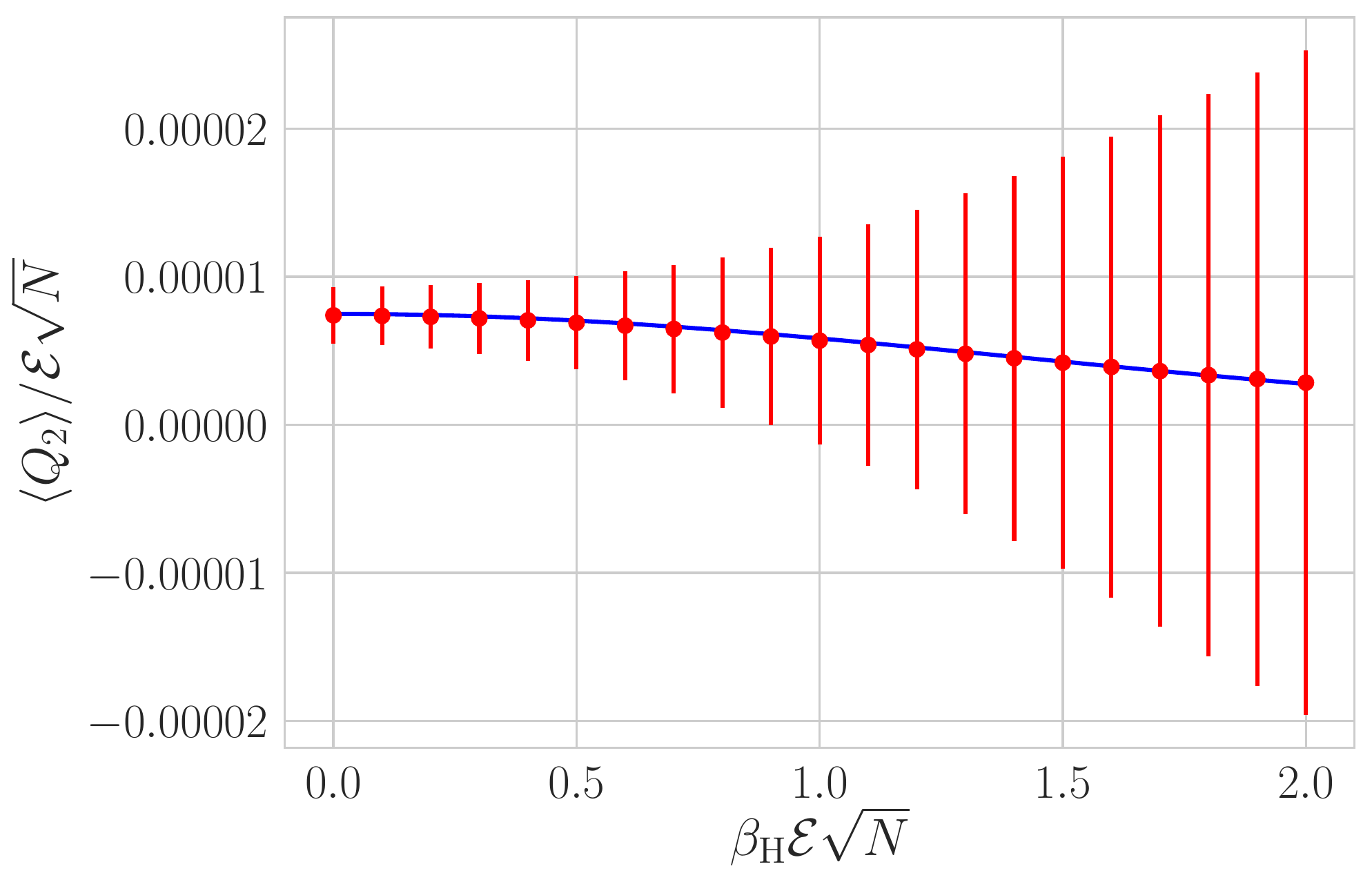}
    \caption{$| \expval{Q_2} |$ vs. $\betaH$ 
    at $\TCold = 0$ and    $\Wb = 2^{-4}\dAvg$}
    \label{fig:Q2_betaH}
  \end{subfigure}
  \caption{\caphead{Magnitude $| \expval{Q_2} |$ 
  of the average heat absorbed during cold thermalization (stroke 2) 
  as a function of 
    (a) the cold-bath bandwidth $\Wb$ (\ref{fig:Q2_Wb}), 
    (b) the cold-bath temperature $\TCold$ (\ref{fig:Q2_TC}), and 
    (c) the hot-bath temperature $\THot = 1 / \betaH$ (\ref{fig:Q2_betaH}):}
    The blue lines represent the magnitude of 
    the analytical prediction~\eqref{eq:Q2_help3}.
    See Sec.~\ref{section:Numerics_main} for 
    other parameters and definitions.
    The analytics match the numerics' shapes,
    and the agreement is fairly close, in the appropriate limits
    (where $\frac{ \Wb }{ \dAvg } \ll 1$ and $\TCold/\dAvg \ll 1$, 
     in the gray shaded regions).
    The analytics systematically underestimate $| \expval{Q_2} |$ 
    at fixed $\Wb$, due to
    the small level repulsion at finite $\Sites$.
    The analytical prediction~\eqref{eq:Q2_help3} 
    substantially underestimates $| \expval{Q_2} |$ 
    when the cold-bath bandwidth is large, $\Wb \gtrsim \dAvg$.
    Such disagreement is expected:
    The analytics rely on $\frac{ \Wb }{ \dAvg }  \ll  1$,
    neglecting chains of small gaps:
    $\delta'_j \, ,  \delta'_{j+1} \, ,  \dots < \Wb$.
    Such chains proliferate as $\Wb$ grows.
    A similar reason accounts for the curve's 
    crossing the origin in Fig.~\ref{fig:Q2_TC}:
    We analytically compute $\expval{Q_2}$ only to second order in $\TCold/\dAvg$.    }
  \label{fig:num_Q2}
\end{figure}

Equation~\eqref{eq:Q2_help3} is compared with numerical simulations in Fig.~\ref{fig:num_Q2}.
In the appropriate regime (wherein $\Wb \ll \dAvg$ and $\TCold \ll \Wb$),
the analytics agree well with the numerics, to within finite-size effects.

In terms of small dimensionless parameters, 
\begin{align}
   \label{eq:Q2_help4}
   \expval{ Q_2 }   & =  \dAvg 
   \left[ - \frac{1}{2}  \left( \frac{ \Wb }{ \dAvg }  \right)^2
           +  \frac{ \pi^2 }{ 6 }  \:  \frac{ 1 }{  ( \betaC  \dAvg )^2 }  \right]
  \left[ 1  -  \frac{ \Sites }{ 4 }  \left( \betaH \HScale \right)^2  \right]
   +  O ( . )   \, .
\end{align}
The leading-order term is second-order.
So is the $\betaC$ correction; but 
$\frac{ 1 }{  ( \betaC  \dAvg )^2 }  
\ll  \left( \frac{ \Wb }{ \dAvg }  \right)^2$, 
by assumption [Eq.~\eqref{eq:Regime}].
The $\betaH$ correction is fourth-order---too small to include.
To lowest order, 
\begin{align}
   \label{eq:EDiff2b}   \boxed{
   \expval{ Q_2 }  \approx
   -  \frac{ \left( \Wb \right)^2 }{ 2 \dAvg }  } \, .
\end{align}

\subsection{Average heat $\expval{ Q_4 }$ absorbed during stroke 4}
\label{section:Q4}

The $\expval{ Q_4 }$ calculation proceeds similarly to
the $\expval{ Q_2 }$ calculation.
When calculating $\expval{ Q_2 }$, however,
we neglected contributions from 
the engine's cold-thermalizing down
two small gaps.
Two successive gaps have a joint probability
$\sim \left( \frac{ \Wb }{ \dAvg }  \right)^2$
of being $< \Wb$ each.
Thermalizing across each gap, the engine absorbs heat $\leq \Wb$.
Each such pair therefore contributes
negligibly to $\expval{ Q_2 }$, as
$\dAvg  O  \left(  \left[  \frac{ \Wb }{ \dAvg }  \right]^3  \right)$.

We cannot neglect these pairs  
when calculating $\expval{ Q_4 }$.
Each typical small gap widens, during stroke 3, 
to size $\sim \dAvg \, .$
These larger gaps are thermalized across during stroke 4,
contributing at the nonnegligible second order, as
\mbox{ $\sim \dAvg O  \left(  \left[  \frac{ \Wb }{ \dAvg }  \right]^2  \right)$}
to $\expval{ Q_4 } \, .$
Chains of $\geq 3$ small MBL gaps contribute negligibly.

The calculation is tedious, appears in~\cite[App. G 5]{NYH_17_MBL},
and yields
\begin{align}
   \label{eq:Q4_Result}  \boxed{
   \expval{ Q_4 }      \;
   \approx  
   \Wb  -  \frac{ 2 \ln 2 }{ \betaC }
   +  \frac{ ( \Wb )^2 }{ 2 \dAvg }
   +  4 \ln 2  \:  \frac{ \Wb }{ \betaC \dAvg }  }  \, .
\end{align}
The leading-order terms are explained heuristically
below Eq.~\eqref{eq:WTotApprox2_Main} in the main text.

The leading-order $\betaC$ correction,
$- \frac{ 2 \ln 2 }{ \betaC }$, shows that
a warm cold bath
lowers the heat required to reset the engine.
Suppose that the cold bath is maximally cold: $T_\CTemp = 0$.
Consider any trial that the engine begins just above a working gap
(an ETH gap $\delta > \Wb$ that narrows to 
an MBL gap $\delta' < \Wb$).
Cold thermalization drops the engine deterministically to the lower level.
During stroke 4, the engine must absorb $Q_4 > 0$ 
to return to its start-of-trial state.
Now, suppose that the cold bath is only cool: $\TCold \gtrsim  0$.
Cold thermalization might leave the engine in the upper level.
The engine needs less heat, on average, to reset
than if $T_\CTemp = 0$.
A finite $T_\CTemp$ therefore detracts from $\expval{ Q_4 }$.
The $+  4 \ln 2  \:  \frac{ \Wb }{ \betaC \dAvg }$
offsets the detracting. However, the positive correction
is smaller than the negative correction, 
as $\frac{ \Wb }{ \dAvg }  \ll  1 \, .$

A similar argument concerns $T_\HTemp  <  \infty$.
But the $\betaH$ correction is too small
to include in Eq.~\eqref{eq:Q4_Result}:
$\expval{ Q_4 }  \approx  \Wb
- \frac{ 2 \ln 2 }{ \betaC }  
+  \frac{ ( \Wb )^2 }{ 2 \dAvg }  \:  e^{ - N ( \betaH \HScale )^2 / 4 }$.

\begin{figure}
  \begin{subfigure}{0.3\textwidth}
    \centering
    \includegraphics[width=\textwidth]{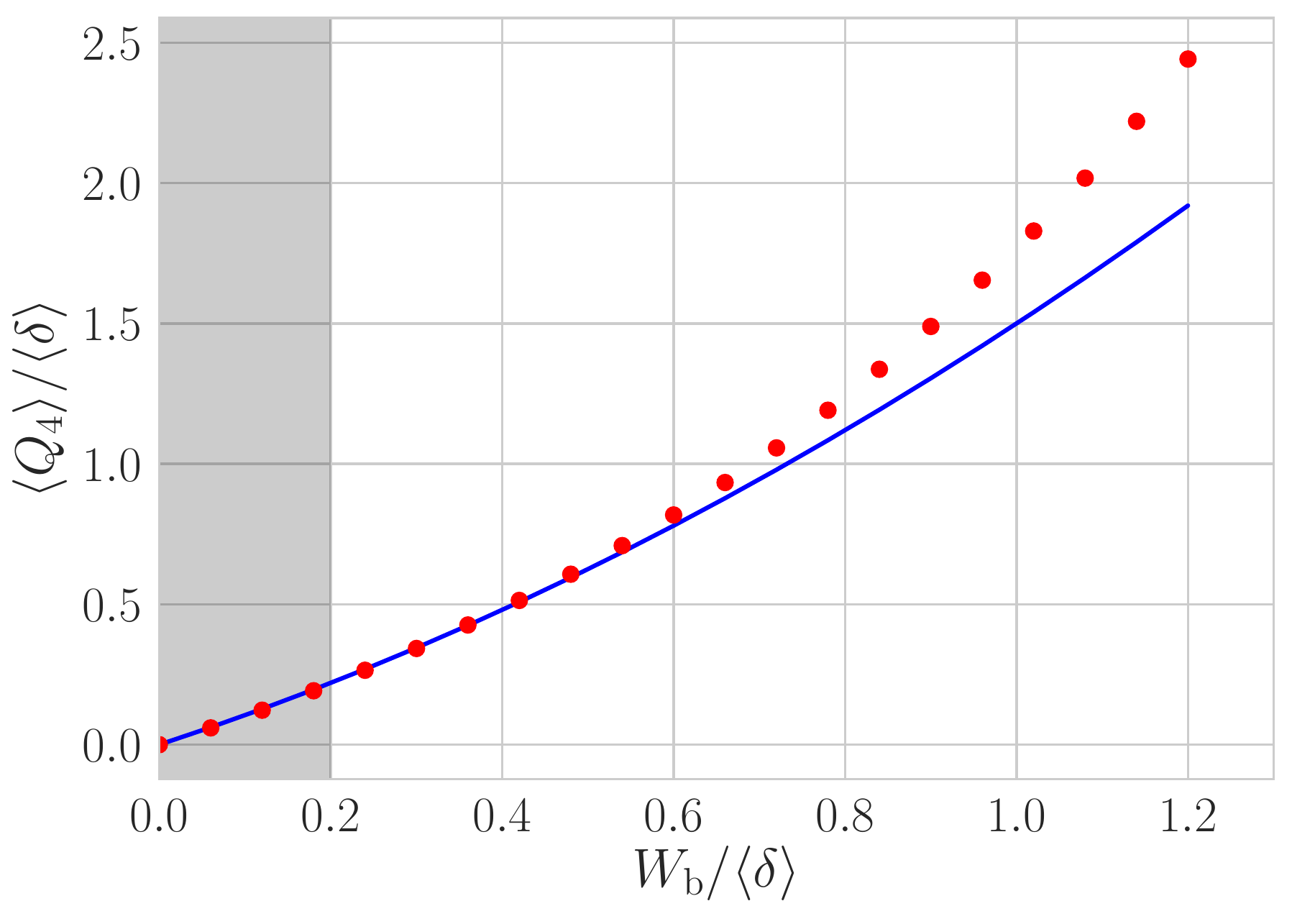}
    \caption{$\expval{Q_4}$ vs. $\Wb$ at 
    $\TCold = 0$ and $\THot = \infty$}
  \end{subfigure}
  \begin{subfigure}{0.3\textwidth}
    \centering
    \includegraphics[width=\textwidth]{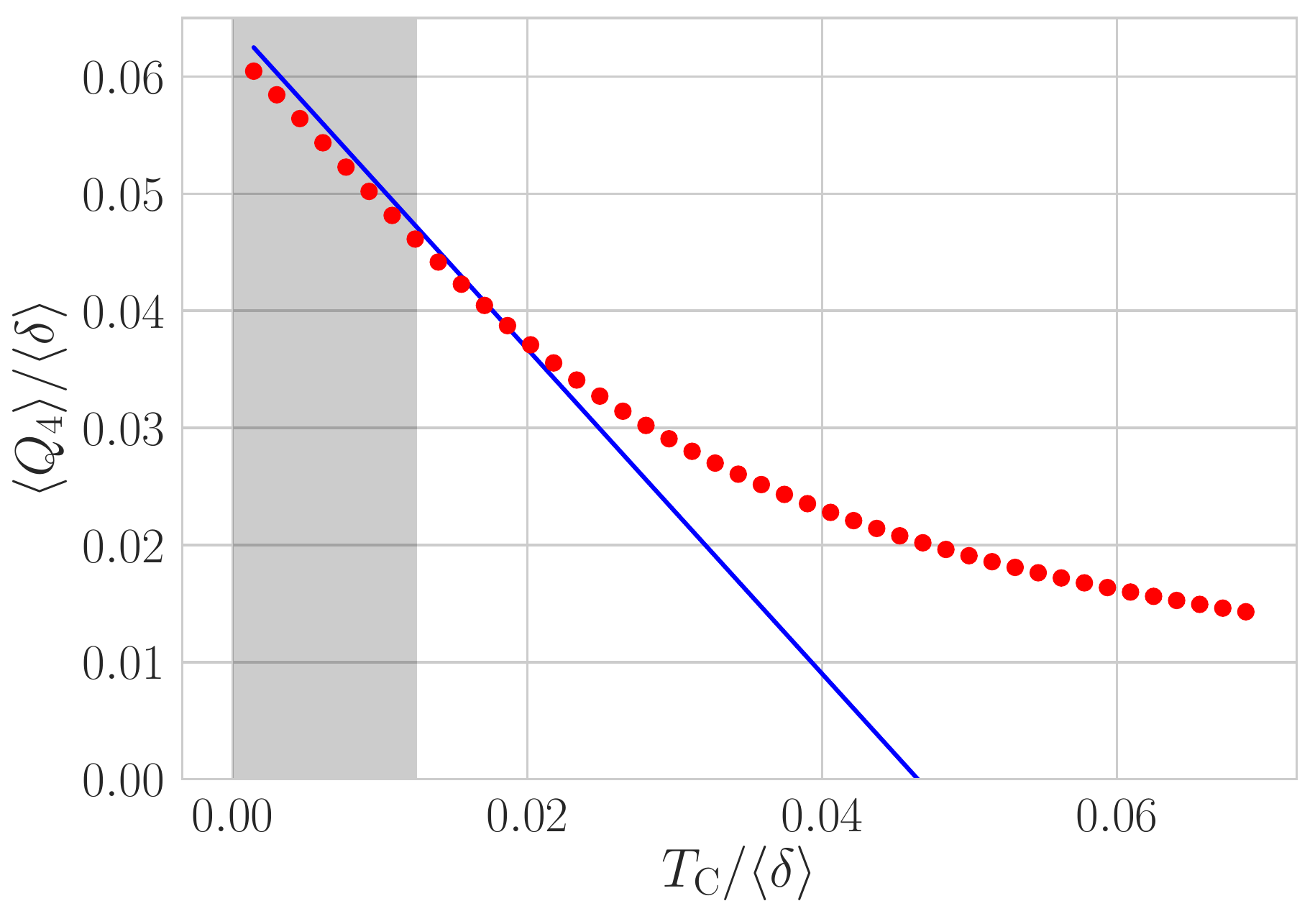}
    \caption{$\expval{Q_4}$ vs. $\TCold$ at 
    $\THot = \infty$ and 
    $\Wb = 2^{-4}\dAvg$}
  \end{subfigure}
  \begin{subfigure}{0.3\textwidth}
    \centering
    \includegraphics[width=\textwidth]{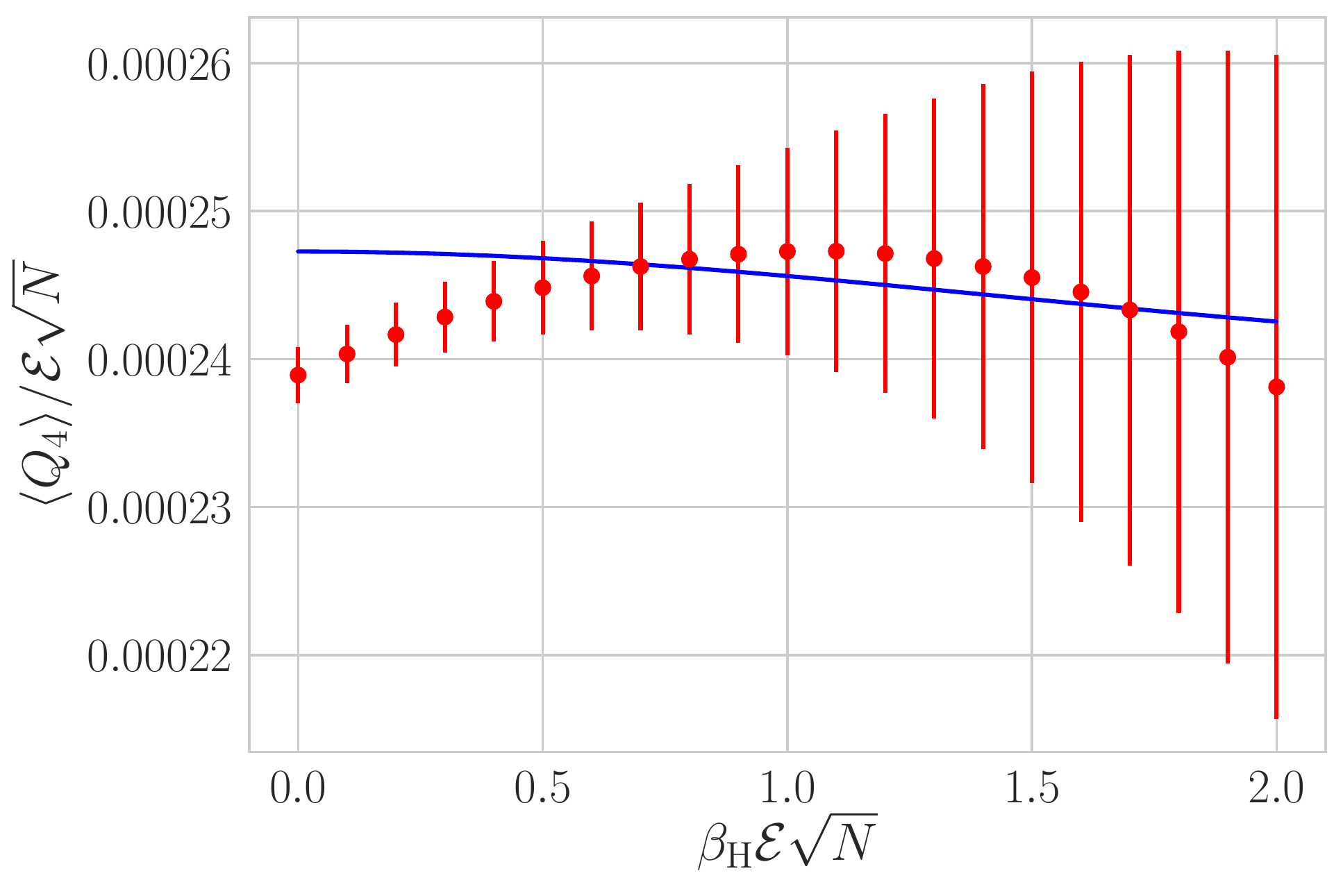}
    \caption{$\expval{Q_4}$ vs. $\betaH$ 
    at $\TCold = 0$ and
    $\Wb = 2^{-4}\dAvg$}
  \label{fig:Q4_betaH}
  \end{subfigure}
  \caption{\caphead{Average heat $\expval{Q_4}$ absorbed
    during hot thermalization (stroke 4) as a function of 
    (a) the cold-bath bandwidth $\Wb$,
    (b) the cold-bath temperature $\TCold$,
    and (c) the hot-bath temperature $\THot = 1 / \betaH$:} 
    The blue lines represent the analytical prediction~\eqref{eq:Q4_Result}, 
    to lowest order in $\TCold$,
with the $\betaH$ dependence of $\expval{ Q_4 }$,
too small a correction to include in Eq.~\eqref{eq:Q4_Result}:
$\expval{ Q_4 }  \approx  \Wb
- \frac{ 2 \ln 2 }{ \betaC }  
+  \frac{ ( \Wb )^2 }{ 2 \dAvg }  \:  e^{ - N ( \betaH \HScale )^2 / 4 }$.
    See Sec.~\ref{section:Numerics_main} for 
    other parameters and definitions.
    The analytics' shapes agree with the numerics',
    and the fit is fairly close, in the appropriate limits
    (where $e^{ - \betaC \Wb }  \ll  1$, $\frac{ 1 }{ \betaC \dAvg }  \ll  1$,
    and $\frac{ \Wb }{ \dAvg }  \ll  1$,    in the gray shaded regions).
    The predictions underestimate $\expval{ Q_4 }$;
    see the Fig.~\ref{fig:num_Q2} caption.
    Figure~\ref{fig:Q4_betaH} suggests that the numerics deviate significantly from the analytics: The numerics appear to depend on $\betaH$ via a linear term absent from the $\expval{ Q_4 }$ prediction. This seeming mismatch appears symptomatic of finite sample and system sizes.
  }
  \label{fig:num_Q4}
\end{figure}

Figure~\ref{fig:num_Q4} shows
Eq.~\eqref{eq:Q4_Result}, to lowest order in $\TCold$,
as well as the $\betaH$ dependence of $\expval{ Q_4 }$.
The analytical prediction is compared with numerical simulations.
The agreement is close, up to finite-size effects, 
in the appropriate regime
($\TCold \ll \Wb \ll \dAvg$).

\subsection{Average per-cycle power $\expval{ W_\tot }$}
\label{section:WTot}

\begin{figure}
  \begin{subfigure}{0.3\textwidth}
    \centering
    \includegraphics[width=\textwidth]{MBLSZ30_wb-WTOT-L12.pdf}
    \caption{$\expval{W_\tot}$ vs. $\Wb$ 
    at $\TCold = 0$ and $\THot = \infty$}
  \end{subfigure}
  \begin{subfigure}{0.3\textwidth}
    \centering
    \includegraphics[width=\textwidth]{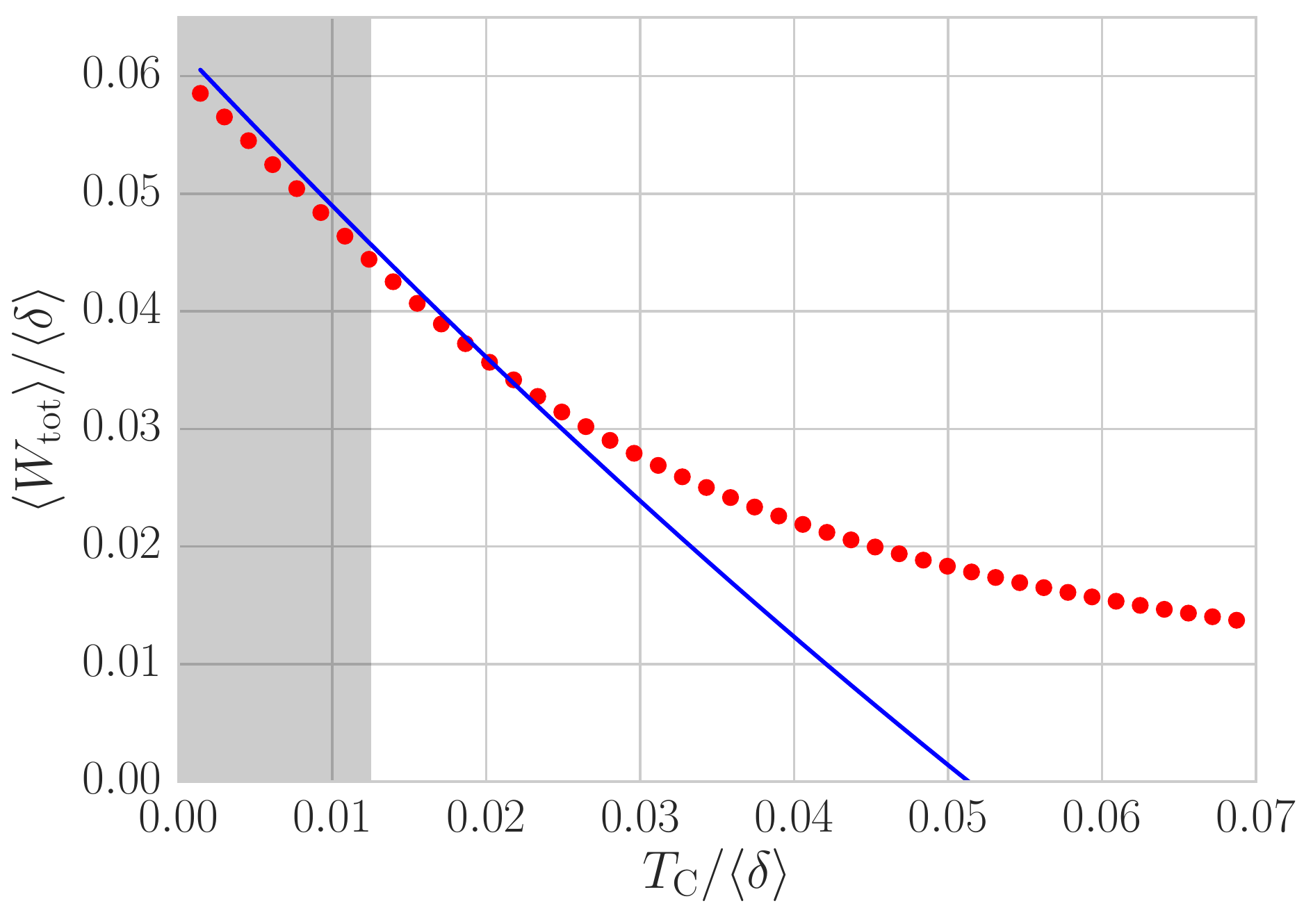}
    \caption{$\expval{W_\tot}$ vs. $\TCold$ 
    at $\THot = \infty$ and $\Wb = 2^{-4}\dAvg$}
  \end{subfigure}
  \begin{subfigure}{0.3\textwidth}
    \centering
    \includegraphics[width=\textwidth]{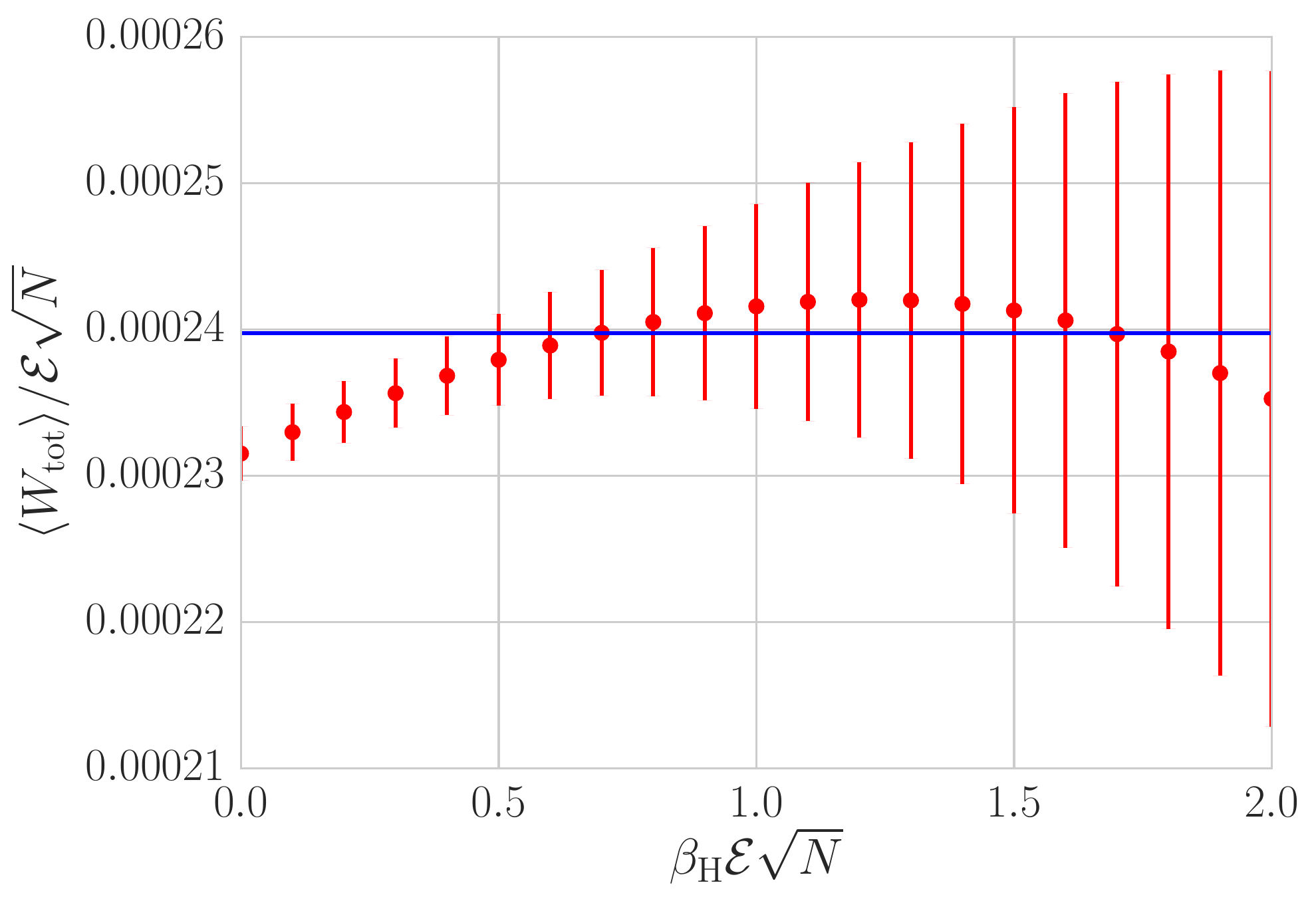}
    \caption{$\expval{W_\tot}$ vs. $\betaH$ 
    at $\TCold = 0$ and $\Wb = 2^{-4}\dAvg$}
  \label{fig:Wtot_betaH}
  \end{subfigure}
  \caption{\caphead{Per-cycle power $\expval{W_\tot}$ as a function of 
    (a) the cold-bath bandwidth $\Wb$, 
    (b) the cold-bath temperature $\TCold$,
    and (c) the hot-bath temperature $\THot = 1 / \betaH$:}
    The blue lines represent the analytical prediction
    $\expval{ W_\tot }  \approx  \Wb  -  \frac{ 2 \ln 2 }{ \betaC }$:
    Eq.~\eqref{eq:WTotApprox2}, 
    to first order in $\frac{ \Wb }{ \dAvg }$ and in $\frac{1}{ \betaC \dAvg }$.
    The analytics largely agree with the numerics
    in the appropriate regime:
    $\frac{ \Wb }{\dAvg} \ll 1$, and 
    $\frac \TCold \dAvg \ll 1$ (in the gray shaded region).
    Outside that regime,
    the analytics underestimate $\expval{ W_\tot }$;
    see Fig.~\ref{fig:num_Q2} for an analysis.
    Figure~\ref{fig:Wtot_betaH} suggests that the numerics depend on $\betaH$
    via a linear term absent from the analytical prediction;
    see the caption of Fig.~\ref{fig:Q4_betaH}.}
  \label{fig:num_WTOT}
\end{figure}

By the first law of thermodynamics,
the net work outputted by the engine equals
the net heat absorbed.
Summing Eqs.~\eqref{eq:Q4_Result} and~\eqref{eq:EDiff2b}
yields the per-trial power, or average work outputted per engine cycle:
\begin{align}
   \label{eq:WTotApprox2}
   \boxed{ \expval{ W_\tot } }
   =  \expval{ Q_2 }  +  \expval{ Q_4 }
   \boxed{ \approx    
   \Wb  -  \frac{ 2 \ln 2 }{ \betaC }
   +  4 \ln 2  \:  \frac{ \Wb }{ \betaC \dAvg } } \, .
\end{align}
The leading-order $\betaH$ correction is negative
and too small to include---of order 
$\dAvg  \left(  \frac{ \Wb }{ \dAvg }  \right)^2 
\Sites  \left( \betaH \HScale  \right)^2  \, .$
Equation~\eqref{eq:WTotApprox2} agrees well with the numerics in the appropriate limits ($\TCold \ll \Wb \ll \dAvg$) and beyond,
as shown in Fig.~\ref{fig:num_WTOT}.
The main text contains
the primary analysis of Eq.~\eqref{eq:WTotApprox2}.
Here, we discuss the $\expval{ Q_2 }$ correction, limiting behaviors,
and scaling.

The negative $\expval{ Q_2 }  =  - \frac{ \left( \Wb \right)^2 }{ \dAvg }$ 
detracts little from the leading term $\Wb$ of $\expval{ Q_4 }$:
$\frac{ ( \Wb )^2 }{ \dAvg } \ll \Wb$, 
since $\frac{ \Wb }{ \dAvg } \ll 1$.
The $\expval{ Q_2 }$ cuts down on the per-trial power little.

The limiting behavior of Eq.~\eqref{eq:WTotApprox2} makes sense:
Consider the limit as $\Wb \to 0$.
The cold bath has too small a bandwidth to thermalize the engine,
so the engine should output no work, on averge.
Indeed, the first and third terms in Eq.~\eqref{eq:WTotApprox2} vanish,
being proportional to $\Wb$.
The second term vanishes
because $\betaC \to \infty$ more quickly than $\Wb \to 0 \, ,$
by Eq.~\eqref{eq:Regime}: The cold bath is very cold.

Equation~\eqref{eq:WTotApprox2} scales with the system size $\Sites$
no more quickly than $\sqrt{\Sites} / 2^\Sites$,
by the assumption 
$\Wb  \ll  \dAvg  \sim  \sqrt{\Sites} / 2^\Sites$.
This scaling makes sense:
The engine outputs work
because the energy eigenvalues meander
upward and downward in Fig.~\ref{fig:Compare_thermo_Otto_fig}
as $H(t)$ is tuned.
In the thermodynamic limit, levels squeeze together.
Energy eigenvalues have little room in which to wander,
and the engine outputs little work.
Hence our parallelization of fixed-length mesoscopic subengines
in the thermodynamic limit (Sec.~\ref{section:Thermo_limit_main}).

\subsection{Efficiency $\eta_\MBL$ in the adiabatic approximation}
\label{section:AdiabaticEta}


The efficiency is defined as
\begin{align}
   \label{eq:EtaDefAgain}
   \eta_\MBL  :=   \frac{ \expval{ W_\tot } }{ \expval{ Q_\In } }  \, .
\end{align}
The numerator is averaged separately from the denominator because
averaging $W_\tot$ over runs of one mesoscopic engine
is roughly equivalent to 
averaging over simultaneous runs of parallel subengines
in one macroscopic engine.
$\frac{ \expval{ W_\tot } }{ \expval{ Q_\In } }$ 
may therefore be regarded as
the $\frac{ W_\tot }{ Q_\In }$ of one
macroscopic-engine trial.

The positive-heat-absorbing-stroke is stroke 4, 
in the average trial:
\begin{align}
   \expval{ Q_\In }  
   =  \expval{ Q_4 }
   =   \expval{ W_\tot }  -  \expval{ Q_2 }
   = \label{eq:Q4Eval}
   \expval{ W_\tot }  
   \left( 1 - \frac{ \expval{ Q_2 } }{ \expval{ W_\tot } }  \right)
   =      \expval{ W_\tot }     \left( 1  +  \phi \right)    \, ,
\end{align}
wherein
\begin{align}
   \label{eq:Xi1}
   \phi   :=   -  \frac{  \expval{ Q_2 } }{ \expval{ W_\tot } }
  \approx  \frac{\Wb }{ 2 \dAvg }  \, .
\end{align}

Substituting from Eq.~\eqref{eq:Q4Eval} 
into Eq.~\eqref{eq:EtaDefAgain} yields
\begin{align}
   \label{eq:ManyBodyEff}
   \boxed{ \eta_\MBL  \approx  }
   \frac{ \expval{ W_\tot } }{   \expval{ W_\tot } ( 1  +  \phi )  }
   \approx   1  -  \phi
   =  \boxed{    1  -  \frac{ \Wb  }{ 2 \dAvg }  } \,  .
\end{align}


Using suboptimal baths diminishes the efficiency. 
Adding $\betaC$-dependent terms
from Eq.~\eqref{eq:WTotApprox2} to $\expval{ W_\tot }$ yields
\begin{align}
  \label{eq:phi-corrections}
   \phi'  =   \frac{ \Wb}{ 2 \dAvg }  
   +  \frac{ \ln 2 }{ \betaC \dAvg }
   -  2 \ln 2  \:  \frac{ \Wb }{ \dAvg }  \:  \frac{1}{ \betaC \dAvg }  \, .
\end{align}
The $\betaH$ correction,
$1 - \frac{ \Wb }{ 2 \dAvg }  \:  e^{ - \Sites ( \betaH \HScale )^2 / 4 }$,
is too small to include.
The correction shares the sign of $\betaH$:
A lukewarm hot bath lowers the efficiency.

\begin{figure}
  \begin{subfigure}{0.3\textwidth}
    \centering
    \includegraphics[width=\textwidth]{MBLSZ30_wb-eta-L12.pdf}
    \caption{$\eta_\MBL$ vs. $\Wb$ at $\TCold = 0$ and $\THot = \infty$}
  \end{subfigure}
  \begin{subfigure}{0.3\textwidth}
    \centering
    \includegraphics[width=\textwidth]{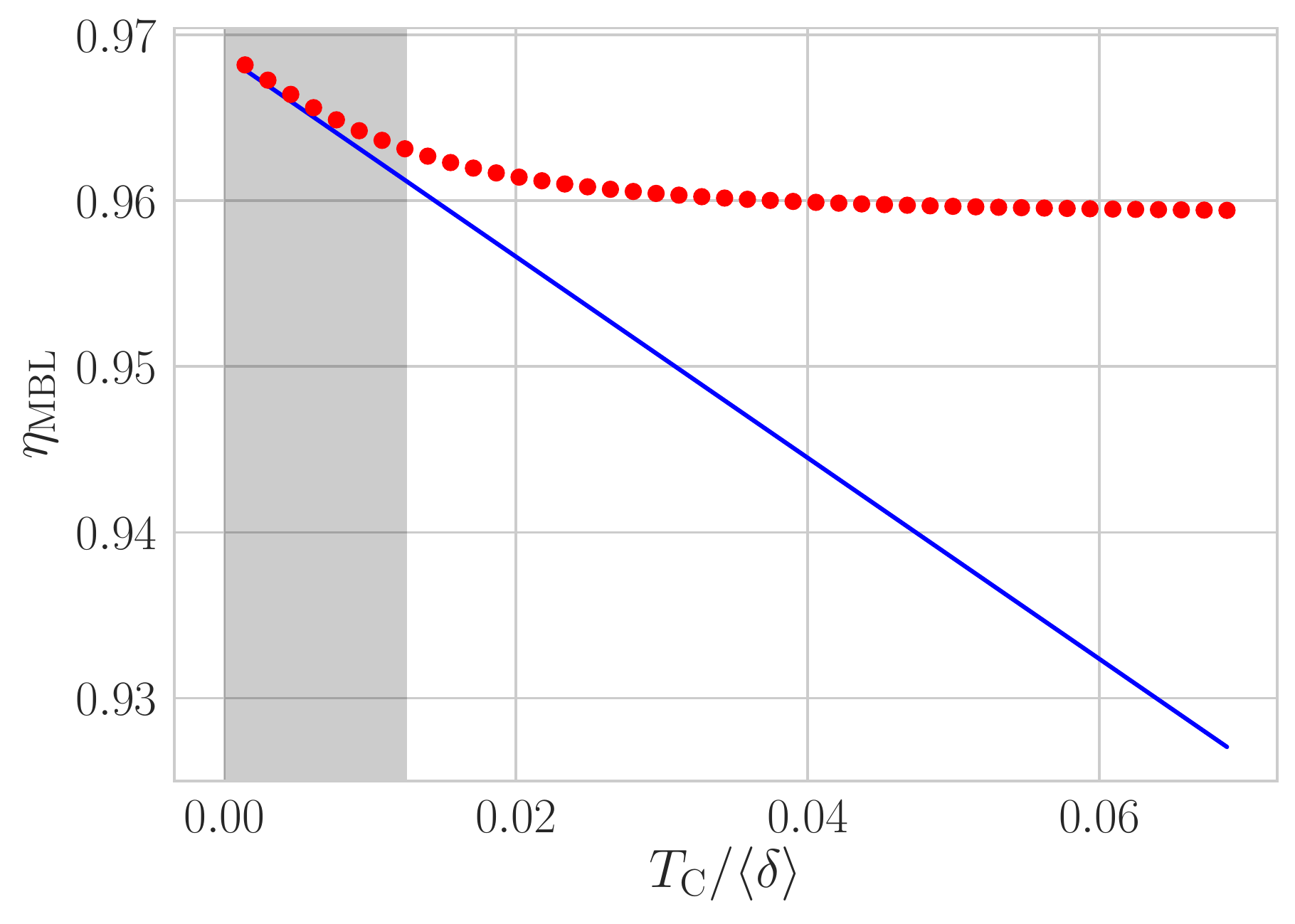}
    \caption{$\eta_\MBL$ vs. $\TCold$ at 
    $\THot = \infty$ and
    $\Wb \approx 10^{-4} \sqrt{\Sites}\HScale \approx 0.04\dAvg$}
  \end{subfigure}
  \begin{subfigure}{0.3\textwidth}
    \centering
    \includegraphics[width=\textwidth]{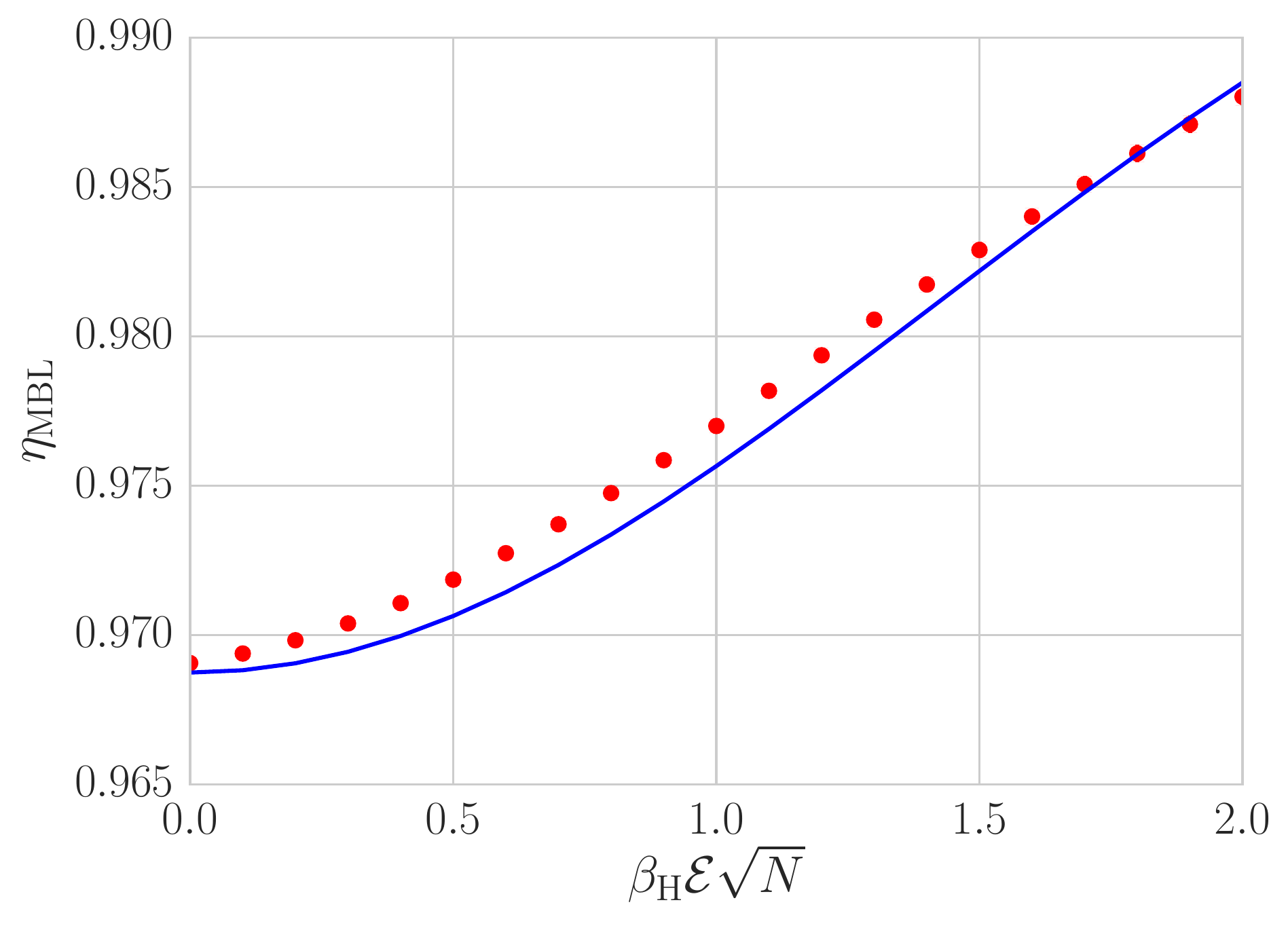}
    \caption{$\eta_\MBL$ vs. $\betaH$ 
    at $\TCold = 0$ and    $\Wb = 2^{-4}\dAvg$}
  \label{fig:Eta_betaH}
  \end{subfigure}
  \caption{\caphead{Efficiency $\eta_\MBL$ as a function of 
    (a) the cold-bath bandwidth $\Wb$, 
    (b) the cold-bath temperature $\TCold$,
    and (c) the hot-bath temperature $\THot = 1 / \betaH$:}
    The blue lines represent the analytical predictions~\eqref{eq:ManyBodyEff}
    and~\eqref{eq:phi-corrections}.
    Figure~\eqref{fig:Eta_betaH} shows 
the leading-order $\betaH$ dependence of $\eta_\MBL$,
a correction too small to include in Eq.~\eqref{eq:phi-corrections}:
$1 - \frac{ \Wb }{ 2 \dAvg }  \:  e^{ - \Sites ( \betaH \HScale )^2 / 4 }$.
See Sec.~\ref{section:Numerics_main} for 
    other parameters and definitions.
    The analytics agree with the numerics fairly well
    in the appropriate regime 
    ($ \frac{ \Wb }{ \dAvg }  \ll 1$, $\frac{ \TCold }{ \dAvg }  \ll  1$,
    and $\sqrt{ \Sites }  \: \THot \HScale  \ll  1$).
    The analytics underestimate $\eta_\MBL$;
    see the Fig.~\ref{fig:num_Q2} caption.
  }
  \label{fig:num_eta}
\end{figure}

Expressions~\eqref{eq:ManyBodyEff} and~\eqref{eq:phi-corrections}
are compared with results from numerical simulations
in Fig.~\ref{fig:num_eta}.
The analytics agree with the numerics in the appropriate regime 
($\TCold \ll \Wb \ll \dAvg$).

\section{Phenomenological model for 
the macroscopic MBL Otto engine}
\label{section:ThermoLimitApp}


The macroscopic MBL Otto engine
benefits from properties of MBL
(Sec.~\ref{section:Thermo_limit_main}),
localization and local level repulsion.
We understand these properties from 
Anderson insulators~\cite{Anderson_58_Absence} and
perturbation theory.
Anderson insulators are reviewed in Sec.~\ref{section:Anderson_Ham}.
Local level repulsion in Anderson insulators~\cite{Sivan_87_Energy} 
in the strong-disorder limit
is reviewed in Sec.~\ref{section:And_Repulsion}.
Section~\ref{section:Phenom_MBL_subapp}
extends local level repulsion to MBL.
Local level repulsion's application to the MBL engine
is discussed in Sec.~\ref{section:Repuls_Eng}.
Throughout this section, $\Sites$ denotes
the whole system's length.

\subsection{Anderson localization}
\label{section:Anderson_Ham}

Consider a 1D spin chain 
or, equivalently, a lattice of spinless fermions.
An Anderson-localized Hamiltonian $H_\Anderson$ 
has almost the form of Eq.~\eqref{eq:SpinHam},
but three elements are removed:
the $t$-dependence,
$Q \LParen h ( \alpha_t )  \RParen$, and
the interaction
($\bm{\sigma}_j  \cdot  \bm{\sigma}_{j + 1}$
is replaced with 
$\sigma_j^+  \,  \sigma_{j + 1}^-   +  \hc$).

Let $\ket{0}$ denote some reference state
in which all the spins point downward
(all the fermionic orbitals are empty).
In this section, we focus, for concreteness, on 
the properties of single-spin excitations 
relative to $\ket{0}$~\cite{Anderson_58_Absence,Sivan_87_Energy}.
The $\ell^\th$ excitation is represented, in fermionic notation, 
as $\sum\nolimits_x \psi_\ell(x)  \,   \sigma^+_{x} |0\rangle$. 
The single-excitation wave functions $\psi_\ell(x)$ are localized:
$x_\ell$ denotes the point at which the probability density
$| \psi_\ell (x) |^2$ peaks.
The wave function decays exponentially with the distance $|x - x_\ell |$
from the peak:
\begin{align}
   \label{eq:Eigenfxn}
   \psi_\ell(x)  \approx  \sqrt{ \frac{2}{ \xi_\Anderson } }  \;
   e^{ - | x - x_\ell | / \xi_\Anderson } \, .
\end{align}
The localization length varies with 
the Hamiltonian parameters as
\begin{align}
   \label{eq:Xi}
   \xi_\Anderson  \sim  \frac{1}{ \ln h }  
\end{align}
at large disorder, whose overall strength is $h$.

%
%
%
\subsection{Local level repulsion in Anderson insulators}
\label{section:And_Repulsion}

We begin with the infinitely localized limit, $h \rightarrow \infty$.
We take $\HScale \rightarrow 0$ to keep 
the Hamiltonian's energy scale finite. 
The hopping terms can be neglected, 
and particles on different sites do not repel.
Single-particle excitations are localized on single sites.
The site-$i$ excitation corresponds to an energy $2 \HScale h h_i$. 
Since the on-site potentials $h \cdot h_i$ are uncorrelated,
neighboring-site excitations' energies are uncorrelated.

Let us turn to large but finite $h$. 
Recall that $h \cdot h_i$ is drawn uniformly at random from $[-h, \, h]$.
The uniform distribution has a standard deviation of
$\frac{h}{ \sqrt{3} }  \gg  1 \, .$
Therefore, $h | h_i - h_{i+1}| \gg 1$
for most pairs of neighboring sites.
The hopping affects these sites' wave functions and energies weakly.
But with a probability $\sim  \frac{1}{h}$, 
neighboring sites have local fields $h \cdot h_i$ and $h \cdot h_{i+1}$ 
such that $h |h_i - h_{i+1}| \alt 1$. 
The hopping hybridizes such sites. 
The hybridization splits the sites' eigenvalues by an amount
$\sim \sqrt{h^2 (h_i - h_{i+1})^2 + \HScale^2 } \geq \HScale$. 

Consider, more generally, two sites separated by a distance $L \, .$
Suppose that the sites' disorder-field strengths are separated by $< 1/h^L$.
(The upper bound approximates the probability amplitude associated with
a particle's hopping the $L$ intervening sites).
The sites' excitation energies and energy eigenfunctions 
are estimated perturbatively.
The expansion parameter is $1 / h \, .$
To zeroth order, the energies are uncorrelated
and (because $h | h_i - h_{i + L} |  <  1 / h^L$) are
split by $< \HScale /h^L \, .$
The eigenfunctions are hybridized at order $L \, .$
The perturbed energies are split by 
$\geq  \HScale/h^L  \sim  \HScale    e^{ - L / \xi_\Anderson }  \, .$ 
[Recall that $\xi_\Anderson  \sim  1/\ln h$, by Eq.~\eqref{eq:Xi}.]

Hence eigenstates localized on nearby sites have correlated energies:
\emph{The closer together sites lie in real space, 
the lower the probability that they correspond to similar energies.}
This conclusion agrees with global Poisson statistics: 
Consider a large system of $\Sites \gg 1$ sites.
Two randomly chosen single-particle excitations 
are typically localized a distance $\sim \Sites$ apart.
The argument above implies only that 
the energies lie $> \HScale  e^{ - \Sites /\xi_\Anderson }$ apart. 
This scale is exponentially smaller (in $\Sites$) than
the average level spacing $\sim \frac{ \HScale  h }{ \Sites }$ 
between single-particle excitations.\footnote{
\label{footnote:SinglePartE}
The average level spacing between single-particle excitations 
scales as $\sim 1 / \Sites$ for the following reason.
The reference state $\ket{0}$ consists of
$\Sites$ downward-pointing spins.
Flipping one spin upward yields a single-particle excitation.
$\Sites$ single-particle-excitation states exist,
as the chain contains $\Sites$ sites.
Each site has an energy $\sim \pm \HScale  h$, to zeroth order,
as explained three paragraphs ago.
The excitation energies therefore fill 
a band of width $\sim  \HScale h \, .$
An interval $\sim \frac{ \HScale  h }{ \Sites }$
therefore separates single-particle-excitation energies,
on average.}

We can quantify more formally the influence of hybridization
on two energies separated by $\omega$
and associated with eigenfunctions 
localized a distance $L$ apart.
The \emph{level correlation function} is defined as
\begin{equation}
   \label{eq:R}
   R(L, \omega)  :=  \frac{1}{ \Sites^2 } \sum_{i, n, n'} 
   |\langle 0 | \sigma^-_i | n \rangle|^2  \,
   |\langle 0 | \sigma^-_{i+L} | n' \rangle|^2  \,
   \delta(E_n - E_{n'} - \omega) - \tilde{\DOS} (\omega)^2  \, .
\end{equation}
The spatially averaged density of states at frequency $\omega$
is denoted by $\tilde{\DOS} (\omega) := \frac{1}{ \Sites } 
\sum_n |\langle 0 | \sigma^-_i | n \rangle|^2  \,\delta(E_n - \omega)$.
$|n\rangle$ and $|n'\rangle$ denote eigenstates,
corresponding to single-particle excitations relative to $|0\rangle$, associated with energies $E_n$ and $E_{n'}$. 
In the Anderson insulator, $R(L, \omega) \approx 0$ 
when $\omega \gg \HScale    e^{ -L/\xi_\Anderson }$: 
Levels are uncorrelated when far apart in space and/or energy. 
When energies are close ($\omega \alt \HScale    e^{ -L/\xi_\Anderson }$), 
$R(L, \omega)$ is negative. These levels repel (in energy space).

\subsection{Generalization to many-body localization}
\label{section:Phenom_MBL_subapp}

The estimates above can be extended 
from single-particle Anderson-localized systems 
to MBL systems initialized in arbitrary energy eigenstates
(or in position-basis product states).
$R(L, \omega)$ is formulated
in terms of matrix elements $\langle 0 | \sigma_i^- | n \rangle$ 
of local operators $\sigma_i^-$. 
The local operators relevant to Anderson insulators 
have the forms of 
the local operators relevant to MBL systems.
Hence $R(L, \omega)$ is defined for MBL
as for Anderson insulators. 
However, $\ket{0}$ now denotes a generic many-body state.

Let us estimate the scale $\J_L$ of 
the level repulsion between MBL energies,
focusing on exponential behaviors.
The MBL energy eigenstates result from
perturbative expansions about Anderson energy eigenstates.
Consider representing the Hamiltonian as a matrix $\mathcal{M}$
with respect to the true MBL energy eigenbasis.
Off-diagonal matrix elements
couple together unperturbed states.
These couplings hybridize the unperturbed states,
forming corrections.
The couplings may be envisioned as
rearranging particles throughout a distance $L$.

MBL dynamics is unlikely to rearrange particles 
across considerable distances, due to localization.
Such a rearrangement is
encoded in an off-diagonal element 
$\mathcal{M}_{ij}$ of $\mathcal{M}$.
This $\mathcal{M}_{ij}$ must be small---suppressed exponentially in $L$.
$\mathcal{M}_{ij}$ also forces
the eigenstates' energies apart,
contributing to level repulsion~\cite[App.~F]{NYH_17_MBL}.
Hence the level-repulsion scale is suppressed exponentially in $L$:
\begin{align}
   \label{eq:J_L}
   \J_L  \sim  \HScale  e^{-L/\zeta }  \, ,
\end{align}
for some $\zeta \, .$
At infinite temperature, $\zeta$ must $<  \frac{1}{ \ln 2 }$ 
for the MBL phase to remain stable~\cite{mbmott}.
Substituting into Eq.~\eqref{eq:J_L} yields $\J_L  <  \frac{ \HScale }{ 2^L }$.
The level-repulsion scale is smaller than the average gap.

The size and significance of $\J_L$ depend on 
the size of $L$.
At the crossover distance $\xi$,
the repulsion $\J_L$ (between 
energy eigenfunctions localized a distance $\xi$ apart)
becomes comparable to
the average gap $\sim \frac{ \HScale }{ 2^{\xi} }$ between 
the eigenfunctions in the same length-${\xi}$ interval:
$\HScale  e^{ - {\xi} / \zeta }  
\sim  \frac{1}{e}  \,  \frac{\HScale}{ 2^{\xi} } \, .$
Solving for the crossover distance yields
\begin{align}
   \label{eq:Xi_MBL}
   \xi  \sim  \frac{1}{ \frac{1}{ \zeta }  -  \ln 2 } \, .
\end{align}
Relation~\eqref{eq:Xi_MBL} provides a definition
of the MBL localization length $\xi \, .$
[This $\xi$ differs from the Anderson localization length $\xi_\Anderson$,
Eq.~\eqref{eq:Xi}.]
Solving for $\zeta$ yields
\begin{align}
   \label{eq:Zeta_xi}
   \zeta  \sim  \frac{1}{ \frac{1}{ \xi }  +  \ln 2 }  \, .
\end{align}

The MBL Otto cycle involves two localization lengths
in the thermodynamic limit.
In the shallowly localized regime, 
$\xi  =  \xi_\Loc \, .$
Each eigenfunction has significant weight on $\xi_\Loc \approx 10$ sites,
in an illustrative example.
In the highly localized regime, $\xi  =  \xi_\VeryLoc \, .$
Eigenfunctions peak tightly: 
$\xi_\VeryLoc  \approx  1 \, .$

Suppose that the particles are rearranged across a large distance $L  \gg  \xi$.
The level-repulsion scale
\begin{align} 
   \label{eq:JFarExpn}  
   & \boxed{ \JFar 
   \sim  \HScale  e^{ - L / \xi }  \;  2^{ - L }  }  \, .
\end{align}
In the MBL engine's very localized regime,
in which $\xi = \xi_\VeryLoc$,
if $L = \xi_\Loc$ equals one subengine's length,
$\JFar  =  \deltaMBL$.

Now, suppose that particles are rearranged across a short distance
$L  \lesssim  \xi$.
Random-matrix theory approximates this scenario reasonably
(while slightly overestimating the level repulsion).
We can approximate the repulsion between 
nearby-eigenfunction energies with
the average gap $\dAvg^\LL$ in the energy spectrum of a length-$L$ system:
\begin{align}
   \label{eq:JCloseExpn}  \boxed{
   \JClose   \sim    \dAvg^\LL
   \sim   \frac{ \HScale }{ 2^L }  } \, .
\end{align}

\subsection{Application of local level repulsion 
to the MBL Otto engine in the thermodynamic limit}
\label{section:Repuls_Eng}

Consider perturbing an MBL system locally.
In the Heisenberg picture, 
the perturbing operator spreads across a distance 
$L(t) \sim \zeta \ln (\HScale t)$~\cite{Nandkishore_15_MBL}.
(See also~\cite{Khemani_15_NPhys_Nonlocal}.)
The longer the time $t$ for which the perturbation lasts,
the farther the influence spreads.

Consider tuning the Hamiltonian infinitely slowly,
to preclude diabatic transitions: $t \to \infty \, .$
Even if the Hamiltonian consists of spatially local terms,
the perturbation to each term
spreads across the lattice.
The global system cannot be subdivided into 
independent subengines. 
The global system's average gap vanishes
in the thermodynamic limit: $\dAvg  \to  0 \, .$
Since $\expval{ W_\tot }  \sim  \Wb  \ll  \dAvg$,
the per-cycle power seems to vanish in the thermodynamic limit:
$\Wb \to 0$.

Now, consider tuning the Hamiltonian at a finite speed $v$.
Dimensional analysis suggests that
the relevant time scale is $t  \sim  \frac{ \HScale }{ v } \, .$
Local perturbations affect a region of length
$\sim  L( \HScale / v ) \sim \zeta \ln (\HScale^2/v)$. 
On a length scale $L( \HScale / v )$,
global level correlations govern the engine's performance 
less than local level correlations do,
i.e., less than $R \LParen L( \HScale / v ), \omega \RParen$ does. 
This correlator registers level repulsion 
at a scale independent of $\Sites$. 
Finite-speed tuning 
renders finite the average gap accessible to 
independent subengines,
the $\dAvg$ that would otherwise close in the thermodynamic limit.
Each mesoscale subengine therefore outputs $\expval{ W_\tot } > 0 \, .$

We can explain the gap's finiteness differently:
Suppose that the engine's state starts some trial 
with weight on the $j^\th$ energy level.
The eigenenergies wiggle up and down during stroke 1.
The $j^\th$ energy may approach the $(j - 1)^\th$.
Such close-together energies likely correspond to far-apart subengines.
If the levels narrowly avoided crossing,
particles would be rearranged across a large distance.
Particles must not be, as subengines must function independently.
Hence the engine must undergo a diabatic transition:
The engine's state must retain its configuration.
The engine must behave as though 
the approaching energy level did not exist.
Effectively removing the approaching level from 
the available spectrum effectively creates a gap in the spectrum.
One can create such an effective gap 
(can promote such diabatic transitions)
by tuning the Hamiltonian at a finite $v$.

\section{Constraint 2 on cold thermalization:
Suppression of high-order-in-the-coupling energy exchanges}
\label{section:Tau_therm_virtual_app}


Section~\ref{section:Times_main} introduces 
the dominant mechanism by which 
the bath changes a subengine's energy.
The energy changes by an amount $\sim \Wb$,
at a rate $\sim \coupling$.
Higher-order processes can change the subengine energy
by amounts $> \Wb$ and operate at rates $O ( \coupling^\ell )$, 
wherein $\ell \geq 2$.
The subengine should thermalize across just small gaps
$\delta \leq \Wb$.
Hence the rate-$\coupling^\ell$ processes must operate much more slowly
than the rate-$\coupling$ processes:
$\coupling$ must be small.
We describe the higher-order processes, 
upper-bound $\coupling$, and lower-bound $\tau_\therm$.

The higher-order processes can be understood as follows.
Let $H_\tot  =  H_\macro(t)  +  H_\bath  +  H_\inter$
denote the Hamiltonian that governs the engine-and-bath composite.
$H_\tot$ generates the time-evolution operator
$U(t)  :=  e^{ -i H_\tot t }$.
Consider Taylor-expanding $U(t)$.
The $\ell^\th$ term is suppressed in $\coupling^\ell$,
contains $2\ell$ fermion operators $c_j$ and $c_{j'}^\dag$,
and contains $\ell$ boson operators $b_\omega$ and $b_{\omega'}^\dag$.
This term encodes 
the absorption, by the bath, of $\ell$ energy quanta of sizes $\leq \Wb$.
The subengine gives the bath a total amount $\sim \ell \Wb$ of heat.
The subengine should not lose so much heat.
Hence higher-order processes should occur much more slowly
than the rate-$\coupling$ processes:
\begin{align}
   \label{eq:High_ord_1}
   \tau_\HighOrd  \gg  \tau_\therm  \, .
\end{align}

Let us construct an expression for the left-hand side.
Which processes most urgently require suppressing?
Processes that change the subengine's energy by $\gtrsim \dAvg$.
Figure~\ref{fig:Compare_thermo_Otto_fig} illustrates why.
If the right-hand leg has length $\gtrsim \dAvg$,
the right-hand leg could be longer than the left-hand leg.
If it were, the trial would yield net negative work, $W_\tot < 0$.
The bath would absorb energy $\dAvg$ from a subengine
by absorbing $\sim \frac{ \dAvg }{ \Wb }$ packets
of energy $\sim \Wb$ each.
Hence the bath would appear to need to flip $\sim L  =  \frac{ \dAvg }{ \Wb }$ spins
to absorb energy $\sim \dAvg$.
(We switch from fermion language to spin language for convenience.)
However, the length-$L$ spin subchain
has a discrete effective energy spectrum.
The spectrum might lack a level associated with the amount
$\text{(initial energy)} - \dAvg$ of energy.
If so, the bath must flip more than $\frac{ \dAvg }{ \Wb }$ spins---local
level correlations suggest 
$\sim \xi_\Loc$ spins (App.~\ref{section:ThermoLimitApp}).
Hence $L  =  \max \left\{  \frac{ \dAvg }{ \Wb } ,  \xi_\Loc  \right\}$.
Energy is rearranged across the distance $L$
at a rate $\propto  \coupling^L$.

Having described the undesirable system-bath interactions,
we will bound $\coupling$ via Fermi's Golden Rule, 
Eq.~\eqref{eq:FGR_Main}.
Let $\Gamma_{fi}  \sim  1 / \tau_\HighOrd$ now denote 
the rate at which 
order-$g^L$ interactions occur.
The bath DOS remains $\DOS_\bath ( E_{if} )  \sim  \frac{1}{ \Wb }$.
Let us estimate the matrix-element size $| \langle f | V | i \rangle |$.
The bath flips each spin at a rate $\coupling$
(modulo a contribution from the bath's DOS).
Flipping one spin costs an amount $\sim \HScale$ of energy, on average.
[$\HScale$ denotes the per-site energy density,
as illustrated in Eq.~\eqref{eq:SpinHam}.]
Hence $L$ spins are flipped at a rate 
$\sim  \HScale \left( \frac{ \coupling }{ \HScale } \right)^L$.
The initial $\HScale$ is included for dimensionality.
We substitute into Fermi's Golden Rule [Eq.~\eqref{eq:FGR_Main}],
then solve for the time:
\begin{align}
   \label{eq:High_ord_2}
   \tau_\HighOrd  
   \sim  \frac{ \Wb  \,  
   \HScale^{ 2 \left( L  -  1 \right) } }{
   \coupling^{ 2 L } }  \,
   \quad \text{wherein} \quad
   L  =  \max \left\{  \frac{ \dAvg }{ \Wb } ,  \:  \xi_\Loc  \right\}  \, .
\end{align}

We substitute from Eqs.~\eqref{eq:High_ord_2} and~\eqref{eq:Tau_therm_main}
into Ineq.~\eqref{eq:High_ord_1}.
Solving for the coupling yields
\begin{align}
   \label{eq:High_ord_3} 
   \coupling  \ll  \HScale\cdot \left(\frac{\deltaMBL}{\HScale}\right)^{ 1 / ( L - 1)}   \, 
   \quad \text{wherein} \quad
   L  =  \max \left\{\frac{ \dAvg }{ \Wb } ,  \:  \xi_\Loc  \right\}  \, .
\end{align}
Substituting back into Eq.~\eqref{eq:Tau_therm_main} yields 
a second bound on $\tau_\therm$:
\begin{align}
   \label{eq:High_ord_4}  \boxed{
   \tau_\therm  \gg \frac{ \Wb}{\deltaMBL^2} 
   \left(  \frac{ \HScale }{ \deltaMBL  }  \right)^{1 / (L - 1) }  \, ,
   \quad \text{wherein} \quad
   L  =  \max \left\{  \frac{ \dAvg }{ \Wb } ,  \:   \xi_\Loc  \right\}   }  \, .
\end{align}

Let us express the bound in terms of localization lengths.
We set $\Wb  \sim  \frac{ \dAvg }{10}$ 
and approximate $L \pm 1 \sim  L  \sim  \xi_\Loc$.
We substitute in for $\dAvg$ from Eq.~\eqref{eq:JCloseExpn}
and for $\deltaMBL$ from Eq.~\eqref{eq:JFarExpn}:
\begin{align}
   \label{eq:High_ord_5}  \boxed{
   \tau_\therm  \gg  \frac{1}{10  \HScale}  \:
   e^{2 \xi_\Loc / \xi_\VeryLoc }  \:
   2^{ 2 \xi_\Loc }  }  \, .
\end{align}
This inequality is looser than Ineq.~\eqref{eq:Markov_main}:
The no-higher-order-processes condition 
is less demanding than Markovianity.

\section{Numerical simulations of the MBL Otto engine}
\label{section:MBLNumApp}

We simulated one 12-site mesoscale engine at half-filling.
(We also studied other system sizes, to gauge finite-size effects.)
Our code is available at \url{https://github.com/christopherdavidwhite/MBL-mobile}.
The random-field Heisenberg Hamiltonian~\eqref{eq:SpinHam}
governed the system.
We will drop the subscript from $H_\Sim(t)$.


Call the times at which the strokes end $t = \tau, \tau', \tau'',$ and $\tau'''$
(see Fig.~\ref{fig:2Level_v3}). 
For each of $N_{\rm reals} \approx 1,000$ disorder realizations, 
we computed the whole density matrix $\rho(t)$ at $t = 0, \tau, \tau', \tau'', \tau'''$. (See App.~\ref{section:MBLNumApp:adiabatic} and~\ref{section:MBLNumApp:coldtherm} for an explanation of how.) 
The engine's time-$t$ internal energy is
$E(t) = \Tr\LParen  H(t)\rho(t)  \RParen \, .$
The quantities of interest are straightforwardly
\begin{align}
  & \langle W_1\rangle = E(0) - E(\tau)  \, ,  \quad
  \langle W_3\rangle = E(\tau''') - E(\tau'')  \, ,  \\
  & \langle Q_2\rangle = E(\tau'')  - E(\tau')  \, ,  \quad \text{and} \quad
  \langle Q_4\rangle = E(0)  - E(\tau''')  \, .
\end{align}
We disorder-averaged these quantities before dividing to compute the efficiency,
$\eta_\MBL  =  1 -  \frac{ \expval{ W_1 }  +  \expval{ W_3 } }{
\expval{ Q_4 } }  \, .$

\subsection{Scaling factor}
\label{section:MBLNumApp:scale}

We wish to keep the DOS constant through the cycle. 
To fix $\DOS(E)$, we rescale the Hamiltonian by a factor $Q \LParen h ( \alpha_t )  \RParen$.
We define $Q^2  \LParen h( \alpha_t )  \RParen$ as
the disorder average of the variance of the unrescaled DOS:
\begin{align}
   \label{eq:Q2_def}
  Q^2  \LParen h( \alpha_t )  \RParen 
  &:=   \Bigg\langle
  \Bigg(\frac 1 {\HDim} \sum_{j = 1}^{\HDim} E_j^2\Bigg)
  - \Bigg( \frac 1 {\HDim}  \sum_{j = 1}^\HDim E_j\Bigg)^2
  \Bigg\rangle_{\rm disorder}   
  =
  \Bigg\langle
  \frac{1}{ \HDim }  \Tr \LParen \tilde{H}^2 ( t ) \RParen 
  - \Bigg(\frac{1}{ \HDim }\Tr \LParen \tilde{H} ( t )  \RParen  \Bigg)^2
  \Bigg\rangle_{\rm disorder}   \, .
\end{align}
The $\tilde{H}(t)$ denotes an unrescaled variation on the random-field Heisenberg Hamiltonian $H(t)$ of Eq.~\eqref{eq:SpinHam}:
\begin{equation}
  \tilde{H}(t) :=  \HScale  \left[
      \sum_{j = 1}^{\Sites - 1}   
      \bm{\sigma}_j  \cdot  \bm{\sigma}_{j+1}
      +  h ( \alpha_t )  \sum_{j = 1}^\Sites
      h_j     \sigma_j^z  \right]  \, .
\end{equation}

To compute $Q^2  \LParen h ( \alpha_t )  \RParen$, 
we rewrite the unrescaled Hamiltonian as
\begin{equation}
  \tilde{H}(t) =   \HScale  \left[
  2\sum_{j = 1}^{\Sites - 1} \left(  \sigma^+_j\sigma^-_{j+1} +  \hc  \right)
  + \sum_{j = 1}^{\Sites - 1} \sigma^z_j\sigma^z_{j+1} 
  + h( \alpha_t ) \sum_{j = 1}^{\Sites} h_j \sigma^z_j  \right] \, .
\end{equation}
We assume that $\Sites$ is even, and we work at half-filling.
The $\frac{ \Sites }{ 2 }$-particle subspace has dimensionality
$\HDim = {\Sites \choose \Sites/2}  \, .$

Let us calculate some operator traces that we will invoke later.
Let $X := \prod_{j = 1}^\Sites\sigma^x$ denote the global spin-flip operator.
For any operator $A$ such that $X^\dag AX = -A$,
\begin{equation}
  \Tr (A) = \Tr \left( X^\dag A X \right) = - \Tr (A)  \, .
\end{equation}
We have used the evenness of $\Sites$,
which implies the invariance of the half-filling subspace under $X$. Also,
$\Tr (A) = 0$.
In particular,
$0 = \Tr ( \sigma^z_j ) = \Tr ( \sigma^z_{j}\sigma^z_{j'}\sigma^z_{j''} )$,
if $j \ne j' \ne j''$.

Traces of products of even numbers of $\sigma^z$ factors
require more thought:
\begin{align}
   \label{eq:TraceHelp1}
  \Tr ( \sigma^z_j\sigma^z_{j+1} ) &=
      (\text{\# states $j, j+1 = \uparrow\uparrow$}) 
      + (\text{\# states $j, j+1 = \downarrow\downarrow$}) 
      - 2(\text{\# states $j, j+1 = \uparrow\downarrow$}) 
      \notag\\
      &= {{\Sites - 2} \choose {\Sites/2 - 2}} + {{\Sites - 2} \choose {\Sites/2}} - 2 {{\Sites - 2} \choose {\Sites/2 - 1}}\notag\\
      &= - \HDim \frac 1 {\Sites - 1}\, .
\end{align}
Similarly,
\begin{align}
   \label{eq:TraceHelp2}
  \Tr \left( [\sigma^+_j\sigma^-_j] [\sigma^-_{j+1}\sigma^+_{j+1}]  \right) 
  &= \Tr \left( [\sigma^-_j\sigma^+_j] [\sigma^+_{j+1}\sigma^-_{j+1}]  \right) 
  = (\text{\# states $j, j+1 = \uparrow\downarrow$}) 
  = {{\Sites - 2} \choose {\Sites/2 - 1}}  \\
  &= \HDim \frac{\Sites}{4(L-1)} \, ,
\end{align}
and 
\begin{align}
   \label{eq:TraceHelp3}
  \Tr \left( \sigma^z_j\sigma^z_{j+1}\sigma^z_{j'}\sigma^z_{j'+1} \right)
  &= (\text{\# states $j, j+1, j', j'+1 = \uparrow\uparrow\uparrow\uparrow$})
  + { 4 \choose 2}(\text{\# states $j, j+1, j', j'+1 = \uparrow\uparrow\downarrow\downarrow$})\notag\\
  &\quad+ (\text{\# states $j, j+1, j', j'+1 = \downarrow\downarrow\downarrow\downarrow$})\notag\\
  &\quad- {4 \choose 1}(\text{\# states $j, j+1, j', j'+1 = \uparrow\uparrow\uparrow\downarrow$})
  - {4 \choose 1}(\text{\# states $j, j+1, j', j'+1 = \uparrow\downarrow\downarrow\downarrow$})\notag\\
  &=  {{\Sites - 4} \choose {\Sites/2 - 4}}
  + 6 {{\Sites - 4} \choose {\Sites/2 - 2}}
  +   {{\Sites - 4} \choose {\Sites/2 }} 
  - 6 {{\Sites - 4} \choose {\Sites/2 - 3}}
  - 6 {{\Sites - 4} \choose {\Sites/2 - 1}}\notag\\
  &= \HDim \frac 3 {(\Sites - 1)(\Sites - 3)}  \, ,
\end{align}
wherein the first equality's combinatorial factors come from 
permutations on sites $j$, $j+1$, $j'$, and $j'+1$.

Assembling these pieces, we find
$\Tr \LParen \tilde{H} (t) \RParen 
  = \HScale  \sum_{j = 1}^{ \Sites -1} \Tr \left( \sigma^z_j\sigma^z_j \right) 
  = -  \HScale \HDim.$
Next, we compute $\Tr \LParen \tilde{H}^2(t) \RParen$:
\begin{align}
  \tilde{H} ^2(t)  &=  \HScale^2  \Bigg[
  4\sum_{j}^{\Sites-1} (\sigma^+_j\sigma^-_j)(\sigma^-_{j+1}\sigma^+_{j+1})
  + 4\sum_{j}^{\Sites-1} (\sigma^-_j\sigma^+_j)(\sigma^+_{j+1}\sigma^-_{j+1}) 
  + \sum_{j, j' = 1}^{\Sites-1}\sigma^z_j\sigma^z_{j+1}\sigma^z_{j'}\sigma^z_{j'+1}
  + h^2 ( \alpha_t )  \sum_{j = 1}^\Sites h_j^2
  \nonumber \\ & \qquad \quad
  + (\text{traceless terms})  \Bigg]  \\
  &=  \HScale^2  \Bigg[
  4\sum_{j}^{\Sites-1} (\sigma^+_j\sigma^-_j)(\sigma^-_{j+1}\sigma^+_{j+1})
  + 4\sum_{j}^{\Sites-1} (\sigma^-_j\sigma^+_j)(\sigma^+_{j+1}\sigma^-_{j+1})
  + \sum_{j = 1}^{\Sites-1}  \id
  + \sum_{j = 1}^{\Sites-2}\sigma^z_j\sigma^z_{j+2}
    \nonumber \\ & \qquad \quad
  + \sum_{j=1}^{\Sites-3}\sum_{j' = j+2}^{\Sites - 1}\sigma^z_j\sigma^z_{j+1}\sigma^z_{j'}\sigma^z_{j'+1} + h(\alpha_t)^2 ( \alpha_t )  \sum_{j = 1}^\Sites h_j^2
  + (\text{traceless terms})  \Bigg]  \, .
\end{align}
We take the trace, using Eqs.~\eqref{eq:TraceHelp1},~\eqref{eq:TraceHelp2}, and~\eqref{eq:TraceHelp3}:
\begin{equation}
  \Tr \LParen  \tilde{H}^2(t)  \RParen 
  = \HDim\Bigg[3\Sites - 1 + \frac{\Sites - 2}{\Sites - 1} + h^2 \sum_{j = 1}^\Sites h_j^2\Bigg] \, .
\end{equation}
We disorder-average by taking $h_j^2 \mapsto \int_0^1 dh_j h_j^2 = \frac 1 3$:
\begin{equation}
  \Big\langle\Tr (H^2(t))\Big\rangle_{\text{disorder}} = \HDim\Bigg[3\Sites - 1 + \frac{\Sites - 2}{\Sites - 1} + \Sites \frac{h^2}{3} \Bigg] \, .
\end{equation}
Substituting into Eq.~\eqref{eq:Q2_def},
we infer the rescaling factor's square:
\begin{equation}
  \label{num:rescale:disorder-averaged}
  Q^2  \LParen h( \alpha_t )  \RParen
  = 3  \Sites - 2 + \frac{\Sites - 2}{\Sites - 1} + \Sites \frac{h^2}{3}  \, .
\end{equation}

Our results are insensitive to the details of $Q$.
The width of the DOS in one disorder realization will differ from the disorder average~\eqref{num:rescale:disorder-averaged}. 
Moreover, that difference will vary as we tune $h(\alpha_t)$, 
because the disorder affects only one term.
The agreement between the analytics,
in which $\DOS (E)$ is assumed to remain constant in $t$,
and the numerics is therefore comforting:
The engine is robust against small variations in the rescaling.

\subsection{Representing states and Hamiltonians}
We structured our software to facilitate a possible extension:
The cold bath might be modeled more realistically,
as coupling to the engine only locally.

We represent the state of one mesoscopic MBL Otto engine with a density matrix
$\rho \in \mathbb{C}^{\HDim\times\HDim} \, ,$
and the Hamiltonian with a matrix
$H \in \mathbb{C}^{\HDim\times\HDim} \, ,$
relative to the basis 
$\Set{ \ket{ s_1 } , \ldots,  \ket{ s_{\HDim }  } } 
=  \Set{  \ket{ \uparrow \ldots \uparrow } ,  \ldots, 
\ket{ \downarrow \ldots \downarrow } }$ 
of products of $\sigma^z$ eigenstates.
We track the whole density matrix, rather than just the energy-diagonal elements, with an eye toward the coherent superpositions that diabatic corrections create.
For an $\Sites$-site chain at half-filling,
$\HDim = {\Sites \choose \Sites/2} 
  \simeq \sqrt{\frac 2 {\pi \Sites}}  \:  2^{\Sites}  \, .$

  \subsection{Strokes 1 and 3: Tuning}
  Simulating diabatic evolution requires a different strategy from simulating adiabatic evolution. We describe the latter in Sec.~\ref{section:MBLNumApp:adiabatic} and the former in Sec.~\ref{section:MBLNumApp:diabatic}.
  
\subsubsection{Adiabatic evolution}
\label{section:MBLNumApp:adiabatic}

The $(l, m )$ entry of the initial-state density matrix is
\begin{equation}
  \rho(0)_{lm} =   \bra{ s_l }  \frac 1 Z e^{-\betaH H(0)}   \ket{ s_m }
  = \frac 1 Z \sum_j e^{-\betaH E_j (0) }  
  \braket{s_l}{E_j(0)}\braket{E_j(0)}{s_m}  \, .
\end{equation}
The $j^\th$ eigenstate of $H(0)$, associated with energy $E_j(0)$,
is denoted by $\ket{E_j(0)}$.
We approximate the time evolution from $0$ to $\tau$ (during stroke 1) as adiabatic.
The evolution therefore does not move weight between levels:
\begin{equation}
  \rho(\tau)_{lm} = \frac{1}{ Z } 
  \sum_j e^{-\betaH E_j (0)} \braket{s_l}{E_j(\tau)}
  \braket{E_j(\tau)}{s_m}  \, .
\end{equation}
If we represented our density matrix relative to an instantaneous energy eigenbasis, simulating the time evolution would be trivial: We would reinterpret the diagonal matrix $\rho$ as being diagonal, with the same elements in a new basis.
However, we wish to represent $\rho(t)$ relative to 
the $\sigma_j^z$ product basis.
This representation enhances the code's flexibility, facilitating the inclusion of diabatic evolutions and a more detailed model of cold thermalization.
To represent $\rho(t)$ relative to the $\sigma_j^z$ product basis, we note that 
\begin{equation}
  \rho(\tau)_{lm} = \sum_{j} \braket{s_l}{E_j(\tau)} \bra{E_j(0)}\rho(0)\ket{E_j(0)} \braket{E_j(\tau)}{s_m}
  = [U(\tau,0)\rho(0) U(\tau,0)^\dag]_{lm}  \, .
\end{equation}
We have defined a time-evolution matrix
$U(\tau, 0) \in \mathbf{C}^{\HDim\times\HDim}$ by
$U(\tau, 0)_{lm} = \sum_j\braket{s_l}{E_j(\tau)} \braket{E_j(0)}{s_m}  \, .$
This matrix is easily computed via
exact diagonalization of $H(0)$ and $H( \tau )$. 

We can compute the density matrix $\rho( \tau'' )$ at the end of stroke 3 
(the tuning from MBL to GOE) 
from the density matrix $\rho ( \tau' )$ at the end of stroke 2 
(the cold-bath thermalization) similarly:
$\rho(\tau'') = U(\tau'', \tau') \rho(\tau') U(\tau'', \tau')^\dag  \, .$
The time-evolution matrix $U(\tau'', \tau') \in \mathbf{C}^{\HDim\times\HDim}$ 
is given by
$U(\tau'', \tau')_{lm} = \sum_j\braket{s_l}{E_j(0)} \braket{E_j(\tau)}{s_m}  \, .$
[Recall that $H ( \tau'' ) = H(0)$ and $H ( \tau' ) = H ( \tau )$.]

\subsubsection{Diabatic (finite-time) evolution}
\label{section:MBLNumApp:diabatic}

We simulate a stepwise tuning, taking
\begin{equation}
  \alpha(t) =  \frac{  \delta t  \,  \lfloor t/\delta t \rfloor  }{T}  \, ,
\end{equation}
wherein  $\delta t$ denotes a time-step size and 
$T \propto (h_\MBL - h_{\mathrm{GOE}})/v$ 
denotes the total tuning time.  
To do this, we compute a time-evolution unitary for the whole stroke
by chaining together the unitaries for each time step. For stroke 1,
\begin{equation}
  U(\tau,0; v,\delta t) 
  =  e^{-iH(\tau - \delta t)\delta t} 
  e^{-iH(\tau - 2\delta t)\delta t} \dots 
  e^{-iH(0)\delta t}  \, ,
\end{equation}
with the number of time steps set by the speed.
We use the time step $\delta t = 0.405 \dAvg$, 
but our results are not sensitive to the time step's size.

\begin{figure}[h]
  \centering
  \includegraphics[width=0.6\textwidth]{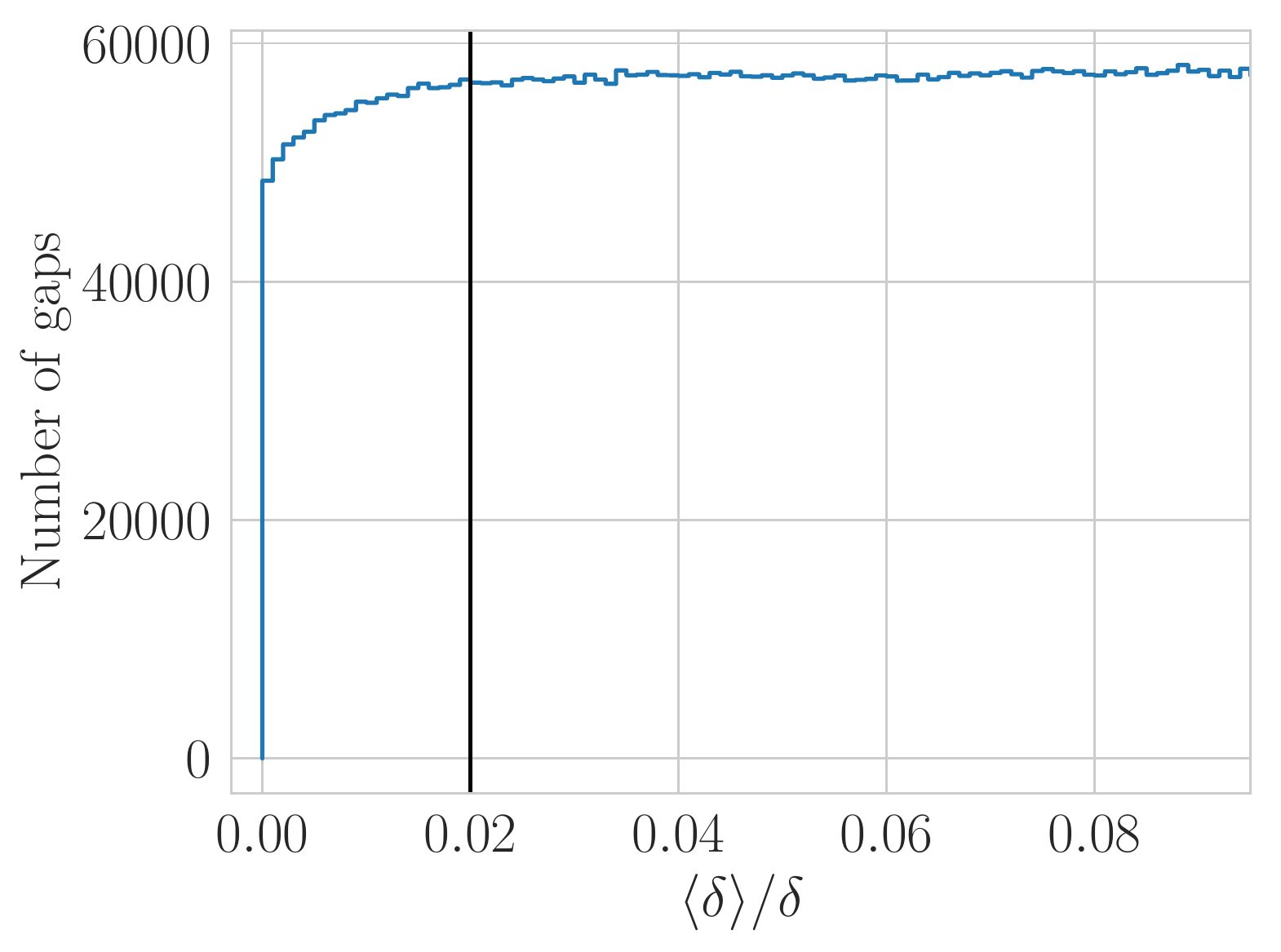}
  \caption{Level-spacing distribution for $10^6$ disorder realizations of the random-field Heisenberg model at disorder width $h = 20$ and 
  system size $\Sites = 8$ (blue line). 
  The vertical black line shows the estimate of the level-repulsion parameter $\deltaMBL$.}
  \label{fig:estimate-delta-}
\end{figure}

In judging the engine's effectiveness at a finite $v$, 
we must estimate the level-repulsion scale $\deltaMBL$.
We do this by diagonalizing $10^6$ disorder realizations 
at the relevant disorder width, $h = 20$, for $\Sites = 8$ sites. 
A histogram of the gaps is plotted in Fig.~\ref{fig:estimate-delta-}.
We then visually estimate the point at which the distribution turns over. 
Our results are not sensitive to this value.

\subsection{Stroke 2: Thermalization with the cold bath}
\label{section:MBLNumApp:coldtherm}

During stroke 2, the system thermalizes with 
a bandwidth-$\Wb$ cold bath.
We make three assumptions.
First, the bandwidth cutoff is hard:
The bath can transfer only amounts $< \Wb$ of energy at a time. 
Therefore, the cold bath cannot move probability mass 
between adjacent levels separated by just one gap $\delta' > \Wb$. 
Second, the bath is Markovian.
Third, the system thermalizes for a long time.
The bath has time to move weight across 
sequences of small gaps $\delta'_j, \delta'_{j + 1}, \ldots < \Wb$.

We can implement thermalization as follows.
First, we identify sequences of levels connected by small gaps.
Second, we reapportion weight amongst the levels 
according to a Gibbs distribution.

%
%
\begin{figure}[tb]
\centering
\includegraphics[width=.4\textwidth, clip=true]{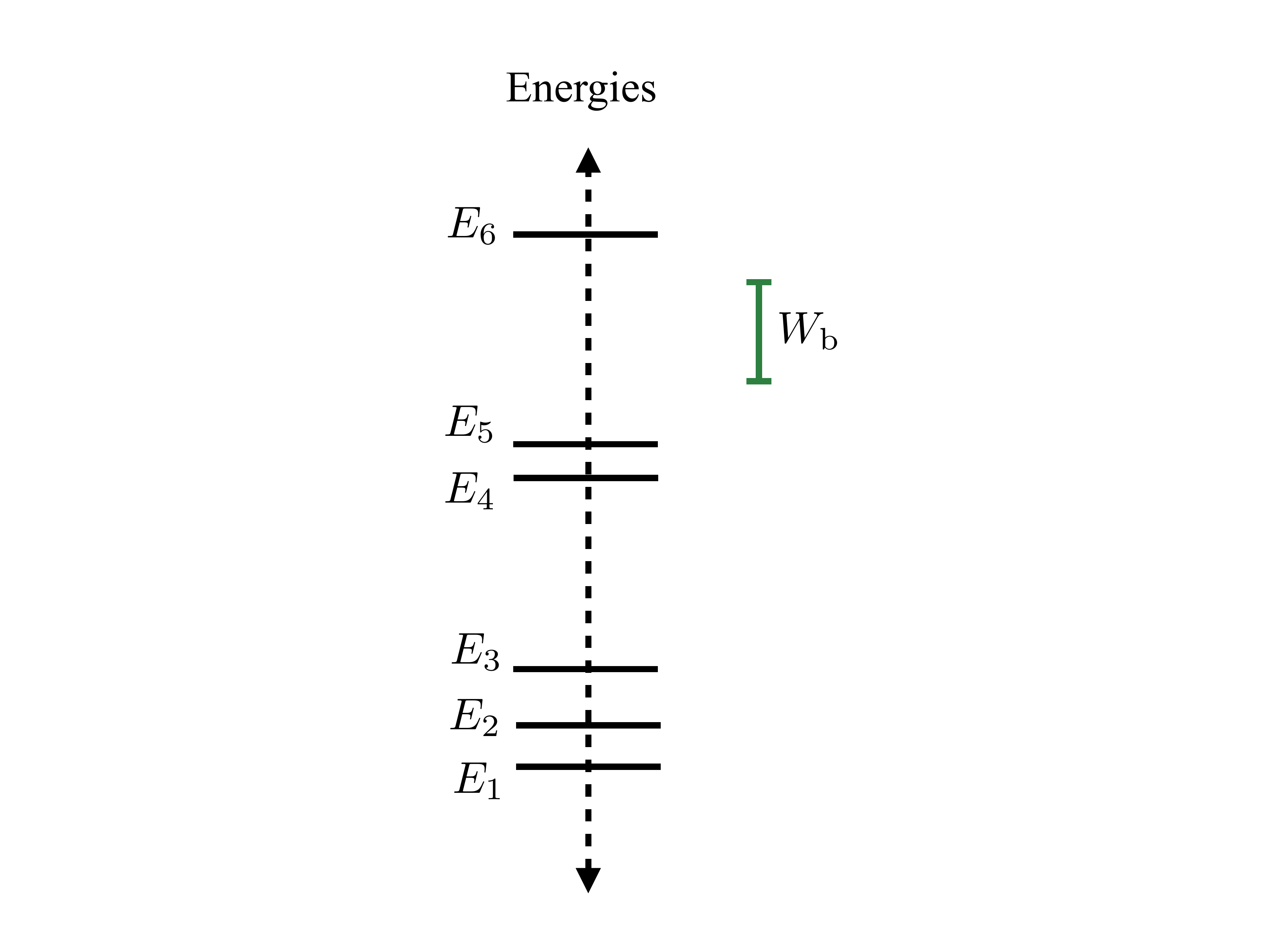}
\caption{\caphead{Energies of a cold-thermalized many-body-localized system:}
We illustrate our implementation of cold thermalization
with this example chain of six energies.
The cold bath has a bandwidth of size $\Wb$, depicted in green.}
\label{fig:Therm_ex}
\end{figure}

Suppose, for example, that the MBL Hamiltonian $H ( \tau )$
contains the following chain of six energies, $E_1, \dots, E_6$,
separated from its surrounding levels by large gaps (Fig.~\ref{fig:Therm_ex}):
\begin{align}
  (E_2 - E_1), (E_3 - E_2) < \Wb  \, ,  \quad
  (E_5 - E_4) < \Wb  \, ,  \quad  \text{and}   \quad
  (E_4 - E_3), (E_6 - E_5) > \Wb  \, . 
\end{align}
We suppress the time arguments to simplify notation.
Before thermalization, 
the density operator is diagonal with respect to the energy basis:
$\rho ( \tau ) = \sum_j \rho_j \ketbra{E_j}{E_j} \, .$
The weight on level $j$ is denoted by $\rho_j$.
Thermalization maps
\begin{align}
  \rho ( \tau ) \mapsto \rho ( \tau' ) & =
      \frac{\rho_1 + \rho_2 + \rho_3}
           {e^{-\betaC E_1} + e^{-\betaC E_2} + e^{-\betaC E_3}}
           \Big(     e^{-\betaC E_1} \ketbra{E_1}{E_1}
                 + e^{-\betaC E_2} \ketbra{E_2}{E_2}
                 + e^{-\betaC E_3} \ketbra{E_3}{E_3}  \Big)  \notag\\
      & \quad +
      \frac{\rho_4 + \rho_5}
           {e^{-\betaC E_4} + e^{-\betaC E_5}}
           \Big(     e^{-\betaC E_4} \ketbra{E_4}{E_4}
               + e^{-\betaC E_5} \ketbra{E_5}{E_5}  \Big) 
      + \rho_6 \ketbra{E_6}{E_6}  \, .
\end{align}

\section{Comparisons with competitor Otto engines}
\label{section:CompetitorApp}


This appendix contains further analysis of the bandwidth engine 
(Sec.~\ref{section:Bandwidth_engine_app})
and introduces an MBL engine tuned between
equal-strength disorder realizations (Sec.~\ref{section:DisorderEngine}).
Section~\ref{section:DisorderEngine} compares with
an MBL engine thermalized with an ordinary-bandwidth cold bath.
The quantum-dot and Anderson-localized engines 
are elaborated on in Apps.~\ref{section:Q_dot_main} and~\ref{section:Anderson_engine_main}.

\subsection{Comparison with bandwidth engine}
\label{section:Bandwidth_engine_app}

Imagine eliminating the scaling factor $Q \LParen h ( \alpha_t ) \RParen$
from the Hamiltonian~\eqref{eq:SpinHam}.
The energy band is compressed and expanded
as the disorder strength $h( \alpha_t )$ is ramped down and up.
The whole band, rather than a gap,
contracts and widens as in Fig.~\ref{fig:Compare_thermo_Otto_fig},
between a size $\sim \HScale \Sites_\macro \, h ( \alpha_0 )$
and a size $\sim  \HScale \Sites_\macro  \,  h ( \alpha_1 )
\gg  \HScale \Sites_\macro \, h ( \alpha_0 )$.
The engine can remain in one phase throughout the cycle.
The cycle does not benefit from 
the ``athermality'' of local level correlations.

Furthermore, this accordion-like motion requires 
no change of the energy eigenbasis's form.
Tuning may proceed quantum-adiabatically:
$v \approx 0$.
The ideal engine suffers no diabatic jumps, losing
$\expval{ W_\diab }_\macro  =  0$.
 
But this engine is impractical:
Consider any perturbation $V$ that fails to commute with 
the ideal Hamiltonian $H(t)$: $[V, H(t)]  \neq  0$.
Stray fields, for example, can taint an environment.
As another example, consider cold atoms in an optical lattice.
The disorder strength is ideally $\HScale h ( \alpha_t )$.
One can strengthen the disorder by
strengthening the lattice potential $U_{\rm lattice}$.
Similarly, one can raise the hopping frequency (ideally $\HScale$)
by raising the pressure $p$.
Strengthening $U_{\rm lattice}$ and $p$
while achieving the ideal disorder-to-hopping ratio
$\frac{ \HScale h(\alpha_t) }{ \HScale }  =  h ( \alpha_t )$
requires fine control.
If the ratio changes from $h ( \alpha_t )$, 
the Hamiltonian $H(t)$ acquires a perturbation $V$
that fails to commute with other terms.

This $V$ can cause diabatic jumps
that cost work $\expval{ W_\diab }_\macro$.
Can the bandwidth engine not withstand several hops---say, 
through $0.02 \HDim_\macro$ levels?

No, because the ground state pulls away from the rest of the spectrum 
as $\Sites_\macro$ grows.
Suppose, for simplicity, that $\TCold = 0$ and $\THot = \infty$.
The bandwidth engine starts stroke 1 
in $\rho(0)  =  \id / \HDim_\macro$.
Diabatic hops preserve $\rho(t)$ during stroke 1, on average:
The engine as likely hops upward as drops.
Cold thermalization drops the engine to the ground state
(plus an exponentially small dusting of higher-level states).
The ground-state energy is generically extensive.
Hence the engine absorbs 
$\expval{ Q_2 }_\macro  \sim  - \Sites_\macro$,
on average.
Suppose that, during stroke 3, the engine jumps up 
through 2\% of the levels.
The engine ends about two standard deviations 
below the spectrum's center,
with average energy $\sim \sqrt{\Sites_\macro }$.
While returning to $\THot = 0$ during the average stroke 4, 
the bandwidth engine absorbs
$\expval{ Q_4 }_\macro  \sim  \sqrt{ \Sites_\macro }$.
The average outputted work 
$\expval{ W_\tot }_\macro  =  \expval{Q_4}_\macro  +  \expval{ Q_2 }_\macro  
\sim  \sqrt{ \Sites_\macro }  -  \Sites_\macro$.
As $\Sites_\macro$ grows, 
$\expval{ W_\tot }_\macro$ 
dips farther below zero.
A few diabatic jumps threaten the bandwidth engine's 
ability to output $\expval{ W_\tot } > 0$.

The bandwidth engine's $v$ must decline 
as $\Sites_\macro$ grows also because 
the typical whole-system gap 
$\dAvg_\macro \sim  \frac{ \HScale }{ \HDim_\macro }$ shrinks.
The smaller the gaps, the greater the likelihood
that a given $v$ induces hops.
As $\dAvg_\macro \to 0$, $v$ must $\to 0$.
The MBL Otto cycle proceeds more quickly,
due to subengines' parallelization.

\subsection{Comparison with MBL engine tuned
between same-strength disorder realizations}
\label{section:DisorderEngine}

Take our MBL Otto cycle, and vary not the disorder strength, 
but the disorder realization during each cycle.
The disorder strength $h(\alpha_t)$ in Eq.~\eqref{eq:SpinHam}
would remain $\gg 1$ and constant in $t$,
while the random variables $h_j$ would change.
Let $\tilde{\Sys}$ denote this constant-$h (\alpha_t)$ engine,
and let $\Sys$ denote the MBL engine.
$\tilde{\Sys}$ takes less advantage of MBL's ``athermality,''
as $\tilde{\Sys}$ is not tuned between 
level-repelling and level-repulsion-free regimes.

Yet $\tilde{\Sys}$ outputs the amount $\expval{ W_\tot }$ of work
outputted by $\Sys$ per cycle, on average.
Because $\Wb$ is small, cold thermalization drops $\tilde{\Sys}$
across only small gaps $\delta' \ll  \dAvg$.
$\tilde{\Sys}$ traverses a trapezoid, 
as in Fig.~\ref{fig:Compare_thermo_Otto_fig}, in each trial.
However, the MBL engine has two advantages:
greater reliability and fewer worst-case (negative-work-outputted) trials.

Both the left-hand gap $\delta$ 
and the right-hand gap $\delta'$ 
traversed by $\tilde{\Sys}$ are Poisson-distributed.
Poisson-distributed gaps more likely assume extreme values
than GOE-distributed gaps:
$P_\MBL^\ParenE ( \delta )  >  P_\ETH^\ParenE ( \delta )$
if $\delta \sim 0$ or $\delta \gg  \dAvg$~\cite{D'Alessio_16_From}.
The left-hand gap $\delta$ traversed by $\Sys$ is GOE-distributed.
Hence the $W_\tot$ outputted by $\tilde{\Sys}$ 
more likely assumes extreme values
than the $W_\tot$ outputted by $\Sys$.
The greater reliability of $\Sys$ may suit $\Sys$ better
to ``one-shot statistical mechanics''~\cite{Dahlsten_11_Inadequacy,delRio_11_Thermo,Aberg_13_Truly,Horodecki_13_Fundamental,Dahlsten_13_Non,Egloff_15_Measure,Brandao_15_Second,Gour_15_Resource,YungerHalpern_16_Beyond,Gour_17_Quantum,Ito_16_Optimal,vanderMeer_17_Smoothed}.
In one-shot theory, predictability of the work $W_\tot$ 
extractable in any given trial
serves as a resource.

Additionally, $\Sys$ suffers fewer worst-case trials than $\tilde{\Sys}$.
We define as \emph{worst-case} a trial in which 
the engine outputs net negative work, $W_\tot < 0$.
Consider again Fig.~\ref{fig:Compare_thermo_Otto_fig}.
Consider a similar figure that depicts the trapezoid traversed
by $\tilde{\Sys}$ in some trial.
The left-hand gap, $\delta$, is distributed as the right-hand gap, $\delta'$, is,
according to $P_\MBL^\ParenE ( \delta )$.
Hence $\delta$ has a decent chance
of being smaller than $\delta'$: $\delta < \delta'$.
$\tilde{\Sys}$ would output $W_\tot < 0$
in such a trial.

Suppose, for simplicity, that 
$T_\HTemp = \infty$ and $T_\CTemp = 0$.
The probability that any given $\Sys$ trial outputs $W_\tot < 0$ is
\begin{align}
   \label{eq:PWorstS0}  
   p_\worst  & \approx  
   \text{(Prob. that the left-hand gap $<$ the right-hand gap)} 
   \\ \nonumber & \quad \times
   \text{(Prob. that the right-hand gap is small enough 
   to be cold-thermalized)}   \\
   \label{eq:PWorstS}
   & \approx  
   \text{(Prob. that the left-hand gap $< \Wb$)}  \times
   \frac{ \Wb }{ \dAvg } \, .
\end{align}
The initial factor is modeled by the area of a region
under the $P_\ETH^\ParenE (\delta)$ curve.
The region stretches from $\delta = 0$ to $\delta = \Wb$.
We approximate the region as a triangle
of length $\Wb$ and height $\frac{\pi}{2}  \,  \frac{ \Wb }{ \dAvg^2 }  \,
e^{ - \frac{ \pi }{ 4 }  \,  \left( \Wb \right)^2 / \dAvg^2 }
\sim  \frac{ \Wb }{ \dAvg^2 }$,
[$\delta \approx \Wb$, 
Eq.~\eqref{eq:P_ETH_Main}, and
$\frac{ \Wb }{ \dAvg }  \ll  1$].
The triangle has an area of 
$\frac{1}{2}  \cdot   \Wb  \cdot
\frac{\pi}{2}  \,  \frac{ \Wb }{ \dAvg^2 }
\sim  \left(  \frac{ \Wb }{ \dAvg }  \right)^2$.
Substituting into Eq.~\eqref{eq:PWorstS} yields
\begin{align}
   \label{eq:PWorstS2}
   p_\worst  
   \sim   \left(  \frac{ \Wb }{ \dAvg }  \right)^3  \, .
\end{align}

Let $\tilde{p}_\worst$ denote
the probability that any given $\tilde{S}$ trial
outputs $W_\tot < 0$.
$\tilde{p}_\worst$ shares the form of Eq.~\eqref{eq:PWorstS}.
The initial factor approximates to
the area of a region under the $P_\MBL^\ParenE (\delta)$ curve.
The region extends from $\delta = 0$ to $\delta = \Wb$.
The region resembles a rectangle
of height $P_\MBL^\ParenE (0)  \approx \frac{1}{ \dAvg }$.
Combining the rectangle's area, $\frac{ \Wb }{ \dAvg }$,
with Eq.~\eqref{eq:PWorstS} yields
\begin{align}
   \label{eq:PWorstSPrime}
   \tilde{p}_\worst  &    \sim  
   \left(  \frac{ \Wb }{ \dAvg }  \right)^2 \, .
\end{align}
Since $\frac{ \Wb }{ \dAvg }  \ll  1$,
$p_\worst  \ll  \tilde{p}_\worst \, .$\footnote{
The discrepancy is exaggerated if
the exponent in Eq.~\eqref{eq:PWorstS2} rises,
if the left-hand $\Sys$ Hamiltonian
is modeled with a Gaussian ensemble other than the GOE.
The Gaussian unitary ensemble (GUE) 
contains an exponent of 4;
the Gaussian symplectic ensemble (GSE),
an exponent of 6.
Different ensembles model different symmetries.}

%
%
%

\subsection{Quantum-dot engine}
\label{section:Q_dot_main}

Section~\ref{section:Order_main} introduced the quantum-dot engine,
an array of ideally independent bits or qubits. 
We add to the order-of-magnitude analysis 
two points about implementations' practicality.
First, the MBL potential's generic nature offers an advantage.
MBL requires a random disorder potential $\{ h ( \alpha_t ) h_j \}$, e.g., 
a ``dirty sample,'' a defect-riddled crystal.
This ``generic'' potential contrasts with the pristine background
required by quantum dots.
Imposing random MBL disorder is expected to be simpler.
On the other hand, a quantum-dot engine does not necessarily need
a small-bandwidth cold bath, $\Wb \ll \dAvg$.

\subsection{Anderson-localized engine}
\label{section:Anderson_engine_main}

Anderson localization follows from
removing the interactions from MBL
(App.~\ref{section:ThermoLimitApp}).
One could implement our Otto cycle with an Anderson insulator
because Anderson Hamiltonians exhibit 
Poissonian level statistics~\eqref{eq:P_MBL_Main}.
But strokes 1 and 3 would require 
the switching off and on of interactions.
Tuning the interaction, as well as the disorder-to-interaction ratio,
requires more effort than tuning just the latter.

Also, particles typically interact in many-body systems.
MBL particles interact; Anderson-localized particles do not.
Hence one might eventually expect less difficulty 
in engineering MBL engines 
than in engineering Anderson-localized engines.

\end{appendices}

\end{document}